\documentclass[twocolumn,preprintnumbers]{revtex4}
\usepackage{amssymb}
\usepackage{amsmath}
\usepackage{amsfonts}
\usepackage{graphicx}
\usepackage{dcolumn}
\usepackage{bm}

\newcommand{\be}{\begin{equation}}
\newcommand{\ee}{\end{equation}}
\newcommand{\bea}{\begin{eqnarray}}
\newcommand{\eea}{\end{eqnarray}}
\newcommand{\bes}{\begin{subequations}\begin{eqnarray}}
\newcommand{\ees}{\end{eqnarray}\end{subequations}}

\begin{document}

\title{Variational Study of Fermionic and Bosonic Systems with Non-Gaussian
States: Theory and Applications}
\author{Tao Shi$^{1}$, Eugene Demler$^{2}$, and J. Ignacio Cirac$^{1}$}
\affiliation{$^{1}$Max-Planck-Institut f\"{u}r Quantenoptik, Hans-Kopfermann-Strasse. 1,
85748 Garching, Germany \\
$^{2}$Department of Physics, Harvard University, 17 Oxford st., Cambridge,
MA 02138}
\date{\today }

\begin{abstract}
We present a new variational method for investigating the ground state and
out of equilibrium dynamics of quantum many-body bosonic and fermionic
systems. Our approach is based on constructing variational wavefunctions
which extend Gaussian states by including generalized canonical
transformations between the fields. The key advantage of such states
compared to simple Gaussian states is presence of non-factorizable
correlations and the possibility of describing states with strong
entanglement between particles. In contrast to the commonly used canonical
transformations, such as the polaron or Lang-Firsov transformations, we
allow parameters of the transformations to be time dependent, which extends
their regions of applicability. We derive equations of motion for the
parameters characterizing the states both in real and imaginary time using
the differential structure of the variational manifold. The ground state can
be found by following the imaginary time evolution until it converges to a
steady state. Collective excitations in the system can be obtained by
linearizing the real-time equations of motion in the vicinity of the
imaginary time steady-state solution. Our formalism allows us not only to
determine the energy spectrum of quasiparticles and their lifetime, but to
obtain the complete spectral functions and to explore far out of equilibrium
dynamics such as coherent evolution following a quantum quench. We
illustrate and benchmark this framework with several examples: a single
polaron in the Holstein and Su-Schrieer-Heeger models, non-equilibrium
dynamics in the spin-boson and Kondo models, the superconducting to charge
density wave phase transitions in the Holstein model.
\end{abstract}

\pacs{}
\maketitle

\section{Introduction}

Many areas of physics face their greatest challenges in understanding
quantum many-body systems in and out of equilibrium. This includes quark
confinement in QCD, non-equilibrium superfluidity in neutron stars, and
strongly correlated electron systems in condensed matter physics. Many
powerful techniques have been developed to analyze strongly correlated
many-body systems, including path integral approach and Feynman diagrams,
effective field theories, large-$N$ expansion, and renormalization group
approach. These methods have been successfully applied to a broad range of
problems, but in many cases the mathematical complexity of the theoretical
techniques makes it difficult to see clearly the underlying physical
phenomena. Hence variational solutions, which allow to unveil fundamental
physical mechanisms with relatively simple wavefunctions, have always been
considered particularly valuable. Variational wavefunctions have been
successfully applied to understand such important physical phenomena as
Bose-Einstein condensation (BEC) \cite{GPE,reviewStringari},
Superconductivity (SC) \cite{BCS}, Superfluidity \cite%
{Originalvariationalapproach}, Quantum Magnetism\cite{RVB}, and the Integer
\cite{Originalpaper} and Fractional \cite{Tsui,Laughlin} Quantum Hall effect
(IQHE and FQHE). Remarkably variational wavefunctions made it possible to
understand not only the ground state properties of these systems but in many
cases also their out of equilibrium dynamics. Choosing appropriate
variational states is a delicate issue however: on the one hand, they should
be sufficiently general to reveal fundamental physical properties, on the
other hand, their structure should be simple enough that one can efficiently
perform many-body computations. This last point strongly restricts the set
of states one can use, as in general the determination of physical
properties starting from a wavefunction requires resources (computation time
/ memory) that increase exponentially with the number of constituents.

Gaussian states \cite{reviewQIT} constitute one of the most successful
variational families. They are given by the exponentials of quadratic
functionals of creation and annihilation operators of the original fields,
and are defined for both bosonic and fermionic systems. They are
characterized by $O(N^{2})$ parameters, where $N$ is the number of modes,
although in the presence of symmetries (e.g. translational invariance) this
number can be dramatically reduced. For Gaussian states the expectation
values of physical observables can be efficiently computed as they obey
Wick's theorem \cite{Wick}, which allows one to reexpress expectation values
of an arbitrary product of mode operators in terms of products of pairs. The
Gross-Pitaevski equation \cite{GPE,reviewStringari,GP} describing BEC and
the dynamics of the condensate is based on a Gaussian wavefunction that is
an exponential of linear functions of mode operators. Gaussian states form
the basis of the BCS theory of SC and have been applied not only to describe
the ground state but also to understand the nature of the phase transition
into the broken symmetry phase as well as the non-equilibrium dynamics of
the order parameter \cite{GLdynamics,Levitov,Yuzbashyan}. Gaussian
variational techniques can be applied to spin models by transforming them to
bosonic or fermionic systems using Holstein-Primakoff \cite{HP},
Jordan-Wigner (Schwinger) \cite{JW,JS}, or the slave boson (fermion)
transformation \cite{SB}. This makes it possible to investigate phenomena
such as (anti-) ferromagnetism, para- and dia-magnetism. Furthermore, the
linearization of time-dependent equations of motion around the Gaussian
state approximating the BCS ground state gives rise to the Bogoliubov-de
Gennes theory for the low energy excitations, which can be used to describe
a large variety of phenomena in superconductors.

While the Gaussian approach has been successful in describing a broad range
of problems, it also has some important limitations. For example, starting
with a model of interacting electrons and phonons it is not possible to use
directly a Gaussian state of electrons and phonons to describe the BCS type
superconducting state. The Gaussian state is factorizable between the
electron and phonon degrees of freedom and can not describe correlations
between electrons and phonons, crucial for understanding phonon induced
attraction between electrons. Only after integrating out the phonons to
obtain a model with explicit electron-electron attraction one can introduce
a familiar BCS type wavefunction. Another important class of systems which
can not be described directly with Gaussian states are Luttinger liquids of
interacting fermions in one dimension. One needs to perform a bosonization,
which can be understood as introducing collective bosonic degrees of
freedom, which then makes it possible to represent the ground state as a
bosonic Gaussian state. The list of important beyond-Gaussian states goes
much longer and includes FQHE systems, the Kondo model and spin-boson
systems \cite{QIM}, and ultracold fermions close to unitarity in the BCS-BEC
crossover regime \cite{UR,Tan1,Tan2,Tan3}. In all these cases, Gaussian
states can not be used directly to describe these paradigmatic many-body
systems because they do not contain sufficient entanglement between
consituent particles. This is particularly striking in the case of
boson-fermion mixtures (either in cold atoms or in the context of
electron-phonon systems), where the Gaussian state is a product of bosonic
and fermionic wavefunctions.

The primary goal of the current paper is to introduce a broad class of
variational wavefunctions, which exhibit strong entanglement between
different microscopic degrees of freedom, yet retain most of the simplicity
of Gaussian wavefunctions. In short, the idea of our approach is to perform
generalized canonical transformations, which introduce correlations and
entanglement between particles. The nature of the appropriate
transformations depends on the system at hand, but the general form is
inspired by canonical transformations in condensed matter physics, including
polaron transformations in electron-phonon systems and flux attachment in
FQHE. After performing the transformation we introduce generalized Gaussian
states. An important new feature of our wavefunctions, which makes them
different from all earlier work, is that we treat the parameters of both the
original unitary transformation and the Gaussian wavefunction as
variational. This gives them sufficient flexibility to describe a variety of
strongly correlated many-body systems ranging from quantum impurity
problems, over to electron-phonon models, and to FQHE systems. Our
variational wavefunctions are well suited to find accurate approximations to
ground states, determine collective excitations, and to describe out of
equilibrium dynamics for certain problems. The simple form of our
variational wavefunctions makes them easy to apply to real problems. We
derive the equations of motion of the variational parameters in imaginary
time for calculating the ground state, and in real time for describing out
of equilibrium dynamics. Expectation values of physical observables can also
be easily computed. We demonstrate the viability of our approach by applying
it to several concrete examples. We analyze the single-polaron problem in
electron-phonon systems described by the Holstein and Su-Schrieffer-Heeger
(SSH) models, and show that the ansatz accurately describes all of the known
physical phenomena. We analyze the spin-boson model (which is directly
related to the Kondo model), and show that our ans\"{a}tz provides an
improvement over methods that have been previously used in the literature
\cite{Silbey,spinco,Hutchison}. In particular we describe the
non-equilibrium quench dynamics of the Kondo impurity spin in the
ferromagnetic regime with easy plane anisotropy. We demonstrate the
existence of finite time crossovers in the dynamics which are consistent
with the equilibrium renormalization group flow. To our knowledge, non
equilibrium dynamics of the Kondo problem in this regime has not been
studied previously. Compared to the usual antiferromagnetic or SU(2)
symmetric ferromagnetic cases this regime requires analyzing the coherent
evolution over longer times, which makes it very challenging to conventional
numerical techniques such as the numerical renormalization group (NRG)
calculation \cite{NRG}. Finally, we analyze the SC-charge density wave (CDW)
phase transition in the Holstein model. We also comment on possible
applications of our formalism to FQHE systems and Luttinger liquids.

This paper is organized as follows. We begin by presenting a general
methodology for studying quantum many-body systems using variational
wavefunctions in Sec. \ref{NGS}. The key element of our method is
understanding the differential structure of the variational manifold \cite%
{tdvp}. In order to describe the ground state, we derive a set of
differential equations for the variational parameters. They correspond to
the projection of the evolution in imaginary time onto the variational
manifold, which ensures that the energy is a monotonically decreasing
function of time. We use a similar technique to describe the real-time
evolution, which ensures the conservation of energy (in time-independent
problems) and other constants of motion. To find the ground state one needs
to solve the equations of motion in imaginary time until the system reaches
a fixed point. By linearizing the equations of motion around the fixed point
of the imaginary time evolution, we derive a set of equations that describe
the low-energy excitations of the theory projected onto the tangent plane of
the variational manifold, so that response and spectral functions can be
computed. This extends the standard Bogoliubov technique to the more general
families of variational states.

To make the presentation more accessible, we first illustrate the method
based on projected equations of motion in the simple case of Gaussian
states. We then introduce canonical transformations to obtain non-Gaussian
states and show how they can be analyzed using projected equations of
motion. The first class of states that we consider is inspired by the flux
attachment idea in the FQHE systems \cite{FA1,FA2,FA3}. It can be used to
describe composite fermions \cite{CF} in FQHE, fragmented condensates \cite%
{Cederbaum1,Cederbaum2}, and entangles boson-fermion mixtures in ultracold
atoms.

In Sec. \ref{SP}, we use a non-Gaussian family inspired by the Lee-Low-Pines
(LLP) transformation \cite{LLP}, and benchmark the variational approach with
a single polaron problem in electron-phonon interacting systems. We analyze
the polaron dispersion in the SSH model using imaginary time evolution and
demonstrate that our approach correctly describes the phase transition in
which the minimum of the dispersion moves from zero to finite momentum \cite%
{SSH}. We also use the real-time evolution method in combination with the
Wei-Norman algebra method (Appendix \ref{AppendixWN}) to determine the
single polaron spectral function.

In Sec. \ref{SB}, we apply our method to the ground state and real-time
dynamics of the spin-boson (Kondo) model. We employ two non-Gaussian
families motivated by the parity conservation and polaron transformations
\cite{Silbey,polaronsc}. In contrast to previous approaches \cite%
{Silbey,spinco} our wavefunctions are constructed using general Gaussian
states, and thus include squeezing as well as displacement of the bosonic
bath modes. This gives us a better value of spin magnetization in the ground
state in comparison to the Silbey-type transformation approach \cite{Silbey}%
. Our results are in excellent agreement with the results of the NRG
calculations \cite{NRG}. Furthermore, we study the spin relaxation and the
dynamics of the bath degrees of freedoms in several regimes of the Kondo
model, and point out a correspondence between nonequilibrium coherent
dynamics and renormalization group flows for equilibrium systems.

In Sec. \ref{SCCDW}, we consider a family of non-Gaussian states inspired by
the polaron transformation and apply the variational principle to study the
transition between the SC and CDW phases in the Holstein model. In agreement
with earlier studies we find that the transition takes place at
half-filling. We determine the SC and CDW order parameters, the density
distributions of the electrons, the displacements and covariances of phonon
modes in both phases.

In Sec. \ref{CO}, we summarize our results and discuss promising directions
for future studies.

\section{Non-Gaussian state approach \label{NGS}}

In this section we develop a variational theory to describe the ground state
and dynamics of a many-body system composed of bosons, fermions, or both. We
consider a system that consists of $N_{b}$ bosonic and $N_{f}$ fermionic
modes. The modes are described in terms of creation and annihilation
operators, $b_{j},b_{j}^{\dagger }$ ($j=1,\ldots ,N_{b}$) for the bosons,
and $c_{j},c_{j}^{\dagger }$ ($j=1,\ldots ,N_{f}$) for the fermions. These
operators fulfill canonical commutation and anti-commutation relations,
respectively. We will use $n_{j}^{b}=b_{j}^{\dagger }b_{j}$ and $%
n_{j}^{f}=c_{j}^{\dagger }c_{j}$ to denote the corresponding number
operators. We will find it convenient to use the quadrature operators, $%
x_{j}=b_{j}+b_{j}^{\dagger }$, $p_{j}=i(b_{j}^{\dagger }-b_{j})$ for the
bosons, and the Majorana operators, $a_{1,j}=c_{j}^{\dagger }+c_{j}$, $%
a_{2,j}=i(c_{j}^{\dagger }-c_{j})$ for the fermions, where the (anti-)
commutation relations are $[x_{i},p_{j}]=2i\delta _{ij}$ and $\{a_{\alpha
,i},a_{\beta ,j}\}=2\delta _{\alpha \beta }\delta _{ij}$, and $\{\ldots \}$
denotes the anti-commutator. In order to shorten the notation we will
collect these operators in column vectors,
\begin{subequations}
\begin{eqnarray}
R &=&(x_{1},\ldots ,x_{N_{b}},p_{1},\ldots ,p_{N_{b}})^{T},  \notag \\
C &=&(c_{1},\ldots ,c_{N_{f}},c_{1}^{\dagger },\ldots ,c_{N_{f}}^{\dagger
})^{T},  \notag \\
A &=&(a_{1,1},\ldots ,a_{1,N_{f}},a_{2,1},\ldots ,a_{2,N_{f}})^{T},
\end{eqnarray}%
where $T$ denotes the transpose. We will denote by $|0\rangle $ the vacuum
state, i.e. the state that fulfills $b_{j}|0\rangle =c_{j}|0\rangle =0$.

We assume that dynamics of the system is described by the Hamiltonian ${H}$
and coupling to external reservoirs can be neglected. Our goal is to find
variational approximations to the wavefunction and energy of the ground
state of this Hamiltonian, as well as efficient description of the system
dynamics. We consider a family of states, $\left\vert \Psi (\xi
)\right\rangle $, where $\xi $ is a short-hand notation for $\xi _{1},\xi
_{2},\ldots $, the set of variational parameters. We will assume that all
the states in the family are normalized, i.e.,
\end{subequations}
\begin{equation}
\langle \Psi (\xi )|\Psi (\xi )\rangle =1
\end{equation}%
for all possible values of the variational parameters $\xi $. Thus, the goal
is to find $\xi _{\mathrm{G}}$ or $\xi (t)$ such that the corresponding
state in the family approximates the ground state or its dynamics,
respectively. In the following we will use the variational principle to
derive a set of differential equations that will allow us to solve these
problems. Once the approximation to the ground state is found, we can
linearize the equations of motion around this state and obtain low energy
excitations.

We begin by summarizing basic features of the generalized Gaussian states, $%
|\Psi _{\mathrm{GS}}\rangle $. They are defined as states that can be
written as exponentials of up to 2-degree polynomials of bosonic mode
operators (note that this includes linear terms), as well as quadratic
fermionic mode operators, acting on the vacuum. By construction such states
are factorizable between fermions and bosons and therefore contain no
correlations between the two types of particles. Furthermore, the
correlation functions fulfill the conditions of Wick's theorem: all
correlation functions can be reduced to combinations of the products of one-
and two- point correlation functions. Intrinsic limitations of the Gaussian
wavefunctions strongly constrain us in the type of states that we can
describe with them. To avoid these limitations we consider a more general
family of wavefunctions

\begin{equation}
|\Psi _{\mathrm{NGS}}\rangle =U_{S}|\Psi _{\mathrm{GS}}\rangle ,
\label{PsiNG}
\end{equation}%
Here $U_{S}$ is a unitary operator, which we will refer to as generalized
canonical transformation. This operator depends on a set of variational
parameters and its primary role is to introduce entanglement between
different fields, in particular between bosons and fermions. Transformation (%
\ref{PsiNG}) allows us to construct states which are no longer constrained
by the Wick's theorem. A practical consideration for the choice of $U_{S}$
is that we should be able to apply variational principle on this state in an
efficient way. Most importantly we want to circumvent the exponential
dependence of the computational resources on the number of particles $%
N_{b,f} $ present in common numerical approaches such as exact
diagonalization.

This Section is divided into four subsections. In the first one we will
review variational principle for both imaginary and real time dynamics. The
former can be used to find the lowest energy state within a family of
variational states. We discuss how to analyze fluctuations around the
variational ground state, thus providing a generalization of the
Bogoliubov-de Gennes theory. Section \ref{Gaussian state} reviews
application of this general technique in the simple case of Gaussian
wavefunctions. While most of these results have been obtained in the
literature before, we provide this discussion in order to make the paper
self-contained and to set the stage for subsequent analysis of non-Gaussian
states. In Sec. \ref{family}, we construct several families of non-Gaussian
states, and show that time-dependent variational method can be efficiently
applied to these states as well. In the last Section we will summarize the
procedure which one needs to follow in order to apply analysis based on
non-Gaussian wavefunctions to a specific problem.

\subsection{Time-dependent variational principle \label{VariationalPrinciple}%
}

In this subsection, we review the time-dependent variational principle \cite%
{tdvp} that will be used throughout the paper. At each infinitesimal
time-step we project the evolution of the wavefunction onto the subspace
tangent to the manifold defining the variational family of states. While
this method is standard for describing the evolution in real time, we also
consider imaginary-time evolution as a way of obtaining a variational
approximation to the ground state within the family $\left\vert \Psi (\xi
)\right\rangle $. At the end of the subsection we will also derive equations
that approximate the low energy dynamics around the variational ground
state. This procedure allows us to obtain elementary excitations around the
ground state.

\subsubsection{Imaginary-time evolution}

Here we derive a set of differential equations for finding an approximation
to the ground state within the family of variational states $\left\vert \Psi
(\xi )\right\rangle $. We first remind the readers that outside of
variational techniques a common approach to finding the ground state is to
start with some initial state $\left\vert \varphi (0)\right\rangle $ and
follow the imaginary time evolution according to
\begin{equation}
\left\vert \varphi (\tau )\right\rangle =\frac{e^{-H\tau }\left\vert \varphi
(0)\right\rangle }{\sqrt{\left\langle \varphi (0)\right\vert e^{-2H\tau
}\left\vert \varphi (0)\right\rangle }}.  \label{IM}
\end{equation}%
As long as there is a non-vanishing overlap of $|\varphi (0)\rangle $ with
the ground state of $H$ and one can accurately compute the evolution of the
wavefunction $e^{-H\tau }\left\vert \varphi (0)\right\rangle $, the ground
state will be obtained from Eq. (\ref{IM}) in the limit $\tau \rightarrow
\infty $. We follow a similar strategy for variational states and implement
the imaginary time evolution within a restricted set of states. Our goal is
to find the lowest energy state in the variational ansatz that we consider.
From Eq. (\ref{IM}) it follows that $\left\vert \varphi (\tau )\right\rangle
$ fulfills%
\begin{equation}
d_{\tau }\left\vert \varphi (\tau )\right\rangle =-(H-\left\langle
H\right\rangle )\left\vert \varphi (\tau )\right\rangle ,  \label{IMM}
\end{equation}%
where $\left\langle H\right\rangle =\left\langle \varphi (\tau )\right\vert
H\left\vert \varphi (\tau )\right\rangle $, and $d_{\tau }$ is a shorthand
notation for the derivative with respect to $\tau $.

In the variational approach we need to project Eq. (\ref{IM}) at every
time-step onto the tangent plane (with respect to the family of states'
manifold). Details of the derivation are given in Appendix \ref%
{AppendixImaginary} and here we only present the final result%
\begin{equation}
d_{\tau }\xi _{i}=\sum_{j=1}\mathbf{G}_{ij}^{-1}\left\langle \Psi
_{j}\left\vert \mathbf{R}_{\Psi }\right\rangle \right.  \label{GME}
\end{equation}%
for the variational parameters $\xi _{i}$. The Gram matrix $\mathbf{G}$ has
elements%
\begin{equation}
\mathbf{G}_{ij}=\langle \Psi _{i}|\Psi _{j}\rangle ,  \label{Gmatrix}
\end{equation}%
where $\left\vert \Psi _{j}\right\rangle =\partial _{\xi _{j}}\left\vert
\Psi (\xi )\right\rangle $ span the tangent plane of the variational state
manifold at $\xi $, which is a subspace of the full many-body Hilbert space.
The vector $\left\vert \mathbf{R}_{\Psi }\right\rangle =-(H-E)\left\vert
\Psi (\xi )\right\rangle $, where%
\begin{equation}
E=\left\langle \Psi (\xi )\right\vert H\left\vert \Psi (\xi )\right\rangle
\label{Energy}
\end{equation}%
is the mean energy of the state $\Psi (\xi )$.

In deriving Eq. (\ref{GME}) we assumed that the Gram matrix is invertible,
namely, the vectors $\Psi _{i}$ spanning the tangent plane are linearly
independent. If this is not the case, one can make $\mathbf{G}$ invertible
by keeping some of the parameters fixed. In Appendix \ref{AppendixImaginary}
we show that, since states in the family are normalized,%
\begin{equation}
d_{\tau }E=-2\langle \mathbf{R}_{\Psi }|\mathbf{P}_{\xi }|\mathbf{R}_{\Psi
}\rangle \leq 0,  \label{dEdtau}
\end{equation}%
where $\mathbf{P}_{\xi }$ is the projector onto the tangent plane. Thus, the
energy $E$ monotonically decreases and reaches a minimum in the limit $\tau
\rightarrow \infty $. We will denote by%
\begin{equation}
|\Psi _{\mathrm{G}}\rangle =\lim_{\tau \rightarrow \infty }\left\vert \Psi
(\xi (\tau ))\right\rangle ,  \label{PsiG}
\end{equation}%
the variational state obtained in that limit, which we expect to approximate
the ground state, and by $\xi _{\mathrm{G}}=\lim_{\tau \rightarrow \infty
}\xi (\tau )$, the corresponding variational parameters \cite{footnote}. The
value of $\langle \mathbf{R}_{\Psi }|\mathbf{R}_{\Psi }\rangle $ is then the
variance of the energy, which should be small if the state we reach is close
to the ground state. Thus, this quantity can be used to estimate the
accuracy of the variational family.

We reminder the readers that the system may have several local minimum, with
different basins of attractions in the imaginary time flow. They may
correspond, for example, to different types of symmetry breaking. Depending
on the initial choice of $\left\vert \varphi (0)\right\rangle $ the long
time limit of imaginary time evolution may be a local minimum, which is not
the global minimum. To find the ground state one needs to compare energies
of different stable points and identify the global minimum. Local minima
that are not global minima of the energy may be interesting in their own
right, e.g., near a first order phase transition when the system may be
"stuck" in a metastable state.

\subsubsection{Real-time evolution}

We can use a similar procedure to approximate the real-time dynamics. We
obtain Eq. (\ref{GME}), but now $\left\vert \mathbf{R}_{\Psi }\right\rangle
=-iH\left\vert \Psi (\xi )\right\rangle $ and the derivative is taken with
respect to the real time, $t$, instead of $\tau $. It follows that%
\begin{equation}
d_{t}E=\left\langle \Psi (\xi )\right\vert d_{t}H\left\vert \Psi (\xi
)\right\rangle \equiv \left\langle d_{t}H\right\rangle  \label{C}
\end{equation}%
for $E$ defined in Eq. (\ref{Energy}). When $H$ is time-independent, Eq. (%
\ref{C}) implies energy conservation, $d_{t}E=0$. In fact, it is well known
\cite{tdvp} that the evolution given by this time-dependent variational
principle is symplectic: for any operator $O$ that satisfies $[O,H]=0$, the
expectation value $\left\langle O\right\rangle _{\xi }=\left\langle \Psi
(\xi )\right\vert O\left\vert \Psi (\xi )\right\rangle $ is conserved by Eq.
(\ref{GME}), i.e., $d_{t}\left\langle O\right\rangle _{\xi }=0$. In
addition, as we show in Appendix \ref{AppendixImaginary}, the real-time
evolution of the variational ground state $|\Psi _{\mathrm{G}}\rangle $
given in (\ref{PsiG}) is
\begin{equation}
\left\vert \Psi _{\mathrm{G}}(t)\right\rangle =e^{-iE_{\mathrm{G}%
}t}\left\vert \Psi _{\mathrm{G}}\right\rangle ,
\end{equation}%
where $E_{\mathrm{G}}=\left\langle \Psi _{\mathrm{G}}\right\vert H\left\vert
\Psi _{\mathrm{G}}\right\rangle $ is the ground state energy.

\subsubsection{Fluctuations \label{fluctuations}}

We can use the differential Eq. (\ref{GME}) to study fluctuations around the
variational ground state, $\left\vert \Psi _{\mathrm{G}}\right\rangle $. Let
us consider the evolution of the system in a state whose variational
parameters are close to those of the ground state. The dynamics in this
state can be described by low energy excitations in the tangent plane of $%
\left\vert \Psi _{\mathrm{G}}\right\rangle $. In order to find the
elementary excitations we linearize Eq. (\ref{GME}) around the equilibrium
positions, $\xi _{\mathrm{G}}$; that is, $\xi =\xi _{\mathrm{G}}+\epsilon $,
so that (see Appendix \ref{AppendixImaginary})%
\begin{equation}
\mathbf{G}d_{t}\epsilon =-i\mathbf{M}\epsilon ,  \label{GBB}
\end{equation}%
where $\mathbf{M}_{ij}=\left\langle \Psi _{i}\right\vert H\left\vert \Psi
_{j}\right\rangle $.

The motion Eq. (\ref{GBB}) can be solved by introducing the vector $\eta =%
\mathbf{G}^{1/2}\epsilon $, where we used that $\mathbf{G}$ is
positive-definite, and thus has a positive square-root. It defines an
orthonormal basis in the tangent plane and satisfies%
\begin{equation}
d_{t}\eta =-i\mathbf{L}\eta ,  \label{dteta}
\end{equation}%
where the hermitian matrix $\mathbf{L}=\mathbf{G}^{-1/2}\mathbf{M\mathbf{G}}%
^{-1/2}$ can be viewed as the Hamiltonian projected onto the tangent
subspace, $H_{\mathbf{P}}$, expressed in the orthonormal basis
\begin{equation}
\left\vert V_{k}\right\rangle =\sum_{j}\left\vert \Psi _{j}\right\rangle (1/%
\sqrt{G})_{jk}.
\end{equation}%
The eigenvalues, $\mu ^{\lambda }$, and eigenvectors, $\eta ^{\lambda }$, of
$\mathbf{L}$ determine the fluctuation spectrum and the nature of the
excitations. The orthonormal basis of the tangent subspace $H_{\mathbf{P}}$
is given by
\begin{equation}
\left\vert \Psi _{\mathrm{ex}}^{\lambda }\right\rangle =\sum_{k}\eta
_{k}^{\lambda }\left\vert V_{k}\right\rangle .  \label{etalambda}
\end{equation}

When discussing states in the tangent plane, $\left\vert V_{i}\right\rangle $%
, it is important to remember that they include both collective modes and
single particle excitations. For Gaussian states they are closely related to
the Bogoliubov excitations. Hence, this theory may be viewed as a
generalization of the Bogoliubov-de Gennes equations. Equation (\ref{GBB})
indicates that the states $\left\vert V_{i}\right\rangle $ themselves are
not eigenstates of $H_{\mathbf{P}}$, and thus they will hybridize in a sense
that $\left\vert V_{i}\right\rangle \rightarrow \left\vert
V_{j}\right\rangle $, with the transition amplitude%
\begin{equation}
\mathcal{A}_{ji}(t)=\left\langle V_{j}\right\vert e^{-iH_{\mathbf{P}%
}t}\left\vert V_{i}\right\rangle =\sum_{\lambda }\eta _{j}^{\lambda }\eta
_{i}^{\lambda \ast }e^{-i\mu ^{\lambda }t}.
\end{equation}%
The spectral function of the excitation $\left\vert V_{k}\right\rangle $ can
be defined as
\begin{eqnarray}
Z_{k}(\omega ) &=&-\frac{1}{\pi }\text{Im}[-i\int_{0}^{+\infty }\mathcal{A}%
_{kk}(t)e^{i\omega t}]  \notag \\
&=&\sum_{\lambda }\left\vert \eta _{k}^{\lambda }\right\vert ^{2}\delta
(\omega -\mu ^{\lambda }).  \label{Zk}
\end{eqnarray}%
In the thermodynamics limit when the number of variational parameters is
infinite we expect to find a continuous spectrum of excitations. Peaks in $%
Z_{k}(\omega )$ can be interpreted as describing collective modes. Positions
of the peaks in $\omega$ correspond to the collective mode energies and
their widths to the inverse of the lifetimes.

\subsection{Gaussian states \label{Gaussian state}}

Mean-field theory has been one of the most successful approaches for
understanding quantum many-body systems. The two most important examples of
this approach are the Bologoliubov theory of superfluidity of weakly
interacting bosons and the BCS theory of superconductivity. At its core the
mean-field theory is a variational approach which uses a family of Gaussian
states $|\Psi _{\mathrm{GS}}\rangle $. In this subsection we summarize
time-dependent variational approach for general Gaussian states. We point
out that time-dependent Gaussian states have been discussed before to
describe a broad range of dynamical phenomena \cite{Levitov,HiggsInSC}.
While results of this section can be found in earlier literature, although
with a different notation and motivation, we present them here for
completeness and as a simple illustration of the general time-dependent
variational theory.

\subsubsection{Definition}

Gaussian states are defined as $\left\vert \Psi _{\mathrm{GS}}\right\rangle
=U_{\mathrm{GS}}\left\vert 0\right\rangle $, where the operators $U_{\mathrm{%
GS}}$ describes a unitary transformation
\begin{equation}
U_{\mathrm{GS}}=e^{i\theta }e^{i\frac{1}{2}R^{T}\sigma \Delta _{R}}e^{-i%
\frac{1}{4}R^{T}\xi _{b}R}e^{i\frac{1}{4}A^{T}\xi _{m}A}.  \label{GS}
\end{equation}%
Here, $\theta $ is a global phase, $\Delta _{R}$ is the displacement vector
of bosons, and $\xi _{b}$ ($\xi _{m}$) are (anti-) symmetric matrices, which
describe correlations of the bosonic (fermionic) modes. We also used the
simplectic matrix%
\begin{equation}
\sigma =\left(
\begin{array}{cc}
0 & {\openone}_{N_{b}} \\
-{\openone}_{N_{b}} & 0%
\end{array}%
\right) ,
\end{equation}%
where ${\openone}_{N_{b}}$ is the $N_{b}\times N_{b}$ identity matrix.

We point out that there is a gauge degree of freedom in the definition of $%
U_{\mathrm{GS}}$, since different Gaussian transformations $U_{\mathrm{GS}}$
and $U_{\mathrm{GS}}V_{\mathrm{GS}}\equiv \tilde{U}_{\mathrm{GS}}$ can
describe the same Gaussian state provided that $V_{GS}\,\left\vert
0\right\rangle =\left\vert 0\right\rangle $. The simplest choice is $V_{%
\mathrm{GS}}=e^{ib^{\dagger }\chi _{b}b}e^{ic^{\dagger }\chi _{f}c}$, where $%
\chi _{b}$, $\chi _{f}$ can be constructed using any Hermitian matrices. If
we show that $\tilde{U}_{\mathrm{GS}}=U_{\mathrm{GS}}V_{\mathrm{GS}}$ also
has a Gaussian form given by Eq. (\ref{GS}), but with different values of
the variational parameters, i.e.,
\begin{equation}
\tilde{U}_{\mathrm{GS}}=e^{i\theta }e^{i\frac{1}{2}R^{T}\sigma \Delta
_{R}}e^{-i\frac{1}{4}R^{T}\tilde{\xi}_{b}R}e^{i\frac{1}{4}A^{T}\tilde{\xi}%
_{m}A},  \label{Ut}
\end{equation}%
then we establish that different $\xi _{b/m}$ can describe the same Gaussian
state. To verify Eq.\ (\ref{Ut}) we check the condition that $\tilde{U}_{%
\mathrm{GS}}$ and $U_{\mathrm{GS}}V_{\mathrm{GS}}$ result in the same
transformations of $\delta R$ and $A$, i.e.,%
\begin{eqnarray}
\tilde{U}_{\mathrm{GS}}^{\dagger }\delta R\tilde{U}_{\mathrm{GS}} &=&V_{%
\mathrm{GS}}^{\dagger }U_{\mathrm{GS}}^{\dagger }\delta RU_{\mathrm{GS}}V_{%
\mathrm{GS}},  \notag \\
\tilde{U}_{\mathrm{GS}}^{\dagger }A\tilde{U}_{\mathrm{GS}} &=&V_{\mathrm{GS}%
}^{\dagger }U_{\mathrm{GS}}^{\dagger }AU_{\mathrm{GS}}V_{\mathrm{GS}},
\label{Re}
\end{eqnarray}%
which, Eq. (\ref{Re}), determines $\tilde{\xi}_{b,m}$ by the relation%
\begin{eqnarray}
e^{\sigma \tilde{\xi}_{b}} &=&\frac{1}{2}e^{\sigma \xi _{b}}W_{b}\left(
\begin{array}{cc}
e^{i\chi _{b}} & 0 \\
0 & e^{-i\chi _{b}^{\ast }}%
\end{array}%
\right) W_{b}^{\dagger },  \notag \\
e^{i\tilde{\xi}_{m}} &=&\frac{1}{2}e^{i\xi _{m}}W_{m}\left(
\begin{array}{cc}
e^{i\chi _{f}} & 0 \\
0 & e^{-i\chi _{f}^{\ast }}%
\end{array}%
\right) W_{m}^{\dagger }.
\end{eqnarray}%
Here, the matrix%
\begin{equation}
W_{b}=\left(
\begin{array}{cc}
{\openone}_{N_{b}} & {\openone}_{N_{b}} \\
-i{\openone}_{N_{b}} & i{\openone}_{N_{b}}%
\end{array}%
\right)  \label{W_b_definition}
\end{equation}%
relates $R$ and $B=(b_{j=1,...,N_{b}},b_{j=1,...,N_{b}}^{\dagger })^{T}$ by $%
R=W_{b}B$, and%
\begin{equation}
W_{m}=\left(
\begin{array}{cc}
{\openone}_{N_{f}} & {\openone}_{N_{f}} \\
-i{\openone}_{N_{f}} & i{\openone}_{N_{f}}%
\end{array}%
\right)
\end{equation}%
relates $A$ and $C$ by $A=W_{m}C$.

The transformations $U_{\mathrm{GS}}$ and $\tilde{U}_{\mathrm{GS}}$ related
by the unitary transformation $V_{\mathrm{GS}}$ define an equivalent class $%
\{U_{\mathrm{GS}}\}$, and the transformations in each class give the same
Gaussian state. As a result, there is some redundancy in the variational
parameters $\xi _{b,m}$. Instead of using the elements $\xi _{b,m}$ as
variational parameter, it is more convenient to use the covariant matrices
(defined below) instead. For each equivalent class $\{U_{\mathrm{GS}}\}$ the
covariant matrices are uniquely defined (note that $\Delta_R$ is also
defined unambiguously).

For bosons, Gaussian states are completely characterized by the displacement
vector $\Delta _{R}$, and the covariant matrix $\Gamma _{b}$ for the
fluctuations $\delta R=R-\Delta _{R}$, defined as
\begin{subequations}
\begin{eqnarray}
\Delta _{Ri} &=&\langle \Psi _{\mathrm{GS}}|R_{i}|\Psi _{\mathrm{GS}}\rangle
, \\
(\Gamma _{b})_{ij} &=&\frac{1}{2}\langle \Psi _{\mathrm{GS}}|\{\delta
R_{i},\delta R_{j}\}|\Psi _{\mathrm{GS}}\rangle ,
\end{eqnarray}%
where both of them take real values. Under the Gaussian state transformation
$U_{\mathrm{GS}}$, the quadrature transforms as $U_{\mathrm{GS}}^{\dagger
}R_{i}U_{\mathrm{GS}}=(\Delta _{R})_{i}+(S_{b}R)_{i}$, where the
Baker-Campbell-Hausdorff (BCH) formula has been used, and the symplectic
matrix $S_{b}=e^{\sigma \xi _{b}}$ fulfills $S_{b}\sigma S_{b}^{T}=\sigma $.
The symmetric covariance matrix $\Gamma _{b}$ fulfills
\end{subequations}
\begin{equation}
\Gamma _{b}=S_{b}S_{b}^{T}  \label{Sb}
\end{equation}%
for pure states, as it is the case here.

In the Gaussian state $\left\vert \Psi _{\mathrm{GS}}\right\rangle $
fermions are characterized by the covariance matrix
\begin{equation}
(\Gamma _{m})_{ij}=\frac{i}{2}\left\langle \Psi _{\mathrm{GS}}\right\vert
[A_{i},A_{j}]\left\vert \Psi _{\mathrm{GS}}\right\rangle .
\end{equation}%
This matrix is real and anti-symmetric, and for pure states it fulfills $%
\Gamma _{m}^{2}=-{\openone}$. By BCH formula, the Majarana operator
transform as $U_{\mathrm{GS}}^{\dagger }A_{i}U_{\mathrm{GS}}=(U_{m}A)_{i}$,
where $U_{m}=e^{i\xi _{m}}$. Since $\xi _{m}$ is an anti-symmetric Hermitian
matrix, $U_{m}$ is an orthogonal matrix. The covariance matrix is related to
$U_{m}$ through $\Gamma _{m}=-U_{m}\sigma U_{m}^{T}$. Sometimes it is more
convenient to use the original creation and annihilation operators, and
define%
\begin{equation}
\Gamma _{f}\equiv \left\langle \Psi _{\mathrm{GS}}\right\vert CC^{\dagger
}\left\vert \Psi _{\mathrm{GS}}\right\rangle =\frac{1}{2}{\openone}%
_{2N_{f}}-i\frac{1}{4}W_{m}^{\dagger }\Gamma _{m}W_{m}.  \label{Gf}
\end{equation}

Wick's theorem can be applied to all Gaussian states, so that higher order
correlations can be expressed in terms of the displacement vector and the
covariant matrices. Note that Eq. (\ref{GS}) does not introduce any
correlation between the bosons and the fermions, so that $\left\vert \Psi _{%
\mathrm{GS}}\right\rangle $ is a product state between bosons and fermions.

\subsubsection{Variational principle}

It is convenient to take the elements of $\Delta _{R}$ and $\Gamma _{b,m}$
as variational parameters. We do not provide a separate derivation of the
equations of motion for Gaussian states but only present the final results.
Readers interested in the derivation can use appendix \ref{AppendixME}, in
which we obtain equations of motion for a broader class of non-Gaussian
states defined in Eq. (\ref{PsiNG}). Gaussian states are a special case of
such states with $U_{S}=1$.

For the imaginary-time evolution we obtain%
\begin{eqnarray}
d_{\tau }\Delta _{R} &=&-\Gamma _{b}h_{\Delta },  \notag \\
d_{\tau }\Gamma _{b} &=&\sigma ^{T}h_{b}\sigma -\Gamma _{b}h_{b}\Gamma _{b},
\notag \\
d_{\tau }\Gamma _{m} &=&-h_{m}-\Gamma _{m}h_{m}\Gamma _{m},  \label{RGGSI}
\end{eqnarray}%
and for the real-time dynamics
\begin{eqnarray}
d_{t}\Delta _{R} &=&\sigma h_{\Delta },  \notag \\
d_{t}\Gamma _{b} &=&\sigma h_{b}\Gamma _{b}-\Gamma _{b}h_{b}\sigma ,  \notag
\\
d_{t}\Gamma _{m} &=&[h_{m},\Gamma _{m}].  \label{RGGSR}
\end{eqnarray}%
Here, the vector $h_{\Delta }=2\delta E/\delta \Delta _{R}$ and the matrices
$h_{b}=4\delta E/\delta \Gamma _{b}$, $h_{m}=4\delta E/\delta \Gamma _{m}$
are determined by the functional derivatives of the mean energy, $%
E=\left\langle H\right\rangle _{\mathrm{GS}}$, corresponding to the Gaussian
state. We note that equations for $\Gamma _{m}$ agree with those in Ref.
\cite{HFGS}.

Solutions of Eqs. (\ref{RGGSI}) in the limit $\tau \rightarrow \infty $
determine the Gaussian mean-field ground state. By solving Eq. (\ref{RGGSR}%
), we can study real-time dynamics in the Gaussian state manifold. Note that
in the standard mean-field (Gross-Pitaevskii) theory for bosons one uses a
coherent state to describe a system in which macroscopic number of bosons
occupy the same single particle state. In fact, Eq. (\ref{RGGSR}) is nothing
but Gross-Pitaevskii equation for the time evolution of the macroscopically
occupied state. Including $\Gamma _{b}$ as variational parameters one can
also describe a squeezed state of bosons, which is usually introduced into
the wavefunction via the Bogoliubov-de Gennes equations. Note that our
approach is more general.

\subsubsection{Fluctuations}

We continue our discussion of the Gaussian states and consider fluctuations
around the variational ground state. In the case of bosons, we have
excitations corresponding to two different directions in the tangent plane:
fluctuations obtained by taking derivatives with respect to $\Delta _{R}$
and with respect to $\Gamma _{b}$. They have the form
\begin{subequations}
\begin{eqnarray}
|V_{j}^{(1)}\rangle &=&U_{\mathrm{GS}}b_{j}^{\dagger }\left\vert
0\right\rangle ,  \label{V1} \\
|V_{ij}^{(2)}\rangle &=&U_{\mathrm{GS}}b_{i}^{\dagger }b_{j}^{\dagger
}\left\vert 0\right\rangle .  \label{V2b}
\end{eqnarray}%
The first type describes single particle excitations, and the second one
corresponds to two particle excitations. In the case of non-interacting
bosons, i.e., when the Hamiltonian $H$ is quadratic in $R$, the Gram matrix $%
\mathbf{G}$ in Eq. (\ref{Gmatrix}) does not connect the single-particle and
two-particle sectors. If we denote by $\lambda ^{(1,2)}$ the eigenvalues
corresponding to the two sectors and by $\eta ^{\lambda ^{(1,2)}}$ the
corresponding eigenvectors of the matrix $\mathbf{L}$ [see Eq. (\ref{dteta}%
)], we will have that for each $\lambda ^{(2)}$ there will exist two $%
\lambda ^{(1)}$ with $\lambda _{i,j}^{(2)}=\lambda _{i}^{(1)}+\lambda
_{j}^{(1)}$, and $\eta _{i,j}^{\lambda ^{(2)}}$, when considered as a
matrix, will have just one non-trivial singular value in its singular value
decomposition. This tells us that quasiparticles do not interact and the
energy of two quasiparticles is the sum of individual energies. In the
presence of interactions we expect that the matrix $\mathbf{L}$ connects the
two sectors, giving rise to a decay of the single particle excitations into
to two particles, something that can be characterized in terms of Eq. (\ref%
{Zk}). For example, such process can describe a decay of one Higgs amplitude
excitation in a strongly correlated superfluid state into a pair of
Goldstone modes [see e.g., \cite{Podolsky}]. The form of the eigenvectors $%
\eta ^{\lambda ^{(2)}}$ describes the nature of interactions between the two
types of fluctuations in Eqs. (\ref{V1}) and (\ref{V2b}).

For Fermions we only have two-particle excitations
\end{subequations}
\begin{equation}
|V_{ij}^{(2)}\rangle =U_{\mathrm{GS}}c_{i}^{\dagger }c_{j}^{\dagger
}\left\vert 0\right\rangle  \label{V2f}
\end{equation}%
in the tangent space. In this case, the eigenvalues and eigenvectors of $%
\eta _{i,j}^{\lambda }$, when considered as a matrix, will determine the two
particle excitations, e.g. particle-hole excitations around the Fermi sea.
For $N_{f}\gg 1$ we expect to recover the standard Bogoliubov-de Gennes
theory. Note that eigenmodes in Eq. (\ref{V2f}) also contain collective
modes, such as the the phase (Goldstone) and amplitude modes for
superconductors, spin waves for magnetically ordered states.

\subsection{Non-Gaussian states \label{family}}

As we discussed earlier, Gaussian states are not sufficiently versatile to
describe many interesting situations. In this subsection, we extend simple
Gaussian states to include a richer structure of entanglement and
correlations between modes, in particular entanglement between the bosonic
and fermionic modes. While we cannot take very general extensions, as we
need a computationally tractable description of many-body systems, we will
see that we can define several broad classes of interesting states. Our main
requirement is that we can compute efficiently quantities that appear in Eq.
(\ref{GME}), i.e., the Gram matrix $\mathbf{G}$, the energy $E$, $\langle
\Psi _{j}|H|\Psi (\xi )\rangle $, as well as $\langle \Psi (\xi )|\Psi (\xi
)\rangle $ to ensure the normalization.

The main idea is to consider states of the form (\ref{PsiNG}), and choose $%
U_{S}$ such that the above quantities can be expressed as expectation values
of operators taken in the Gaussian state, $\left\vert \Psi _{\mathrm{GS}%
}\right\rangle $. In particular operators which we need to compute in order
to solve dynamically Eq. (\ref{GME}) will be usually of the following types:
they contain polynomials of $R$ and $A$, or exponentials of some of these
polynimals. As we discuss below for several useful choices of $U_{S}$ such
expectation values can be reduced to those in the Gaussian states, which
makes it possible to compute them efficiently.

In this subsection, we will introduce five families of Non-Gaussian states $%
U_{S=1,\ldots ,5}$ for problems dealing with fermionic systems, bosonic
systems, and Bose-Fermi mixtures.

\subsubsection{Fermionic systems}

For purely fermionic systems we define%
\begin{eqnarray}
U_{1} &=&\bar{U}_{\mathrm{GS}}\,U_{\mathrm{FA}},  \notag \\
U_{\mathrm{FA}} &=&e^{i\frac{1}{2}\sum_{ij}\omega _{ij}^{f}\text{:}%
n_{i}^{f}n_{j}^{f}\text{:}},  \label{FluxAttachment}
\end{eqnarray}%
where $\bar{U}_{\mathrm{GS}}=e^{iC^{\dagger }\bar{\xi}_{f}C/2}$ and $:\ldots
:$ denotes the normal ordering with respect to the vacuum state. This
transformation has two types of new variational parameters: $\omega ^{f}$
and $\bar{\xi}_{f}$. We remind the readers that $U_{1}$ acts on the Gaussian
state itself [see Eq. (\ref{PsiNG})]. So these new parameters should be
considered together with $\xi _{m}$ introduced in Eq. (\ref{GS}). Unitarity
requires that $\omega ^{f}$ is a real symmetric matrix and $\bar{\xi}_{f}$ a
hermitian matrix. A useful way of understanding $\bar{U}_{\mathrm{GS}}$ is
that it provides a transformation from the original single particle basis to
localized Wannier orbitals. Then $U_{\mathrm{FA}}$ acts locally in space. To
understand the physical meaning of $U_{\mathrm{FA}}$ we observe that
\begin{equation*}
U_{\mathrm{FA}}^{\dagger }c_{j}U_{\mathrm{FA}}=e^{i\sum_{i}\omega
_{ij}^{f}n_{i}^{f}}c_{j}.
\end{equation*}%
This can be understood as a particle in the orbital $i$ contributing a phase
$\omega _{ij}^{f}$ to a particle in the orbital $j$. Taking $\omega
_{ij}^{f} $ of the form $n_{v}\arg (z_{i}-z_{j})$, where $n_{v}$ is an
integer even number and $z_{i}=x_{i}+iy_{i}$, we observe that this is
equivalent to the flux attachment procedure in which $n_{v}$- vortices are
attached to every fermion. Hence transformation (\ref{FluxAttachment})
enables the description of FQHE systems in the spirit of composite-fermions
\cite{FA1,FA2,FA3}. While the exact form of the energetically optimal
functions $\omega _{i-j}^{f}$ may differ, we expect that in FQHE-like
systems variational parameters $\omega _{i-j}^{f}$ develop branch cuts in
the 2D plane, such as $w_{ij}^{f}\propto n_{v}$Im$\ln (z_{i}-z_{j})$. This
should be contrasted to $\omega _{ij}^{f}$ which are analytic functions in
the 2D plane, which we expect to apply to non-topological systems with
time-reversal symmetry.

\subsubsection{Bosonic systems}

For purely bosonic systems, we can define a transformation%
\begin{equation}
U_{2}=\bar{U}_{\mathrm{GS}}\exp (i\frac{1}{2}\sum_{ij}\omega _{ij}^{b}\text{:%
}n_{i}^{b}n_{j}^{b}\text{:}),
\end{equation}%
where $\omega ^{b}$ is a real symmetric matrix, and $\bar{U}_{\mathrm{GS}%
}=e^{iR^{T}\sigma \bar{\Delta}_{R}/2}e^{-iR^{T}\bar{\xi}_{b}R/4}$ is defined
by a real vector $\bar{\Delta}_{R}$ and a real symmetric matricx $\bar{\xi}%
_{b}$. Similar to the fermionic case, this transformation can be used to
describe the flux attachment for bosons. The FQHE of bosons, i.e., the
half-filled phase in the rotating BEC systems, can be investigated using
this transformation.

We also introduce a non-unitary transformation%
\begin{equation}
U_{3}=\frac{1}{\sqrt{\mathcal{N}}}e^{\lambda P}  \label{U3}
\end{equation}%
defined by the variational parameter, $\lambda $, the parity operator $%
P=\exp (i\pi \sum_{j}b_{j}^{\dagger }b_{j})$, and the normalization factor $%
\mathcal{N}$. This transformation creates superposition%
\begin{equation}
\left\vert \Psi _{\mathrm{NGS}}\right\rangle =U_{3}\left\vert \Psi _{\mathrm{%
GS}}^{(+)}\right\rangle =\sum_{s=\pm }u_{s}\left\vert \Psi _{\mathrm{GS}%
}^{(s)}\right\rangle
\end{equation}%
of two Gaussian states%
\begin{equation}
\left\vert \Psi _{\mathrm{GS}}^{(s)}\right\rangle =e^{i\frac{1}{2}%
sR^{T}\sigma \Delta _{R}}e^{-i\frac{1}{4}R^{T}\xi _{b}R}\left\vert
0\right\rangle ,
\end{equation}%
where the amplitudes $u_{s=\pm }$ satisfy the relation $\tanh \lambda
=u_{-}/u_{+}$. In particular, in the limits $\lambda \rightarrow \pm \infty $
we have states with even and odd parity, something which is not possible
with Gaussian states alone. They correspond to certain types of fragmented
consdensates, which can be constructed as superpositions of two Gaussian
states but not as a single Gaussian state. For physical applications of such
states we refer the readers to Refs. \cite{Cederbaum1,Cederbaum2}.

\subsubsection{Bose-Fermi mixtures}

For bosons interacting with fermions we define two types of transformations%
\begin{equation}
U_{4}=e^{i\frac{1}{2}C^{\dagger }\bar{\xi}_{f}C}e^{i\frac{1}{2}%
\sum_{ij}\omega _{ij}^{f}\text{:}n_{i}^{f}n_{j}^{f}\text{:}}e^{i\sum_{ij}%
\bar{\omega}_{ij}R_{i}n_{j}^{f}},  \label{U4_equation}
\end{equation}%
and%
\begin{eqnarray}
&&U_{5}=e^{i\frac{1}{2}C^{\dagger }\bar{\xi}_{f}C}e^{i\frac{1}{2}R^{T}\sigma
\bar{\Delta}_{R}}e^{-i\frac{1}{4}R^{T}\bar{\xi}_{b}R}  \label{U5_equation} \\
&&\times \exp [i\sum_{ij}(\frac{1}{2}\omega _{ij}^{f}\text{:}%
n_{i}^{f}n_{j}^{f}\text{:}+\frac{1}{2}\omega _{ij}^{b}\text{:}%
n_{i}^{b}n_{j}^{b}\text{:}+\omega _{ij}^{bf}n_{i}^{b}n_{j}^{f})],  \notag
\end{eqnarray}%
where $\bar{\omega}$ and $\omega ^{bf}$ are real matrices.

For the special case $\bar{\omega}=0$, transformation $U_{4}$ reduces to $%
U_{1}$. For the case $\omega ^{f}=0$, transformation $U_{4}$ is the polaron
transformation used for describing electron-phonon systems. While in the
usual treatments of polaronic phenomena in electron-phonon systems
variational parameters in $\bar{\xi}_{f}$ are taken as time independent, we
will allow all parameters of $U_4$ and $U_5$ to change during the imaginary
or real time evolutions. Note that $U_{4}$ generates entanglement between
fermions and bosons. It is characterized by the non-vanishing cubic
correlations $\left\langle \delta R_{i}A_{j}A_{k}\right\rangle $. In Secs. %
\ref{SB} and \ref{SCCDW}, we will illustrate the application of this
variational polaron transformation using two concrete problems:
spin-relaxation in the spin-boson model and the SC-CDW phase transition in
the Holstein model.

The transformation $U_{5}$ can be employed to study the quantum phases of
Bose-Fermi mixtures. The entanglement properties between the bosonic and
fermionic modes are characterized by the non-trivial quartic correlations $%
\left\langle \delta R_{i}\delta R_{j}\delta R_{k}\delta R_{l}\right\rangle $%
, $\left\langle A_{i}A_{j}A_{k}A_{l}\right\rangle $, and $\left\langle
\delta R_{i}\delta R_{j}A_{k}A_{l}\right\rangle $. For the special case $%
\omega ^{bf}=0$, $U_{5}$ reduces to $U_{1}U_{2}$. A physical motivation for
introducing states described by $U_{5}$ with $\omega _{bf}\neq 0$ comes from
the Lee-Low-Pines (LLP) transformation \cite{LLP} in impurity problems. In
Sec. \ref{SP} we will illustrate how to apply the variational principle to
the family of states obtained using Eq. (\ref{U5_equation}) and study the
single polaron problem in both Holstein and SSH models. Our analysis goes
beyond computing the dispersion of the polaron and allows us to obtain
complete spectral functions. We also discuss a single-polaron phase
transition in the SSH model.

In the following we will show how one can efficiently compute quantities
that appear in the time evolution described by Eq. (\ref{GME}) for the
families of non-Gaussian states introduced in this section. Since $U_{4,5}$
already include $U_{1,2}$ we only need to carry out this task for $U_{3,4,5}$%
. Note that all states are normalized, since all the $U$s are unitary except
for $U_{3}$, for which we explicitly included the normalization factor in
Eq. (\ref{U3}).

\subsubsection{Efficient computations for non-Gaussian states \label%
{efficient}}

In order to analyze the dynamics described by Eqs. (\ref{GME}) for the three
types of non-Gaussian states, $U_{3,4,5}\left\vert \Psi _{\mathrm{GS}%
}\right\rangle $, we need to be able to compute the Gram matrix and the
overlap $\langle \Psi _{j}|\mathbf{R}_{\Psi }\rangle $. Here, we show how to
compute these quantities analytically for any Hamiltonian, $H$, that is a
polynomial \textit{poly}$(R,C)$ in terms of $R$ and $C$. We will explain the
main steps in this section, and in Appendices \ref{Appendixtangent}-\ref%
{AppendixMF} we provide a more detailed derivation.

Firstly, we show that the tangent vectors $\left\vert \Psi _{j}\right\rangle$
can be written as
\begin{equation}
U_{S}\exp [i\sum_{j}(\alpha _{j}n_{j}^{f}+\beta _{j}n_{j}^{b}+\gamma
_{j}R_{j})]\text{\textit{poly}}(R,C)\left\vert \Psi _{\mathrm{GS}%
}\right\rangle .  \label{Texp}
\end{equation}%
This is obtained as follows. We write%
\begin{equation}
\partial _{\tau }\left\vert \Psi _{\mathrm{NGS}}\right\rangle =U_{S}\left[
\left( \partial _{\tau }U_{\mathrm{GS}}\right) U_{\mathrm{GS}}^{-1}+O\right]
U_{\mathrm{GS}}\left\vert 0\right\rangle ,  \label{Ltv}
\end{equation}%
where $O=U_{S}^{-1}\partial _{\tau }U_{S}$. In Appendix \ref{Appendixtangent}%
, we prove that $(\partial _{\tau }U_{\mathrm{GS}})U_{\mathrm{GS}}^{-1}$
only contains constant, linear, and quadratic terms in the Bose and Fermi
creation and annihilation operators, and $O$ is the sum of operators of the
form%
\begin{equation}
\exp [i\sum_{j}(\alpha _{j}n_{j}^{f}+\beta _{j}n_{j}^{b}+\gamma _{j}R_{j})]%
\text{\textit{poly}}(R,C)  \label{ExpPL}
\end{equation}%
for $U_{S}=U_{3,4,5}$. As a result, the tangent vectors $\partial _{\xi
_{j}}\left\vert \Psi _{\mathrm{NGS}}\right\rangle $ are composed of terms
like those appearing in Eq. (\ref{Texp}). Note that for the special case $%
\bar{\xi}_{f,b}=\bar{\Delta}_{R}=0$ in $U_{4,5}$, the tangent vectors%
\begin{equation}
\partial _{\xi _{j}}\left\vert \Psi _{\mathrm{NGS}}\right\rangle
=U_{S}poly(R,C)\left\vert \Psi _{\mathrm{GS}}\right\rangle ,
\end{equation}%
i.e., $\alpha _{j}=\beta _{j}=\gamma _{j}=0$ in Eq. (\ref{Texp}).

Secondly, we notice that the right hand side of Eq. (\ref{GME}) is
determined by $H(R,C)\left\vert \Psi _{\mathrm{NGS}}\right\rangle $. In
Appendix \ref{Appendixtangent}, we show that for $U_{S}=U_{3,4,5}$ and the
Hamiltonian $H$ with the polynomial form \textit{poly}$(R,C)$,\ the state $%
H(R,C)\left\vert \Psi _{\mathrm{NGS}}\right\rangle $ is also composed of
terms as in Eq. (\ref{Texp}).

Finally, the Gram matrix and the overlap $\left\langle \Psi _{j}\left\vert
\mathbf{R}_{\Psi }\right\rangle \right. $ are determined by the expectation
values%
\begin{equation}
\left\langle e^{i\sum_{j}(\beta _{j}n_{j}^{b}+\gamma _{j}R_{j})}\text{poly}%
(R)\right\rangle _{\mathrm{GS}}\left\langle e^{i\sum_{j}\alpha _{j}n_{j}^{f}}%
\text{poly}(C)\right\rangle _{\mathrm{GS}}  \label{mExpPL}
\end{equation}%
on the Gaussian state $\left\vert \Psi _{\mathrm{GS}}\right\rangle $. In
Appendices \ref{AppendixMB} and \ref{AppendixMF}, we show how to evaluate
them analytically with the help of Gaussian techniques \cite{QO,Db,Df}.

As an example, in Appendices \ref{AppendixME} and \ref{AppendixME2}, we
derive the equations of motion for $\Delta _{R}$ and $\Gamma _{b,m,f}$
characterizing the Gaussian part in the non-Gaussian state $%
U_{4,5}\left\vert \Psi _{\mathrm{GS}}\right\rangle $ with $\bar{\xi}_{f,b}=%
\bar{\Delta}_{R}=0$, where Eqs. (\ref{RGGSI}) and (\ref{RGGSR}) are
reproduced in the Gaussian limit $U_{S}=I$.

\subsubsection{Fluctuations}

As the non-Gaussian states introduced here are constructed on top of the
Gaussian ones, they contain the latter. Thus, among the tangent vectors
there will be terms of the form%
\begin{equation}
|V_{b}^{(1)}\rangle =U_{S}U_{\mathrm{GS}}b_{j}^{\dagger }|0\rangle ,
\end{equation}%
as well as
\begin{subequations}
\begin{eqnarray}
|V_{b}^{(2)}\rangle &=&U_{S}U_{\mathrm{GS}}b_{i}^{\dagger }b_{j}^{\dagger
}|0\rangle ,  \notag \\
|V_{f}^{(2)}\rangle &=&U_{S}U_{\mathrm{GS}}c_{i}^{\dagger }c_{j}^{\dagger
}|0\rangle .
\end{eqnarray}%
They describe single-, and two-particle excitations in the rotated frame
defined by the transformation $U_{S}U_{\mathrm{GS}}$. As before, the tangent
vectors do not contain states with an odd number of fermionic excitations
due to the fermionic-superselection rule, but their properties can be
studied following the approach presented after equation (\ref{Hbar_def})
below (see also discussion in Appendix E).

The spectrum of $\mathbf{L}$ gives information about quasiparticles, such as
their energies, quasiparticle weight, and lifetime. All information about
quasiparticles is contained in the spectral function $Z_{k}(\omega )$. The
non-Gaussian character of the state is reflected in the fact that the
tangent space contains states with several types of excitations, i.e., the
three-particle states
\end{subequations}
\begin{subequations}
\begin{eqnarray}
|V_{b}^{(3)}\rangle &=&U_{S}U_{\mathrm{GS}}b_{i}^{\dagger }b_{j}^{\dagger
}b_{k}^{\dagger }|0\rangle ,  \notag \\
|V_{bf}^{(3)}\rangle &=&U_{S}U_{\mathrm{GS}}b_{i}^{\dagger }f_{j}^{\dagger
}f_{k}^{\dagger }|0\rangle ,
\end{eqnarray}%
and the four-particle states
\end{subequations}
\begin{subequations}
\begin{eqnarray}
|V_{b}^{(4)}\rangle &=&U_{S}U_{\mathrm{GS}}b_{i}^{\dagger }b_{j}^{\dagger
}b_{k}^{\dagger }b_{l}^{\dagger }|0\rangle ,  \notag \\
|V_{bf}^{(4)}\rangle &=&U_{S}U_{\mathrm{GS}}b_{i}^{\dagger }b_{j}^{\dagger
}f_{k}^{\dagger }f_{l}^{\dagger }|0\rangle ,  \notag \\
|V_{f}^{(4)}\rangle &=&U_{S}U_{\mathrm{GS}}f_{i}^{\dagger }f_{j}^{\dagger
}f_{k}^{\dagger }f_{l}^{\dagger }|0\rangle .
\end{eqnarray}%
Our analysis includes interactions among all of the excitations $|V^{(1\sim
4)}\rangle $, which appear in the tangent space. These interactions lead to
the decay of quasi-particles and collective excitations.

Finally, we can use unitaries $U_{S}=U_{1,2,4,5}$, which minimize the energy
within the family of Gaussian states to define a new Hamiltonian in the
rotating frame as%
\begin{equation}
\bar{H}=U_{\mathrm{GS}}^{\dagger }U_{S}^{\dagger }HU_{S}U_{\mathrm{GS}}.
\label{Hbar_def}
\end{equation}%
We can use the quadratic expansion of $\bar{H}$ to study fermionic
quasiparticles in the ground state (the procedure for calculating $h_{{b,m}}$
which define an effective quadratic Hamiltonian is presented in Appendix \ref%
{AppendixME}). We note that we could also analyze interactions between these
quasiparticles perturbatively by expanding $\bar{H}$ beyond quadratic order
and using standard field theoretical techniques, such as Green's function or
the renormalization approaches \cite{polaronsc}.

\subsection{Summary of Section \protect\ref{NGS}}

In this subsection, we formalized the time dependent variational theory for
several families of non-Gaussian states. This approach can be used to study
the ground state and real-time dynamics of many-body systems that contain
both fermions and bosons. Here, we briefly summarize the procedure:

(i) Choose the appropriate transformation $U_{S}$ and use physical intuition
and symmetries to set some of the parameters equal to zero.

(ii) For the selected $U_{S}$, compute analytically the Gram matrix and the
overlap $\langle \Psi _{j}|\mathbf{R}_{\Psi }\rangle $ using the methods
presented in Sec. \ref{efficient} as a function of variational parameters.

(iii) Solve differential Eq. (\ref{GME}) until the system reaches the steady
state solution $\xi _{\mathrm{G}}$. In this fixed point compute the variance
of the energy and verify that the selected family of variational states is
appropriate.

(iv) To analyze elementary excitations around the ground state use the
formalism of linearized equations of motion from Sec. \ref{fluctuations}.
This means determining and diagonalizing matrix $\mathbf{L}$. Properties of
the single and two particle excitations can be analyzed using the effective
Hamiltonian obtained from Eq. (\ref{Hbar_def}) and discussion in Appendix %
\ref{AppendixME}.

(v) Use the variational ansatz to study real time dynamics. Applicability of
the considered class of wavefunctions can be estimated every step by
computing the norm of $||(1-\mathbf{P}_{\xi })\,|\mathbf{R}_{\Psi }\rangle
||^{2}$.

In the next sections we illustrate the general discussion presented in this
section with several concrete examples. When possible, we will provide a
comparison between our results and previously published ones to benchmark
the variational methods.


\section{Analysis of polarons in the Holsten and Su-Schrieffer-Heeger models \label{SP}}

In this section, we apply the non-Gaussian state approach developed in Sec. %
\ref{NGS} to investigate the problem of an individual electron interacting
with a phonon bath, the so-called polaron model. Although this type of
systems has been studied in condensed matter physics for more than sixty
years since the pioneering papers of Landau, Pekar, ans Fr\"{o}hlich, there
are still many interesting not fully understood questions. We focus on the
paradigmatic cases of the Holstein and Su-Schrieffer-Heeger (SSH)\ models.
We demonstrate that variational approach gives the dispersion of the
polaronic quasiparticle which is in agreement with the results of earlier
studies \cite{SSH,MA}. In particular, we observe that a single polaron phase
transition in the SSH model \cite{SSH} can be described very accurately by
the non-Gaussian state when combined with the LLP transformation. Furthemore
we study the real time evolution of polarons starting from a state in which
an electron creation operator is applied to a phonon vacuum. This analysis
allows us to extract the full spectral function of the polaron, which is
difficult to obtain using the Monte Carlo approach. We will present results
for the time dependent mean quadratures and the squeezing of the phonons.

The general lattice model for the electron phonon system is given by
\end{subequations}
\begin{eqnarray}
H &=&\sum_{nm}t_{nm}c_{n}^{\dagger }c_{m}+\sum_{q}\omega _{q}b_{q}^{\dagger
}b_{q}  \notag \\
&&+\sum_{nm,q}c_{n}^{\dagger }c_{m}[g_{nm}(q)b_{q}+g_{mn}^{\ast
}(-q)b_{-q}^{\dagger }],  \label{Hsp}
\end{eqnarray}%
where $t_{nm}\equiv t_{n-m}$ is the electron hopping amplitude between sites
$n$ and $m$, $\omega _{q}$ is the frequency of the phonon with momentum $q$,
and for translationally invariant systems the electron-phonon coupling $%
g_{nm}(q)=e^{iq(n+m)/2}\tilde{g}_{n-m}(q)$. Two paradigmatic cases, the
Holstein and the SSH models, describe two qualitatively different cases of
electron-phonon coupling. The former corresponds to phonons coupling to the
on-site energy of electrons and the latter describes phonons modulating
electron tunneling, i.e.,%
\begin{equation}
\tilde{g}_{l}(q)=\left\{
\begin{array}{c}
\frac{g}{\sqrt{N_{b}}}\delta _{l0}\text{, Holstein model,} \\
\frac{2ig}{\sqrt{N_{b}}}\delta _{l,\pm 1}\sin \frac{q}{2}\text{, SSH model.}%
\end{array}%
\right.
\end{equation}

The Hamiltonian (\ref{Hsp}) conserves the total electron number $%
N_{e}=\sum_{k}c_{k}^{\dagger }c_{k}$. In this section we concentrate on the
single electron subspace, i.e., $N_{e}=1$. The Hamiltonian (\ref{Hsp}) does
not conserve the phonon number. Hence even though there is only one electron
in the system, many phonons may be excited either in the ground state or
during real-time evolution.

To understand the character of the phonon dressing of a single electron, we
perform a unitary transformation of the Hamiltonian $H_{\mathrm{LLP}}=U_{%
\mathrm{LLP}}^{\dagger }HU_{\mathrm{LLP}}$ with $U_{\mathrm{LLP}%
}=e^{-iQ_{b}X}$, where $Q_{b}=\sum_{q}qb_{q}^{\dagger }b_{q}$ is the total
momentum operator of the phonons and $X=\sum_{n}nc_{n}^{\dagger }c_{n}$ is
the coordinate operator of the electron. The LLP transformation belongs to
the class $U_{5}$ introduced in Sec. \ref{NGS}. The LLP transformation
accomplishes two important goals. Firstly it separates explicitly the total
conserved momentum of the system. Secondly it can be understood as going to
the frame co-moving with the electron. The LLP transformed Hamiltonian is
\begin{eqnarray}
H_{\mathrm{LLP}} &=&\sum_{k}c_{k}^{\dagger }c_{k}\sum_{\delta }t_{\delta
}e^{-i(k-Q_{b})\delta }+\sum_{q}\omega _{q}b_{q}^{\dagger }b_{q}
\label{Hamiltonian_LLP} \\
&&+\sum_{k}c_{k}^{\dagger }c_{k}\sum_{l,q}[\tilde{g}_{l}(q)e^{-i(k-\frac{q}{2%
}-Q_{b})l}b_{q}+\mathrm{H.c.}].  \notag
\end{eqnarray}%
Note that in (\ref{Hamiltonian_LLP}) occupation numbers $c_{k}^{\dagger
}c_{k}$ are integrals of motion and can be related to the conserved total
momentum of the system. Hence for the polaron with momentum $k$, the ground
state can be described by the variational state $c_{k}^{\dagger }\left\vert
0\right\rangle \otimes \left\vert \Psi _{\mathrm{GS}}\right\rangle _{b}$,
where the Gaussian state%
\begin{equation}
\left\vert \Psi _{\mathrm{GS}}\right\rangle _{b}=e^{i\theta _{0}}e^{i\frac{1%
}{2}R^{T}\sigma \Delta _{R}}e^{-i\frac{1}{4}R^{T}\xi _{b}R}\left\vert
0\right\rangle _{b}
\end{equation}%
of the phonons is the approximate ground state of the Hamiltonian%
\begin{eqnarray}
\bar{H}_{k} &=&\sum_{q}\omega _{q}b_{q}^{\dagger }b_{q}+\sum_{\delta
}t_{\delta }e^{-i(k-Q_{b})\delta }  \notag \\
&&+\sum_{l,q}e^{-i(k-\frac{q}{2}-Q_{b})l}\tilde{g}_{l}(q)b_{q}+\mathrm{H.c.}.
\label{HSP}
\end{eqnarray}%
Here $R=(x_{q},p_{q})^{T}$ is the quadrature defined in the basis of
momentum eigenstates. The real time dynamics of a state with a well defined
total momentum $k$ can also be studied using the ansatz $c_{k}^{\dagger
}\left\vert 0\right\rangle \otimes \left\vert \Psi _{\mathrm{GS}%
}(t)\right\rangle _{b}$. We remind the readers that the factorization of the
wavefunction is only present after the LLP transformation. In the
\textquotedblleft original frame\textquotedblright , i.e., with the bare
electron and phonon operators, this state displays strong entanglement
between the electron and phonons. When the initial state of the system is
not an eigenstate of the total momentum, it should be expanded in momentum
eigenstates and the dynamics in each $k$-sector should be studied separately
(see e.g., \cite{fabianblochoscillations}). The state $\left\vert \Psi _{%
\mathrm{NGS}}\right\rangle =U_{\mathrm{LLP}}(c_{k}^{\dagger }\left\vert
0\right\rangle \otimes \left\vert \Psi _{\mathrm{GS}}\right\rangle _{b})$ in
the original representation is in the non-Gaussian state family (\ref{PsiNG}%
), where the transformation $U_{\mathrm{LLP}}$ does not have any variational
parameters. Hence the ground state properties and real time dynamics can be
studied by Eqs. (\ref{IMNGS}) and (\ref{RENGS}) from Appendix \ref%
{AppendixME}.

\subsection{Ground state properties and single polaron phase transitions}

In this subsection, we study the ground state properties of polarons by
solving Eq. (\ref{RGGSI}) [or equivalently Eq. (\ref{IMNGS}) with $O_{\Delta
}=0$, $O_{b}=0$], where the vector $h_{\Delta }=2\delta E_{k}/\delta \Delta
_{R}$ and the matrix $h_{b}=4\delta E_{k}/\delta \Gamma _{b}$ can be
obtained from the expectation value of the energy%
\begin{equation}
E_{k}=\left\langle \bar{H}_{k}\right\rangle _{\mathrm{GS}}=\text{ }%
_{b}\left\langle \Psi _{\mathrm{GS}}\right\vert \bar{H}_{k}\left\vert \Psi _{%
\mathrm{GS}}\right\rangle _{b}.
\end{equation}%
A detailed calculation of the last expression for $\bar{H}_{k}$ from Eq. (%
\ref{HSP}) is given in Appendix \ref{AppendixMB}. Here we only summarize the
result. We find
\begin{widetext}
\begin{eqnarray}
E_{k} &=&\frac{1}{4}\Delta _{R}^{T}\mathbf{\omega }\Delta _{R}+\frac{1}{4}tr(%
\mathbf{\omega }\Gamma _{b})+\sum_{\delta }t_{\delta }e^{-ik\delta }\frac{%
s_{0}e^{-\frac{1}{2}\Delta _{R}^{T}\tilde{\Gamma}_{B}^{-1}(1-f_{0})\Delta
_{R}}}{\sqrt{\det (\Gamma _{B}/2)}}  \notag \\
&&+2\text{Re}\sum_{\delta }e^{-ik\delta }\frac{s_{0}e^{-\frac{1}{2}\Delta
_{R}^{T}\tilde{\Gamma}_{B}^{-1}(1-f_{0})\Delta _{R}}}{\sqrt{\det (\Gamma
_{B}/2)}}\Delta _{R}^{T}\tilde{\Gamma}_{B}^{-1}\mathbf{g}-\frac{1}{2}%
\sum_{q}\omega _{q},  \label{ESP}
\end{eqnarray}
\end{widetext}where, in the basis of momentum eigenstates the frequency
matrix is diagonal $\mathbf{\omega }={\openone}_{2}\otimes diag(\omega _{q})$%
, the matrices%
\begin{eqnarray}
\Gamma _{B} &=&\sqrt{1-f_{0}}\Gamma _{b}\sqrt{1-f_{0}}+1+f_{0},  \notag \\
\tilde{\Gamma}_{B} &=&(1-f_{0})\Gamma _{b}+1+f_{0}
\end{eqnarray}%
are determined by $f_{0}={\openone}_{2}\otimes diag(e^{iq\delta })$, and $%
\mathbf{g}=(1,i)^{T}\otimes \tilde{g}_{\delta }(q)e^{iq\delta /2}$. As shown
in Appendix \ref{AppendixMB}, the sign $s_{0}$ can be determined by the
Takagi diagonalization \cite{Takagi} of the symmetric matrix $\Gamma _{B}$.

We can use Eq. (\ref{ESP}) to find the vector $h_{\Delta }$ and the matrix $%
h_{b}$ that enter Eq. (\ref{RGGSR})
\begin{widetext}
\begin{eqnarray}
h_{\Delta } &=&\mathbf{\omega }\Delta _{R}-2\sum_{\delta }t_{\delta
}e^{-ik\delta }\frac{s_{0}e^{-\frac{1}{2}\Delta _{R}^{T}\tilde{\Gamma}%
_{B}^{-1}(1-f_{0})\Delta _{R}}}{\sqrt{\det (\Gamma _{B}/2)}}\tilde{\Gamma}%
_{B}^{-1}(1-f_{0})\Delta _{R}  \notag \\
&&+4\text{Re}\sum_{\delta }e^{-ik\delta }\frac{s_{0}e^{-\frac{1}{2}\Delta
_{R}^{T}\tilde{\Gamma}_{B}^{-1}(1-f_{0})\Delta _{R}}}{\sqrt{\det (\Gamma
_{B}/2)}}[1-\tilde{\Gamma}_{B}^{-1}(1-f_{0})\Delta _{R}\Delta _{R}^{T}%
\mathbf{]}\tilde{\Gamma}_{B}^{-1}\mathbf{g}  \label{YSP}
\end{eqnarray}%
and%
\begin{eqnarray}
h_{b} &=&\mathbf{\omega }-2\sum_{\delta }t_{\delta }e^{-ik\delta }\frac{%
s_{0}e^{-\frac{1}{2}\Delta _{R}^{T}\tilde{\Gamma}_{B}^{-1}(1-f_{0})\Delta
_{R}}}{\sqrt{\det (\Gamma _{B}/2)}}W_{1}  \notag \\
&&-4\text{Re}\sum_{\delta }e^{-ik\delta }\frac{s_{0}e^{-\frac{1}{2}\Delta
_{R}^{T}\tilde{\Gamma}_{B}^{-1}(1-f_{0})\Delta _{R}}}{\sqrt{\det (\Gamma
_{B}/2)}}(\Delta _{R}^{T}\tilde{\Gamma}_{B}^{-1}\mathbf{g}%
W_{1}+W_{2}+W_{2}^{T})  \label{OSP}
\end{eqnarray}
\end{widetext}are determined by Eqs. (\ref{hY}) and (\ref{ESP}), where%
\begin{eqnarray}
W_{1} &=&\tilde{\Gamma}_{B}^{-1}(1-f_{0})[1-\Delta _{R}\Delta _{R}^{T}\tilde{%
\Gamma}_{B}^{-1}(1-f_{0})],  \notag \\
W_{2} &=&\tilde{\Gamma}_{B}^{-1}\mathbf{g}\Delta _{R}^{T}\tilde{\Gamma}%
_{B}^{-1}(1-f_{0}).
\end{eqnarray}

By solving the equations of motion (\ref{RGGSI}) with $h_{\Delta }$ and $%
h_{b}$ given by Eqs. (\ref{YSP}) and (\ref{OSP}), we obtain the values of $%
\Delta _{R}$ and $\Gamma _{b}$ for the Gaussian ground state $\left\vert
\Psi _{\mathrm{GS}}\right\rangle _{b}$ in the limit $\tau \rightarrow \infty
$. With\ the steady state solution, the energy (\ref{ESP}) determines the
dispersion relation $E_{k}$ of the polaron with momentum $k$. Polaronic
suppression of the quasiparticle weight is given by%
\begin{eqnarray}
Z_{k} &=&\left\vert \left\langle 0\right\vert c_{k}\left\vert \Psi _{\mathrm{%
NGS}}\right\rangle \right\vert ^{2}=\left\vert \left\langle 0\left\vert \Psi
_{\mathrm{GS}}\right\rangle _{b}\right. \right\vert ^{2}  \notag \\
&=&\frac{e^{-\frac{1}{2}\Delta _{R}^{T}(\Gamma _{b}+{\openone}%
_{2N_{b}})^{-1}\Delta _{R}}}{\sqrt{\det [(\Gamma _{b}+{\openone}_{2N_{b}})/2]%
}}.
\end{eqnarray}

To understand the character of the variational solution it is useful to
consider the polaron wavefunction in the original basis $|\Psi _{NGS}\rangle
=U_{LLP}\,(c_{k}^{\dagger }\left\vert 0\right\rangle \otimes \left\vert \Psi
_{\mathrm{GS}}\right\rangle _{b})$. We note that the LLP Hamiltonian (\ref%
{Hamiltonian_LLP}) does not conserve phonon momentum, hence the state $%
\left\vert \Psi _{\mathrm{GS}}\right\rangle _{b}$ is a superposition of
different momentum eigenstates $\left\vert \Psi _{\mathrm{GS}}\right\rangle
_{b}=\sum_{q}\phi _{k}(q)|\Psi _{q}(k)\rangle _{b}$. Here $|\Psi
_{q}(k)\rangle _{b}$ is a phonon state which has net phonon momentum $q$
(the wavefunction $\Psi _{q}(k)$ depends on $k$, but its specific form is
not important for our argument below). We recall that the LLP transformation
simply shifts the electron momentum by the amount equal to the total
momentum of the phonons, therefore $|\Psi _{NGS}\rangle =\sum_{q}\phi
_{k}(q)c_{k-q}^{\dagger }\left\vert 0\right\rangle \otimes |\Psi
_{q}(k)\rangle _{b}$. We use $c_{k-q}^{\dagger }=\frac{1}{\sqrt{N_{0}}}%
\sum_{j_{0}}e^{i(k-q)j_{0}}c_{j_{0}}^{\dagger }|0\rangle $ and $%
\sum_{q}e^{-iqj_{0}}\phi _{k}(q)|\Psi _{q}(k)\rangle _{b}=e^{-i\hat{Q}%
_{b}j_{0}}\left\vert \Psi _{\mathrm{GS}}\right\rangle _{b}$ where $\hat{Q}%
_{b}$ is the operator of the total phonon momentum. Then we find
\begin{eqnarray}
|\Psi _{\mathrm{NGS}}\rangle &=&\frac{1}{\sqrt{N_{b}}}%
\sum_{j_{0}}e^{ikj_{0}}\,|\Phi _{j_{0}}\rangle ,  \notag \\
|\Phi _{j_{0}}\rangle &=&c_{j_{0}}^{\dagger }\left\vert 0\right\rangle
\otimes e^{-i\hat{Q}_{b}j_{0}}\left\vert \Psi _{\mathrm{GS}}\right\rangle
_{b}.
\end{eqnarray}%
The physical interpretation of $|\Phi _{j_{0}}\rangle $ is a polaron
centered on site $j_{0}$. By analyzing $e^{-i\hat{Q}_{b}j_{0}}\left\vert
\Psi _{\mathrm{GS}}\right\rangle _{b}$ we can understand the corresponding
phonon configuration
\begin{eqnarray}
\tilde{\Delta}_{R} &=&\text{ }_{b}\left\langle \Psi _{\mathrm{GS}%
}\right\vert e^{iQ_{b}j_{0}}\tilde{R}e^{-iQ_{b}j_{0}}\left\vert \Psi _{%
\mathrm{GS}}\right\rangle _{b}  \notag \\
&=&\frac{1}{2}W_{b}V_{F}W_{b}^{\dagger }\Delta _{R}  \label{Rj}
\end{eqnarray}%
of phonon fields $\tilde{R}=(x_{j},p_{j})^{T}$ in the coordinate space.
Here,
\begin{equation}
V_{F}=\left(
\begin{array}{cc}
v & 0 \\
0 & v^{\ast }%
\end{array}%
\right) ,
\end{equation}%
the matrix $W_{b}$ was defined in Eq. (\ref{W_b_definition}), and the
Fourier transform is represented in the matrix form by $v$ with the element $%
v_{d,q}=e^{idq}/\sqrt{N_{0}}$, and $d=j-j_{0}$ is the the distance between
the electron and the local phonon mode at the position $j$. The covariance
matrix%
\begin{eqnarray}
\tilde{\Gamma}_{b} &=&\text{ }_{b}\left\langle \Psi _{\mathrm{GS}%
}\right\vert e^{iQ_{b}j_{0}}\frac{1}{2}\{\delta \tilde{R},\delta \tilde{R}%
^{T}\}e^{-iQ_{b}j_{0}}\left\vert \Psi _{\mathrm{GS}}\right\rangle _{b}
\notag \\
&=&\frac{1}{4}W_{b}V_{F}W_{b}^{\dagger }\Gamma _{b}W_{b}V_{F}^{\dagger
}W_{b}^{\dagger }  \label{Gj}
\end{eqnarray}%
describes the squeezing of phonons around the electron, where the
fluctuation field $\delta \tilde{R}=\tilde{R}-\tilde{\Delta}_{R}$.

\begin{figure}[tbp]
\includegraphics[width=0.9\linewidth]{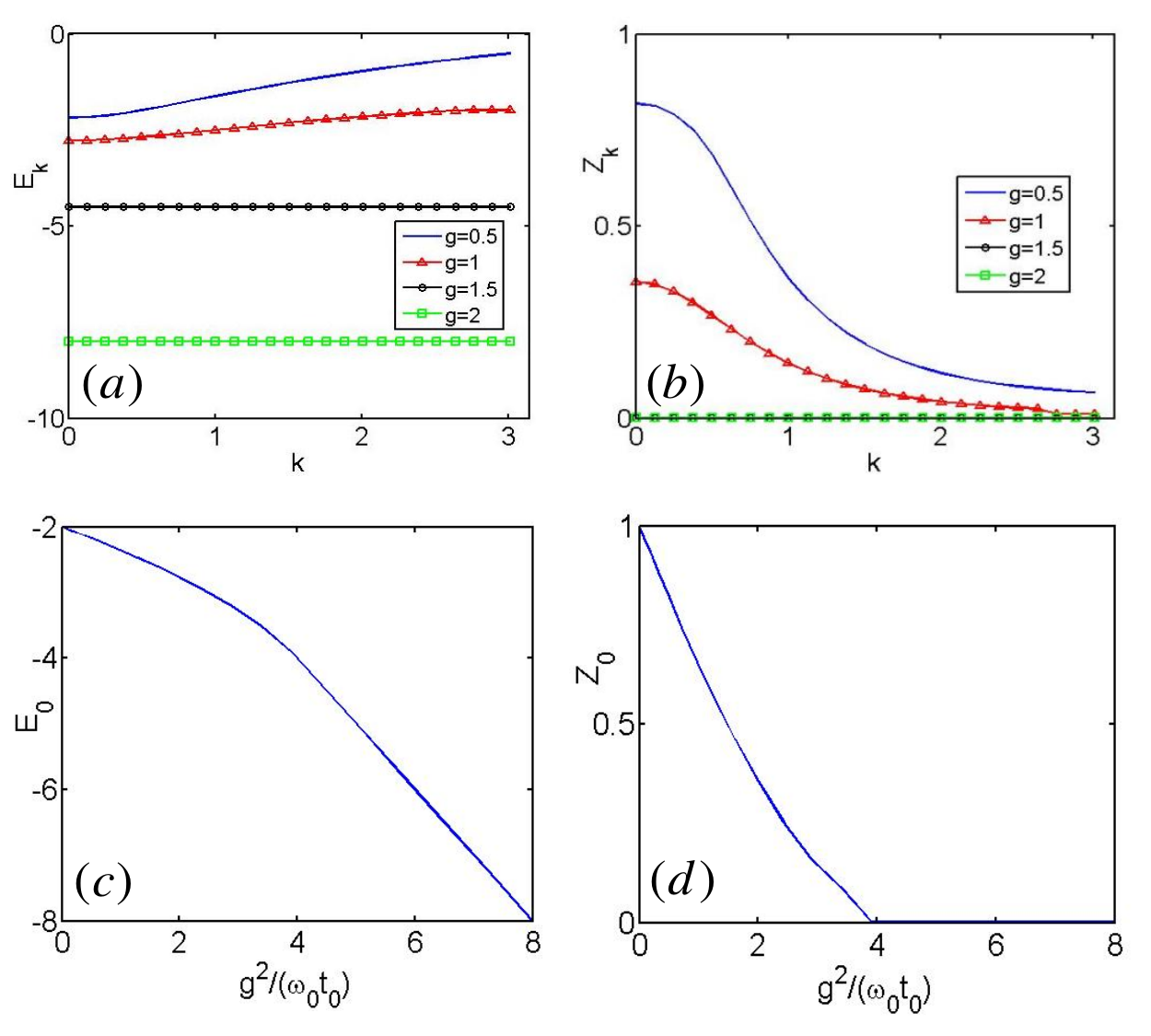}
\caption{The dispersion relation and single particle weight of polarons in
the 1D Holstein model. Variational analysis was done in a system with 50
sites, $\protect\omega_0=0.5$, and the hopping constant $t_0$ taken as the
unit of energy. (a)-(b) Dispersion relations and single particle weights for
different coupling constants. (c)-(d) The energy and the single particle
weight for the polaron with zero momentum.}
\label{H1DEZ}
\end{figure}

In Fig. \ref{H1DEZ}, we present results for polarons in the one dimensional
Holstein model: the dispersion $E_{k}$ and single-particle residue $Z_{k}$.
Note that our analysis gives the lowest energy state for a given total
momentum $k$, which is an integral of motion of the system. The true ground
state corresponds to finding the energy minimum with respect to $k$. In our
analysis we consider only nearest neighbor hopping of electrons, i.e., $%
t_{l}=-t_{0}\delta _{l,\pm 1}$ and we set $t_{0}=1$. We also neglect the
dispersion of phonons, namely, we consider Einstein phonons with frequency $%
\omega _{0}=0.5t_{0}$. From Fig. \ref{H1DEZ}a-b we observe that an increase
in the electron-phonon interaction leads to a strong flattening of the band
and suppression of the quasiparticle weight $Z_{k}$. The momentum dependence
of $Z_{k}$ in Fig. \ref{H1DEZ}b indicates that polaronic dressing is
enhanced at higher momenta.

The bandwidth of the polaron is primarily determined by the second term in
Eq. (\ref{HSP}). Polaronic reduction of the bandwidth (which can be
understood as the effective mass becoming heavier) comes from the $\langle
e^{iQ_{b}\delta }\rangle _{GS}$ factor in the second term in Eq. (\ref{HSP}).

\begin{figure}[tbp]
\includegraphics[width=0.9\linewidth]{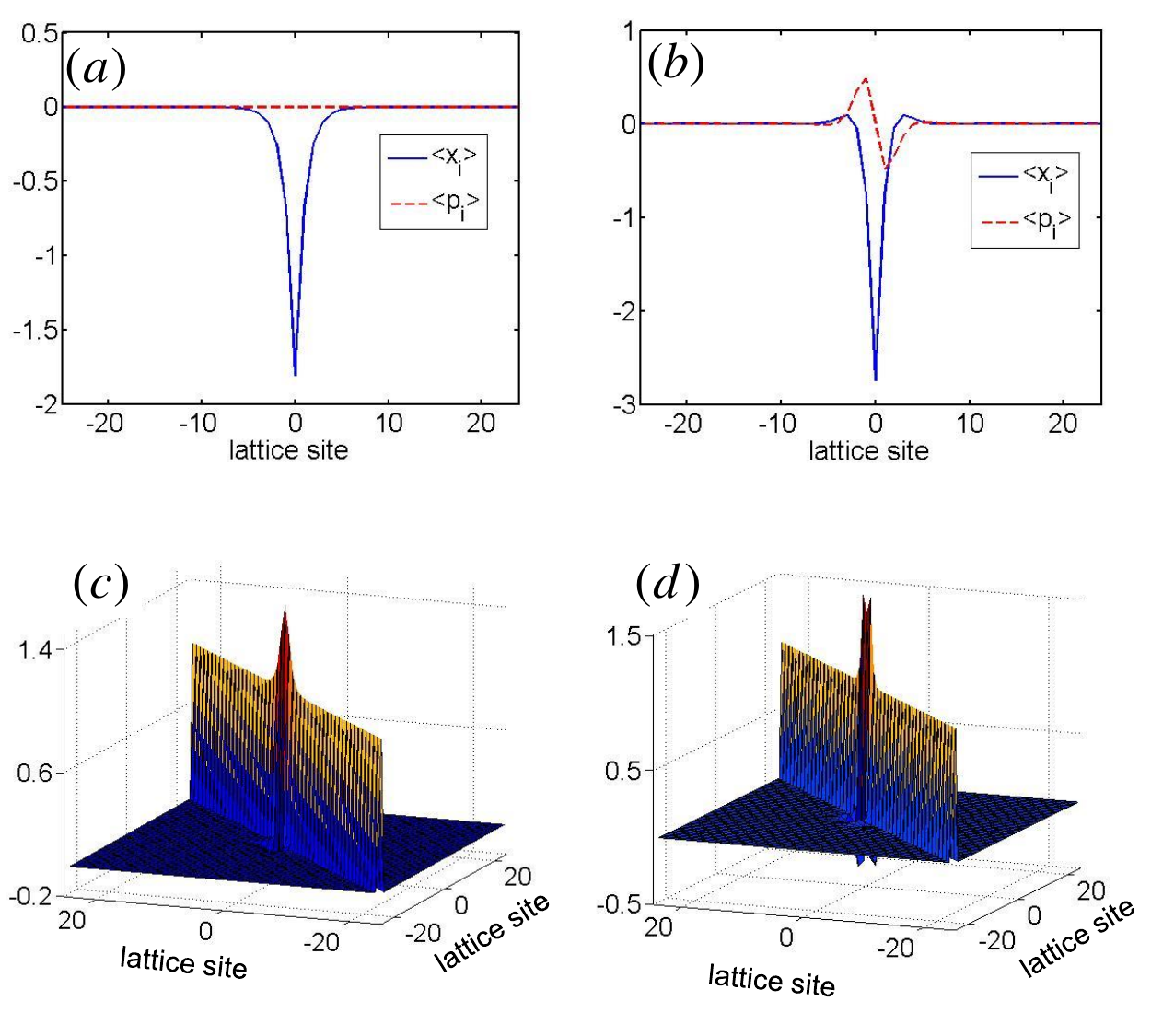}
\caption{The average values of quadratures and correlation functions $%
\left\langle \protect\delta x_{i} \protect\delta x_{j}\right\rangle$ of
phonons in the Holstein model with 50 sites, where $\protect\omega_0=0.5$, $%
g=1$, and the hopping constant $t_0$ is taken as the unit. (a)-(b) The
average values of quadratures along the chain for $k=0$ and $k=\protect\pi/2$%
. (c)-(d) The correlation functions $\left\langle \protect\delta x_{i}
\protect\delta x_{j}\right\rangle$ of displacements at positions $i$ and $j$
for $k=0$ and $k=\protect\pi/2$.}
\label{H1DRG}
\end{figure}

In Figs. \ref{H1DEZ}c-d, we show the energy $E_{0}$ and the single particle
weight $Z_{0}$ for the polaron with momentum $k=0$. These two properties of
the Holstein polaron have been studied in earlier papers using several
techniques: the self-consistent Born approximation, the Lang-Firsov (LF)
approach \cite{LF}, Diagrammatic Monte Carlo (DMC) calculations \cite{DMC},
the momentum average (MA) method \cite{MA}, and the numerical minimization
based on the Toyozawa ans\"{a}tz (TA) \cite{TA1,TA2}.

We emphasize that $E_{0}$ and $Z_{0}$ in Figs. \ref{H1DEZ}c-d agree with the
results from DMC and MA quantitatively \cite{MA}. In the TA, the coherent
and squeezing properties of phonons around the electron in the co-moving
frame can also be studied variationally, where the variational parameters
are obtained by the brute-force minimization of the ground state energy.
Compared with TA, the imaginary time evolution of non-Gaussian states is
more efficient in finding the optimal variational parameters. Thus the
general Gaussian ansatz from Eq. (\ref{GS}) can be used to analyze phonon
squeezing at large distances from the impurity.

The figure \ref{H1DRG} shows the spatial structure of the polaron with
momenta $k=0$ and $\pi /2$ when the coupling constant $g=1$. We present both
the displacement (\ref{Rj}) and the squeezing (\ref{Gj}) of the phonons
around the electron. Note that for a given total momentum of the polaron
they only depend on the distance to the electron, hence we set the electron
position to be $j_{0}=0$. We find that the canonical phonon momentum $%
\left\langle p\right\rangle $\ vanishes on all sites when the total momentum
of the polaron $k=0$. In Figs. \ref{H1DRG}c-d, the correlation functions $%
\left\langle \delta x_{i}\delta x_{j}\right\rangle $ for $k=0$ and $\pi /2$
show that phonons around the electron are squeezed along the direction of
the canonical momentum in the phase space, i.e., $\left\langle \delta
x_{j}^{2}\right\rangle >1$ for $j$ close to $j_{0}$.

\begin{figure}[tbp]
\includegraphics[width=0.9\linewidth]{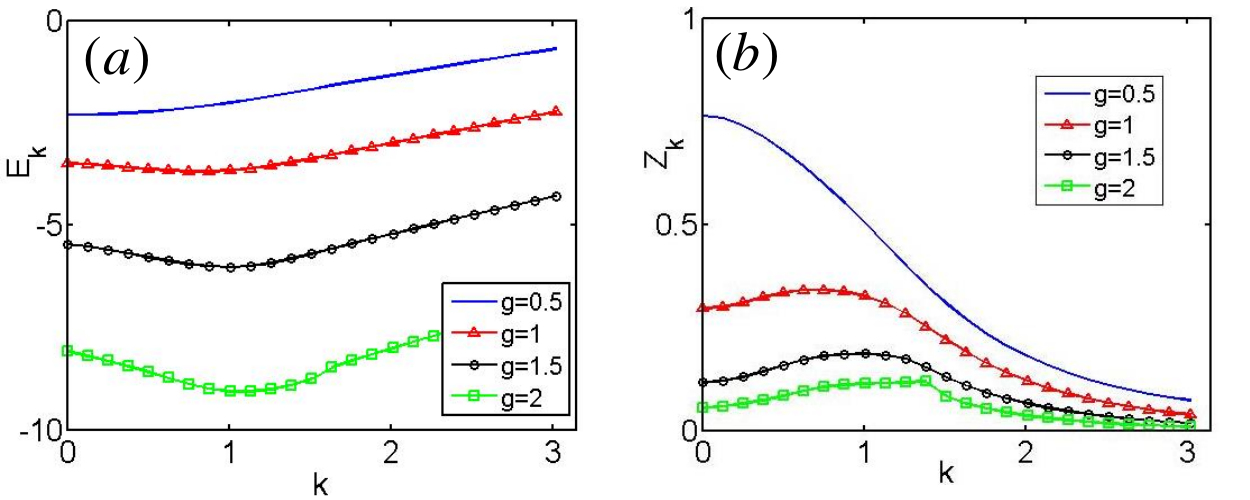}
\caption{The dispersion relation and single particle weight of the 1D SSH
model with 50 sites, where $\protect\omega_0=0.5$ and the hopping constant $%
t_0$ is taken as the unit.}
\label{S1DEZ}
\end{figure}

The remarkable \textquotedblleft single-polaron phase
transition\textquotedblright\ takes place when the electron-phonon
interaction depends on the momenta of the electron and the phonons, as is
the case for the 1D SSH model. For the SSH model with $t_{l}=-t_{0}\delta
_{l,\pm 1}$ and Einstein phonon frequency $\omega _{q}=\omega _{0}=0.5t_{0}$%
, Fig. \ref{S1DEZ} displays the dispersion relation $E_{k}$ and the single
particle weight $Z_{k}$ of the lowest polaron band. In agreement with
earlier studies we find that when the interaction $g$ exceeds a certain
critical value $g_{c}$, the lowest energy state of the polaron is at a
finite momentum $k\neq 0$.

Previously this phase transition has been studied by MA and three numerical
methods \cite{SSH}: DMC, exact diagonalization (ED), and bold DMC. To
understand the origin of the transition using our LLP+Gaussian approach we
observe that the momentum dependence of $E_{k}$ in Eq. (\ref{HSP}) comes
from both the second and the third terms. The former corresponds to the
polaronically dressed electron hopping and the latter comes from the
electron-phonon interaction. The competition between the two terms gives
rise to the polaron dispersion minimum moving away from $k=0$ for large
interaction strengths.

\begin{figure}[tbp]
\includegraphics[width=0.9\linewidth]{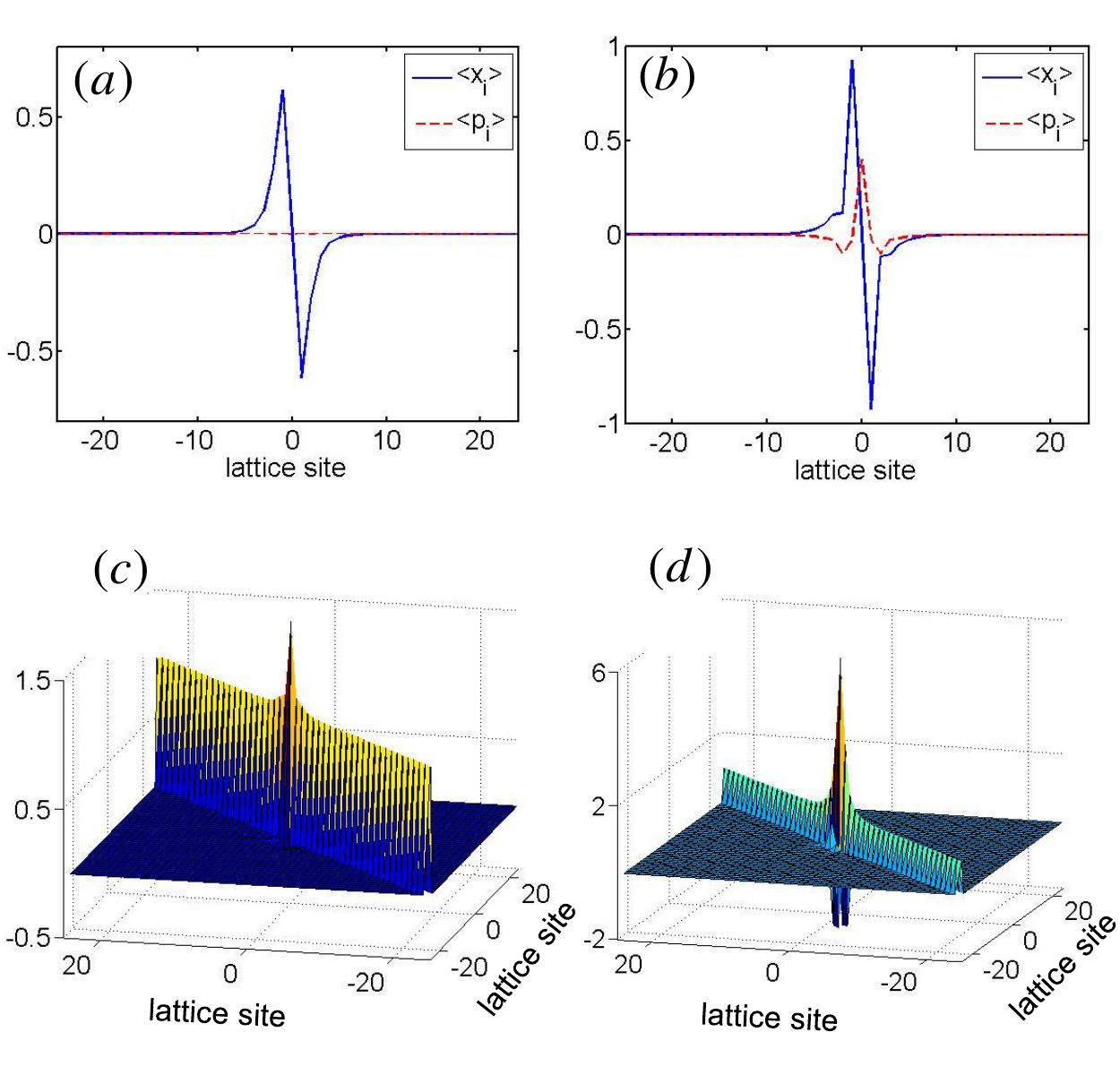}
\caption{The average values of quadratures and correlation functions $%
\left\langle \protect\delta x_{i} \protect\delta x_{j}\right\rangle$ of
phonons in the SSH model with 50 sites, $\protect\omega_0=0.5$ and the
hopping constant $t_0$ taken as the unit of energy. (a)-(b) The average
values of quadratures in the ground states with $k=0$ and $k=0.88$ for $%
g=0.5 $ and $g=1 $, respectively. (c)-(d) The correlation functions $%
\left\langle \protect\delta x_{i} \protect\delta x_{j}\right\rangle$ of
displacements at positions $i$ and $j$ in the ground states with $k=0$ and $%
k=0.88$ for $g=0.5 $ and $g=1$, respectively.}
\label{S1DRG}
\end{figure}

The figure \ref{S1DRG} compares the structure of the ground state polarons
at the two sides of the transition. Parts a) and c) correspond to $g=0.5$ , $%
\omega _{0}=0.5$, and the momentum $k=0$ of the ground state at this
interaction strength. Parts b) and d) correspond to the polaron for $g=1$, $%
\omega _{0}=0.5$, and the momentum $k=0.88$ of the ground state for this
interaction strength. We show both the average value of the phonon
displacements $\tilde{\Delta}_{R}$ and the correlation functions $%
\left\langle \delta x_{i}\delta x_{j}\right\rangle $. Phonon squeezing is
significantly enhanced for larger values of the coupling constant.

\begin{figure}[tbp]
\includegraphics[width=0.9\linewidth]{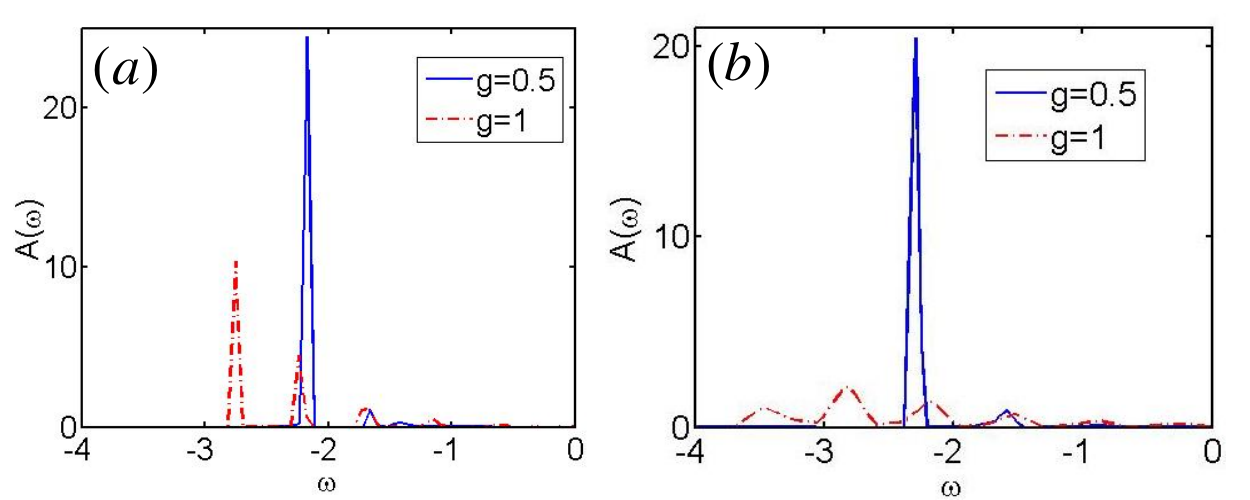}
\caption{The spectral functions of the 1D Holstein model and the 1D SSH
model, where the system size $N=50$, the phonon frequency $\protect\omega%
_0=0.5$, and $t_0$ taken as the unit of energy. (a) The polaron momentum $%
k=0 $ for $g=0.5$ (solid blue curve) and $g=1$ (dashed red curve) in the
Holstein model; (b) The polaron momentum $k=0$ for $g=0.5$ (solid blue
curve) and $k=0.88$ for $g=1$ (dashed red curve) in the SSH model.}
\label{SF}
\end{figure}

\subsection{Real time dynamics}

\begin{figure*}[tbp]
\begin{center}
\includegraphics[width=0.7\linewidth]{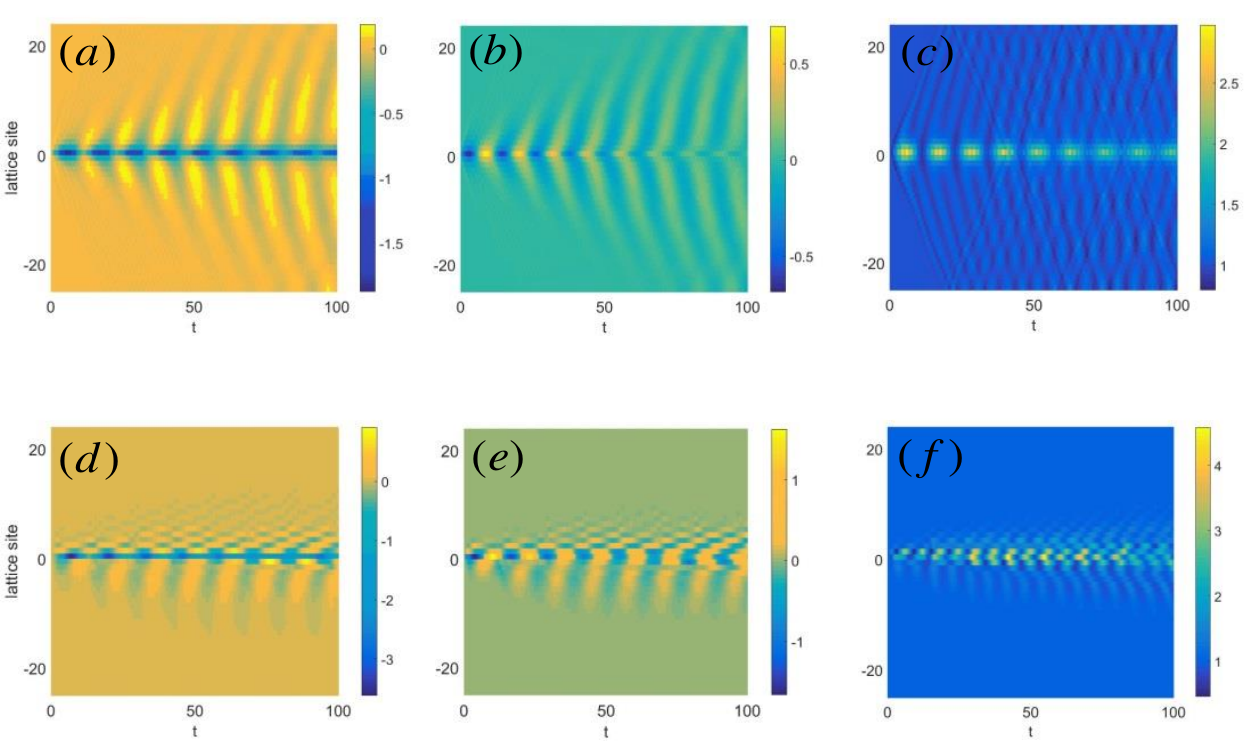}
\end{center}
\caption{The time evolution of average values $\tilde{\Delta}_R$ and
diagonal elements $\left\langle \protect\delta x_{i}^{2}\right\rangle $ of $%
\tilde{\Gamma}_{b}$ for the Holstein model, where $\protect\omega_0=0.5$, $%
g=1$, and $t_0$ is taken as the unit. (a)-(c) The average values $%
\left\langle x_i \right\rangle $, $\left\langle p_i \right\rangle $, and $%
\left\langle \protect\delta x_{i}^{2}\right\rangle $ for $k=0$; (d)-(f) The
average values $\left\langle x_i \right\rangle $, $\left\langle p_i
\right\rangle $, and $\left\langle \protect\delta x_{i}^{2}\right\rangle $
for $k=\protect\pi/2$.}
\label{RGtH1D}
\end{figure*}

We now discuss how to compute the polaron spectral function%
\begin{equation}
\mathcal{A}(\omega )=-\frac{1}{\pi }\text{Im}G_{R}(\omega ),  \label{A}
\end{equation}%
where $G_{R}(\omega )=\int dtG(t)e^{i\omega t}$ is the Fourier transform of
the retarded Green function%
\begin{equation}
G_{R}(t)=-i\left\langle 0\right\vert c_{k}e^{-iHt}c_{k}^{\dagger }\left\vert
0\right\rangle \theta (t).  \label{GR_polaron_definition}
\end{equation}%
Applying the LLP transformation $U_{\mathrm{LLP}}$ to the definition of the
retarded Green's function (\ref{GR_polaron_definition}) we find
\begin{equation}
G_{R}(t)=-i\left\langle 0\right\vert e^{-i\bar{H}_{k}t}\left\vert
0\right\rangle \theta (t),  \label{GR}
\end{equation}%
where $\bar{H}_{k}$ is given in equation (\ref{HSP}).

In the co-moving frame, the real-time evolution $\left\vert \Psi
(t)\right\rangle =e^{-iH_{\mathrm{p}}t}\left\vert 0\right\rangle $ is
approximated by a Gaussian state $\left\vert \Psi _{\mathrm{GS}%
}(t)\right\rangle $ obeying the Schr\"{o}dinger equation%
\begin{equation}
i\partial _{t}\left\vert \Psi _{\mathrm{GS}}(t)\right\rangle =\mathbf{P}%
_{\xi }\bar{H}_{k}\left\vert \Psi _{\mathrm{GS}}(t)\right\rangle  \label{PSE}
\end{equation}%
projected onto the tangent space. Since $\left\vert \Psi _{\mathrm{GS}%
}(t)\right\rangle $ is a Gaussian state, the tangent vectors only contain $%
U_{\mathrm{GS}}\left\vert 0\right\rangle $, $U_{\mathrm{GS}}b_{q}^{\dagger
}\left\vert 0\right\rangle $, and $U_{\mathrm{GS}}b_{q_{1}}^{\dagger
}b_{q_{2}}^{\dagger }\left\vert 0\right\rangle $. After projecting onto the
tangent space (see Eqs. (\ref{UR}) and (\ref{hY}) for details), equation of
motion (\ref{PSE}) becomes%
\begin{equation}
i\partial _{t}\left\vert \Psi _{\mathrm{GS}}(t)\right\rangle =\bar{H}_{%
\mathrm{MF}}\left\vert \Psi _{\mathrm{GS}}(t)\right\rangle ,  \label{SESP}
\end{equation}%
where the normal ordering expansion can be used to construct the mean field
Hamiltonian%
\begin{equation}
\bar{H}_{\mathrm{MF}}=\frac{1}{4}\text{:}\delta R^{T}h_{b}\delta R\text{:}+%
\frac{1}{2}\delta R^{T}h_{\Delta }+E_{k}  \label{H_MF_polaron}
\end{equation}%
with $h_{\Delta }$ and $h_{b}$ given in Eqs. (\ref{YSP}) and (\ref{OSP}).
The first term in $\bar{H}_{\mathrm{MF}}$ is normal ordered with respect to
the squeezed vacuum, i.e., the coherent part has been removed using $\delta
R=R-\Delta _{R}$. As shown in Appendix \ref{AppendixME}, the projected Schr%
\"{o}dinger equation can be used to derive equations describing the real
time evolution of $\Delta _{R}$ and $\Gamma _{b}$. The result is shown in
Eq. (\ref{RGGSR}) [or equivalently Eq. (\ref{RENGS}) with $h_{\Delta
}^{t}=h_{\Delta }$ and $h_{b}^{t}=h_{b}$].

\begin{figure*}[tbp]
\begin{center}
\includegraphics[width=0.7\linewidth]{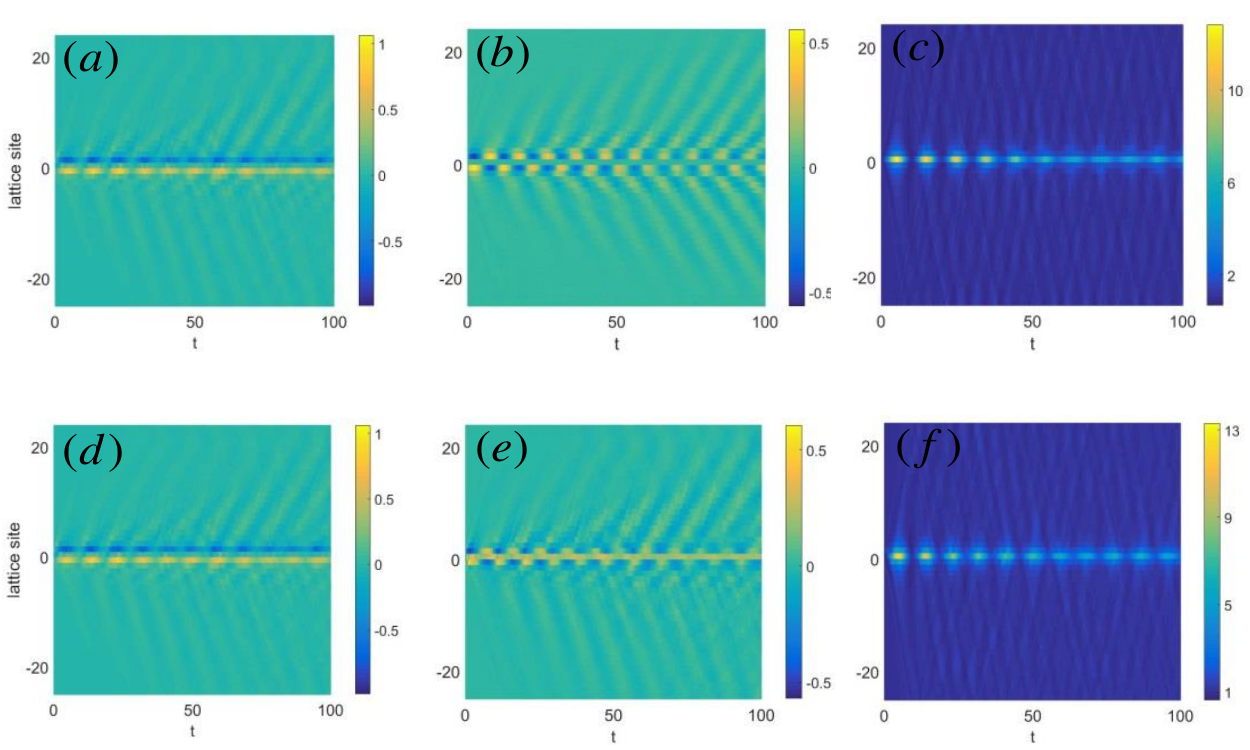}
\end{center}
\caption{The time evolution of average values $\tilde{\Delta}_{R}$ and
diagonal elements $\left\langle \protect\delta x_{i}^{2}\right\rangle $ of $%
\tilde{\Gamma}_{b}$ for the SSH model, where $\protect\omega_0=0.5$, $g=1 $,
and $t_0$ is taken as the unit. (a)-(c) The average values $\left\langle x_i
\right\rangle $, $\left\langle p_i \right\rangle $, and $\left\langle
\protect\delta x_{i}^{2}\right\rangle $ for $k=0$; (d)-(f) The average
values $\left\langle x_i \right\rangle $, $\left\langle p_i \right\rangle $,
and $\left\langle \protect\delta x_{i}^{2}\right\rangle $ for $k=0.88$.}
\label{RGtS1D}
\end{figure*}

One of the challenges in computing $G_{R}(t)$ is that it is defined as an
overlap of the two wavefunctions: $\left\langle 0\left\vert \Psi _{\mathrm{GS%
}}(t)\right\rangle \right. $. Therefore to obtain $G_{R}(t)$ we need to
compute the time dependent overall phase $\theta _{0}(t)$ in $\left\vert
\Psi _{\mathrm{GS}}(t)\right\rangle $. In principle, this calculation can be
done using equation (\ref{H_MF_polaron}). It is more instructive however to
use a different representation of the Gaussian transformation $U_{\mathrm{GS}%
}(t)$, which allows to keep track of the explicit time evolution of the
variational wavefunction. We use Wei-Norman algebra \cite{WN} to write the
transformation in the form%
\begin{equation}
U_{\mathrm{GS}}(t)=e^{i\theta _{0}(t)}e^{i\frac{1}{2}R^{T}\sigma \Delta
_{R}}e^{b^{\dagger }\Lambda _{1}b^{\dagger }}e^{b^{\dagger }\Lambda
_{2}b}e^{b\Lambda _{3}b},  \label{UW}
\end{equation}%
where $b=(b_{q_{1}},...,b_{q_{N}})^{T}$ should be understood as a vector. In
Appendix \ref{AppendixWN}, we use the projected Schr\"{o}dinger Eq. (\ref%
{SESP}) to obtain the following equations for the real-time evolution%
\begin{eqnarray}
\partial _{t}\theta _{0} &=&-\delta E_{k}-\frac{1}{2}tr\mathbf{\omega }%
_{b}-tr(\mathbf{\varpi }^{\dagger }\Lambda _{1}),  \notag \\
i\partial _{t}\Lambda _{1} &=&\frac{1}{2}\mathbf{\varpi }+\mathbf{\omega }%
_{b}\Lambda _{1}+\Lambda _{1}\mathbf{\omega }_{b}^{T}+2\Lambda _{1}\mathbf{%
\varpi }^{\dagger }\Lambda _{1}  \label{TL}
\end{eqnarray}%
of the global phase $\theta _{0}$ and the symmetric matrix $\Lambda _{1}$.
In Eq. (\ref{TL}) we used $\delta E_{k}=E_{k}-tr(h_{b}\Gamma _{b})/4-\Delta
_{R}^{T}h_{\Delta }/4$, and matrices $\mathbf{\omega }_{b}$ and $\mathbf{%
\varpi }$ are defined using the single-particle Hamiltonian%
\begin{equation}
\left(
\begin{array}{cc}
\mathbf{\omega }_{b} & \mathbf{\varpi } \\
\mathbf{\varpi }^{\dagger } & \mathbf{\omega }_{b}^{T}%
\end{array}%
\right) =\frac{1}{2}W_{b}^{\dagger }h_{b}W_{b}
\end{equation}%
in the bosonic Nambu representation $(b,b^{\dagger })^{T}$. Equation (\ref%
{TL}) determines the evolution of the global phase.

In terms of $\theta _{0}$ and $\Lambda _{1}$, the retarded Green's function
is%
\begin{equation}
G_{R}(t)=-ie^{i\theta _{0}(t)}e^{-\frac{1}{2}\Delta _{b}^{\dagger }\Delta
_{b}}e^{\Delta _{b}^{\dagger }\Lambda _{1}\Delta _{b}^{\ast }}\theta (t),
\label{GRt}
\end{equation}%
where the column vector $\Delta _{b}=\left\langle b\right\rangle _{\mathrm{GS%
}}$. The Fourier transform of Eq. (\ref{GRt}) gives the spectral function (%
\ref{A}). In Figs. \ref{SF}a-b, we show the spectral functions for the 1D
Holstein and SSH models respectively. The solid blue and dashed red curves
in Fig. \ref{SF}a display spectral functions of the polaron at $k=0$ for
coupling constants $g=0.5$ and $g=1$. The solid blue and dashed red curves
in Fig. \ref{SF}b display the spectral functions of the polaron with $k=0$
for $g=0.5$ and $k=0.88$ for $g=1$. Note that $k=0.88$ corresponds to the
ground state of the SSH polaron for $g=1$. An important feature of the
spectral function is the presence of several shake-off peaks in the
spectrum. The maximum value of the spectral function does not necessarily
correspond to the lowest energy peak (different peaks are sometimes referred
to as different polarons bands).

The time dependent non-Gaussian state%
\begin{eqnarray}
\left\vert \Psi _{\mathrm{NGS}}\right\rangle &=&U_{\mathrm{LLP}%
}c_{k}^{\dagger }\left\vert 0\right\rangle \left\vert \Psi _{\mathrm{GS}%
}(t)\right\rangle  \notag \\
&=&\sum_{j_{0}}e^{ikj_{0}}c_{j_{0}}^{\dagger }\left\vert 0\right\rangle
e^{-iQ_{b}j_{0}}\left\vert \Psi _{\mathrm{GS}}(t)\right\rangle
\label{Psi_NGS_t}
\end{eqnarray}%
can be used to analyze the time evolution of all physical observables. We
compute the phonon quadratures and correlation functions using equations (%
\ref{Rj}) and (\ref{Gj}). Note that the phonon parameters described by the
state (\ref{Psi_NGS_t}) only depend on the distance between the phonon site $%
j$ and the electron coordinate $j_{0}$. Thus, it is sufficient to consider a
single term in (\ref{Psi_NGS_t}) with one specific $j_{0}$, which we will
set to be at the origin, $j_{0}=0$. In Fig. \ref{RGtH1D}, we show the time
evolution of the phonon quadratures $\tilde{\Delta}_{R}$ and the diagonal
elements $\left\langle \delta x_{i}^{2}\right\rangle $ of the matrix $\tilde{%
\Gamma}_{b}$ for the Holstein model with $\omega _{0}/t_{0}=0.5$ and $%
g/t_{0}=1$. We consider the cases with polaron momenta $k=0$ (the first row)
and $k=\pi /2$ (the second row). In Fig. \ref{RGtS1D}, we show the time
evolution of the phonon quadratures $\tilde{\Delta}_{R}$ and the diagonal
part of the phonon correlations $\left\langle \delta x_{i}^{2}\right\rangle $
for the SSH model with $\omega _{0}/t_{0}=0.5$ and $g/t_{0}=1$. We again set
the electron to be at $j_{0}=0$ and choose polaron momenta $k=0$ (the first
row) and $k=0.88$ (the second row).

\subsection{ Summary of Section \protect\ref{SP}}

We used the non-Gaussian state approach to study the ground state properties
and real time dynamics of polarons. We computed their dispersion,
quasiparticle weight, and obtained full spectral functions. We discussed the
quantum phase transition for SSH polarons, which corresponds to the lowest
energy state of the polaron changing from $k=0$ to finite momentum. What
makes the single polaron problems special is that the LLP transformation
does not involve any variational parameters. Thus we could directly apply
the Gaussian state variational approach to $\bar{H}_{k}$, which describes a
polaron in the co-moving frame. Excellent agreement between our results and
those from earlier studies suggest that a combination of the LLP
transformation and the Gaussian state approach is a powerful theoretical
tool for describing polaronic systems. We point out that similar approach
has also been successfully applied to describe polarons in cold atoms BECs
\cite{Shchadilova} earlier. In the next two examples we consider more
challenging systems in which we need to consider canonical transformations
with time dependent variational parameters in the analysis of both the
ground state and non-equilibrium dynamics.

\section{Nonequilibrium dynamics in Spin-Boson and Kondo models \label{SB}}

In this section we investigate the ground state properties and real-time
dynamics of the spin-boson problem using variational non-Gaussian approach.
The spin-boson problem describes a two level system, i.e. a spin, coupled to
a reservoir of bosonic modes:%
\begin{equation}
H_{\mathrm{SB}}=\frac{\Delta }{2}\sigma _{x}+\sum_{k}\varepsilon
_{k}b_{k}^{\dagger }b_{k}-\frac{1}{2}\sigma
_{z}\sum_{k}g_{k}(b_{k}+b_{k}^{\dagger }).  \label{hsb}
\end{equation}%
Here, $\varepsilon _{k}$ is the dispersion of the boson modes, $g_{k}$ is
their coupling to the spin, and we will use $N_{b}$ to denote the total
number of modes. This section is organized as follows. In the subsection \ref%
{SB} A we review the relation between the spin-boson model and the fermionic
Kondo model \cite{TLS}. This connection relies on bosonizing the 1D Fermi
gas, which can then be mapped onto a spin-boson model with Ohmic
dissipation, which will be the focus of our discussion. The
ferromagnetic/antiferromagnetic phase transition in the Kondo model
corresponds to the localization/delocalization transition in the spin boson
model \cite{TLS}. When presenting the results of our analysis we will
usually do it in the language of the Kondo model since we expect this system
to be more familiar to the readers. In subsections \ref{SB} B-D we introduce
two types of non-Gaussian transformations for analyzing the spin-boson
model. While the two transformations appear to be very different, we show
that they describe the same class of variational wavefunctions. We derive
the equations of motion for the variational parameters for both imaginary
and real time evolution. In the subsection \ref{SB} E we present numerical
results first for the ground state and then for the relaxation dynamics. One
of the surprising findings of our analysis is how well the real time
dynamics follows the RG flow of the equilibrium system. For example, we find
that a system that has ferrmagnetic couplings but in the course of RG flow
parameters flow to the AF regime, exhibits the same ferro to antiferro
crossover in its dynamics.

\subsection{Relation to Kondo physics}

The spin-boson Hamiltonian (\ref{hsb}) is closely related to the Kondo model%
\begin{eqnarray}
H_{\mathrm{K}} &=&\sum_{k\sigma }kc_{k\sigma }^{\dagger }c_{k\sigma }+\frac{%
J_{\perp }}{2}[\sigma _{+}\psi _{\downarrow }^{\dagger }(0)\psi _{\uparrow
}(0)+\mathrm{H.c.}]  \notag \\
&&+\frac{J_{\parallel }}{4}\sigma _{z}[\psi _{\uparrow }^{\dagger }(0)\psi
_{\uparrow }(0)-\psi _{\downarrow }^{\dagger }(0)\psi _{\downarrow }(0)],
\label{hk}
\end{eqnarray}%
where the impurity spin couples anisotropically to the fermionic bath with
strengths $J_{\perp }$ and $J_{\parallel }$. In Eq. (\ref{hk}) we use
fermionic operators at point $x=0$ defined as $\psi _{\sigma
}(x)=\sum_{k}c_{k\sigma }e^{ikx}/\sqrt{L}$ , where $c_{k\sigma }$ are
annihilation operators for fermions with momentum $k$ and spin $\sigma $,
and $L$ is the system size. The connection between the two models (\ref{hsb}%
) and (\ref{hk}) is established by the bosonization dictionary \cite{BF}
\begin{eqnarray}
\psi _{\sigma }(x) &=&\frac{1}{\sqrt{L}}F_{\sigma }e^{i\frac{2\pi N_{\sigma }%
}{L}x}\text{:}\exp i\phi _{\sigma }(x)\text{:},  \notag \\
\rho _{\sigma }(x) &=&\psi _{\sigma }^{\dagger }(x)\psi _{\sigma }(x)=\frac{1%
}{2\pi }\partial _{x}\phi _{\sigma }(x),
\end{eqnarray}%
where the field%
\begin{equation}
\phi _{\sigma }(x)=\sum_{q>0}\frac{e^{-ql_{c}/2}}{\sqrt{n_{q}}}(b_{q\sigma
}e^{iqx}+b_{q\sigma }^{\dagger }e^{-iqx})
\label{Phi_definition_bosonization}
\end{equation}%
is defined by the bosonic annihilation and creation operators $b_{q\sigma }$
and $b_{q\sigma }^{\dagger }$, the integer $n_{q}=qL/(2\pi )$, and the
short-distance cut-off is $l_{c}$. The Klein factor $F_{\sigma }$ obeys the
relations $F_{\sigma }\left\vert N_{\sigma }\right\rangle =\left\vert
N_{\sigma }-1\right\rangle $, $F_{\sigma }^{\dagger }F_{\sigma }=F_{\sigma
}F_{\sigma }^{\dagger }=1$, and $\{F_{\sigma },F_{\sigma ^{\prime
}}^{\dagger }\}=2\delta _{\sigma \sigma ^{\prime }}$, where $\left\vert
N_{\sigma }\right\rangle $ denotes the eigenstate of the number operator $%
N_{\sigma }=\sum_{k}$:$c_{k\sigma }^{\dagger }c_{k\sigma }$:.

The Kondo Hamiltonian can be expressed using the bosonic operators $H_{%
\mathrm{K}}=H_{\text{\textrm{charge}}}+H_{\mathrm{spin}}$, where we
separated the charge part $H_{\text{\textrm{charge}}}=\sum_{k}kb_{kc}^{%
\dagger }b_{kc}$ and the spin part%
\begin{eqnarray}
H_{\mathrm{spin}} &=&\sum_{k}kb_{ks}^{\dagger }b_{ks}+\frac{J_{\parallel }}{4%
\sqrt{2}\pi }\sigma _{z}\partial _{x}\phi _{s}(0)  \notag \\
&&+\frac{J_{\perp }}{4\pi l_{c}}[\sigma _{+}e^{i\sqrt{2}\phi _{s}(0)}+%
\mathrm{H.c.}].
\end{eqnarray}%
Here, $\sigma _{+}$ is redefined as $\sigma _{+}F_{\downarrow }^{\dagger
}F_{\uparrow }\rightarrow \sigma _{+}$, and the charge and spin fields $\phi
_{c(s)}=(\phi _{\uparrow }\pm \phi _{\downarrow })/\sqrt{2}$:%
\begin{equation}
\phi _{c(s)}=\sum_{q>0}\frac{e^{-ql_{c}/2}}{\sqrt{n_{q}}}%
(b_{q,c(s)}e^{iqx}+b_{q,c(s)}^{\dagger }e^{-iqx})
\end{equation}%
are determined by $b_{k,c(s)}=(b_{k\uparrow }\pm b_{k\downarrow })/\sqrt{2}$%
. In the Hamiltonian $H_{\mathrm{K}}$, the charge part $H_{\text{\textrm{%
charge}}}$ is decoupled from the spin Hamiltonian $H_{\mathrm{spin}}$, and
the impurity spin only couples to the spin density excitation in the bath.
In the following, we focus on the spin dynamics governed by the interacting
Hamiltonian $H_{\mathrm{spin}}$.

Under the unitary transformation $U_{\gamma }=e^{i\sqrt{2}\gamma \sigma
_{z}\phi _{s}(0)/2}$, the Hamiltonian $\bar{H}_{\gamma }=U_{\gamma
}^{\dagger }H_{\mathrm{spin}}U_{\gamma }$ in the new basis becomes%
\begin{eqnarray}
\bar{H}_{\gamma } &=&\sum_{k}kb_{ks}^{\dagger }b_{ks}-i\frac{1}{2}\sigma
_{z}\sum_{k>0}g_{k}^{\gamma }(b_{ks}-b_{ks}^{\dagger })  \notag \\
&&+\frac{J_{\perp }}{4\pi l_{c}}[\sigma _{+}e^{i\sqrt{2}(1-\gamma )\phi
_{s}(0)}+\mathrm{H.c.}]+E_{0},  \label{hg}
\end{eqnarray}%
where the energy $E_{0}=(\pi \gamma -J_{\parallel }/2)\gamma
\sum_{k}e^{-kl_{c}}/L$, and the coupling constant%
\begin{equation}
g_{k}^{\gamma }=(2\pi \gamma -\frac{J_{\parallel }}{2})\sqrt{\frac{k}{\pi L}}%
e^{-kl_{c}/2}.
\end{equation}%
For the choice $\gamma =1$, the Hamiltonian (\ref{hg}) is exactly the
spin-boson model (\ref{hsb}) with the interaction $g_{k}=g_{k}^{\gamma }$,
where $b_{k}=ib_{ks}$ and $\Delta =J_{\perp }/(2\pi l_{c})$. At the Toulouse
point $J_{\parallel }=4\pi \gamma $ and $\gamma =1-1/\sqrt{2}$, the
Hamiltonian is exactly solvable by the refermionization technique \cite%
{BF,RF}. The equivalence of the two models established via $U_{\gamma =1}$
allows us to related states in the Kondo and spin-boson models as $%
\left\vert \Psi _{\mathrm{NGS}}^{\mathrm{K}}\right\rangle =U_{\gamma
=1}\left\vert \Psi _{\mathrm{NGS}}^{\mathrm{SB}}\right\rangle $.

\subsection{Two non-Gaussian transformations \label{TwoNG}}

In this subsection, we introduce two types of non-Gaussian transformations
for constructing variational states $\left\vert \Psi _{\mathrm{NGS}%
}\right\rangle $ which can be used to describe the spin-boson model (\ref%
{hsb}).

\textit{Unitary transformation based on parity conservation}. \newline
We observe that the the Hamiltonian (\ref{hsb}) conserves the parity $P_{%
\mathrm{ex}}=e^{i\pi N_{\mathrm{ex}}}$, where the excitation number $N_{%
\mathrm{ex}}$ is defined as
\begin{equation}
N_{\mathrm{ex}}=\frac{1}{2}(\sigma _{x}+1)+\sum_{k}b_{k}^{\dagger }b_{k}.
\end{equation}%
We define the unitary transformation $U_{\mathrm{parity}}=e^{S_{\mathrm{%
parity}}}$ with%
\begin{equation}
S_{\mathrm{parity}}=i\frac{\pi }{2}(\sigma _{z}-1)\sum_{k}b_{k}^{\dagger
}b_{k}.  \label{S1SB}
\end{equation}%
Note that this transformation, which we will call parity transformation,
belongs to the class $U_{5}$ and has no variational parameters.

Under the parity transformation, the Hamiltonian $\bar{H}_{\mathrm{parity}%
}=U_{\mathrm{parity}}^{\dagger }H_{\mathrm{SB}}U_{\mathrm{parity}}$ becomes%
\begin{eqnarray}
\bar{H}_{\mathrm{parity}} &=&\frac{\Delta }{2}\sigma _{x}e^{i\pi
\sum_{k}b_{k}^{\dagger }b_{k}}+\sum_{k}\varepsilon _{k}b_{k}^{\dagger }b_{k}
\notag \\
&&-\frac{1}{2}\sum_{k}g_{k}(b_{k}+b_{k}^{\dagger }).  \label{H1SB}
\end{eqnarray}%
Similarly to the LLP transformation the impurity spin degree of freedom has
been effectively eliminated using the parity integral of motion. Indeed,
while Eq. (\ref{H1SB}) still contains the spin operator $\sigma _{x}$, this
operator now commutes with the Hamiltonian and is therefore conserved. It is
easy to see that $\sigma _{x}$ in the last equation corresponds to the
parity operator in the original Hamiltonian $U_{\mathrm{parity}}^{\dagger
}P_{\mathrm{ex}}U_{\mathrm{parity}}=-\sigma _{x}$. 

For the sake of comparison to the Selbey-type transformation discussed
below, we present variational wavefunctions that obey parity conservation
without performing $U_{\mathrm{parity}}$ explicitly. We define $\left\vert
\pm \right\rangle =(\left\vert \uparrow \right\rangle \pm \left\vert
\downarrow \right\rangle )/\sqrt{2}$ as spin eigenstates of $\sigma _{x}$
with eigenvalues $\pm 1$ and observe that in the even subspace ($P_{\mathrm{%
ex}}=1$), any state can be written as
\begin{eqnarray}
\left\vert \Psi _{\mathrm{NGS}}\right\rangle &=&\left\vert -\right\rangle
\left\vert \mathrm{even}\right\rangle +\left\vert +\right\rangle \left\vert
\mathrm{odd}\right\rangle  \notag \\
&=&\frac{1}{\sqrt{2}}(\left\vert \uparrow \right\rangle \left\vert \Psi
^{+}\right\rangle -\left\vert \downarrow \right\rangle \left\vert \Psi
^{-}\right\rangle ),  \label{pp}
\end{eqnarray}%
where and $\left\vert \Psi ^{\pm }\right\rangle =\left\vert \mathrm{even}%
\right\rangle \pm \left\vert \mathrm{odd}\right\rangle $. Similarly, in the
odd subspace ($P_{\mathrm{ex}}=-1$) all states have the form%
\begin{eqnarray}
\left\vert \Psi _{\mathrm{NGS}}\right\rangle &=&\left\vert -\right\rangle
\left\vert \mathrm{odd}\right\rangle +\left\vert +\right\rangle \left\vert
\mathrm{even}\right\rangle  \notag \\
&=&\frac{1}{\sqrt{2}}(\left\vert \uparrow \right\rangle \left\vert \Psi
^{+}\right\rangle +\left\vert \downarrow \right\rangle \left\vert \Psi
^{-}\right\rangle ).  \label{pm}
\end{eqnarray}

We employ the Gaussian ans\"{a}tz for $\left\vert \Psi ^{+}\right\rangle
=\left\vert \Psi _{\mathrm{GS}}^{\mathrm{parity}}\right\rangle _{b}$, which
leads to $\left\vert \Psi ^{-}\right\rangle =\exp (i\pi
\sum_{k}b_{k}^{\dagger }b_{k})\left\vert \Psi ^{+}\right\rangle $. Then we
have the Gaussian states
\begin{equation}
\left\vert \Psi ^{\pm }\right\rangle =e^{i\theta _{0}}e^{\pm i\frac{1}{2}%
R^{T}\sigma \Delta _{R}}e^{-i\frac{1}{4}R^{T}\xi _{b}R}\left\vert
0\right\rangle .
\end{equation}%
We observe that the two Eqs. (\ref{pp}) and (\ref{pm}) can be combined into
a single non-Gaussian ans\"{a}tz
\begin{equation}
\left\vert \Psi _{\mathrm{NGS}}\right\rangle =U_{\mathrm{parity}}\left\vert
\mp \right\rangle \left\vert \Psi _{\mathrm{GS}}^{\mathrm{parity}%
}\right\rangle _{b},  \label{p1}
\end{equation}

From the Hamiltonian (\ref{H1SB}), we notice that in the even (odd)
subspace, i.e., $\sigma _{x}=-1$ ($1$), $\left\langle e^{i\pi
\sum_{k}b_{k}^{\dagger }b_{k}}\right\rangle $ tends to be positive
(negative) to minimize the ground state energy. However, as we show in Eq. (%
\ref{B0}) in Appendix \ref{AppendixMB}, the expectation value%
\begin{equation}
\left\langle e^{i\pi \sum_{k}b_{k}^{\dagger }b_{k}}\right\rangle _{\mathrm{GS%
}}=e^{-\frac{1}{2}\Delta _{R}^{T}\Gamma _{b}^{-1}\Delta _{R}}
\end{equation}%
is always positive for a Gaussian state. Thus we expect state (\ref{p1}) to
be a good variational wavefunction only in the even subspace ($P_{\mathrm{ex}%
}=1$). To study the ground state and real-time dynamics in the odd subspace (%
$P_{\mathrm{ex}}=-1$), we could use the non-Gaussian state%
\begin{equation}
\left\vert \Psi _{\mathrm{NGS}}\right\rangle =U_{\mathrm{parity}}\left\vert
+\right\rangle U_{3}\left\vert \Psi _{\mathrm{GS}}^{\mathrm{parity}%
}\right\rangle _{b},
\end{equation}%
where $U_{3}$ is applied to tune the weights of the excitations with even
and odd numbers in the Gaussian state $\left\vert \Psi _{\mathrm{GS}}^{%
\mathrm{parity}}\right\rangle _{b}$ such that for $| \tilde{\Psi} \rangle =
U_3 \left\vert \Psi _{\mathrm{GS}}^{\mathrm{parity} }\right\rangle _{b} $
\begin{equation}
\left\langle \tilde{\Psi} \,| \, e^{i\pi \sum_{k}b_{k}^{\dagger }b_{k}} \,|
\, \tilde{\Psi} \,\right\rangle =\left\langle U_{3}^{\dagger }e^{i\pi
\sum_{k}b_{k}^{\dagger }b_{k}}U_{3}\right\rangle _{\mathrm{GS}}
\end{equation}%
can be negative. In this paper we only discuss the ground state and spin
dynamics in the even subspace ($P_{\mathrm{ex}}=1$) and relegate analysis of
the odd sector to future publications.

\textit{Approach based on partial polaron transformation}. \newline
Another approach to constructing variational non-Gaussian states for the
spin-boson model is motivated by Silbey's partial polaron transformation
\cite{Silbey}. We consider the ans\"{a}tz%
\begin{equation}
\left\vert \Psi _{\mathrm{NGS}}\right\rangle =U_{\mathrm{polaron}}\left\vert
\Psi _{\mathrm{GS}}^{\mathrm{polaron}}\right\rangle _{s}\left\vert \Psi _{%
\mathrm{GS}}^{\mathrm{polaron}}\right\rangle _{b},  \label{pp2}
\end{equation}%
where the polaron transformation $U_{\mathrm{polaron}}=e^{S_{\mathrm{polaron}%
}}$ belongs to the class $U_{4}$ with
\begin{equation}
S_{\mathrm{polaron}}=iR^{T}\lambda \sigma _{z}.  \label{polaron}
\end{equation}%
This transformation contains $2N_{b}$ variational parameters in the vector $%
\lambda =(\lambda _{x,k},\lambda _{p,k})^{T}$. Since $U_{\mathrm{polaron}}$
preserves the parity $P_{\mathrm{ex}}$, the Gaussian state $\left\vert \Psi
_{\mathrm{GS}}^{\mathrm{polaron}}\right\rangle _{b}=e^{i\theta _{0}}e^{-i%
\frac{1}{4}R^{T}\xi _{b}R}\left\vert 0\right\rangle $ is a squeezed state
with an even number of bosonic excitations in the bath, and the spin state $%
\left\vert \Psi _{\mathrm{GS}}^{\mathrm{polaron}}\right\rangle
_{s}=\left\vert \pm \right\rangle $ determines the parity $P_{\mathrm{ex}%
}=\mp 1$.

We focus on the even subspace ($P_{\mathrm{ex}}=1$), where the non-Gaussian
ans\"{a}tz%
\begin{eqnarray}
\left\vert \Psi _{\mathrm{NGS}}\right\rangle &=&U_{\mathrm{polaron}%
}\left\vert -\right\rangle \left\vert \Psi _{\mathrm{GS}}^{\mathrm{polaron}%
}\right\rangle _{b}  \notag \\
&=&\frac{1}{\sqrt{2}}(\left\vert \uparrow \right\rangle \left\vert \Psi
^{+}\right\rangle -\left\vert \downarrow \right\rangle \left\vert \Psi
^{-}\right\rangle )  \label{p2}
\end{eqnarray}%
is determined by $\left\vert \Psi ^{\pm }\right\rangle =e^{\pm iR^{T}\lambda
}\left\vert \Psi _{\mathrm{GS}}^{\mathrm{polaron}}\right\rangle _{b}$.
Comparing the states (\ref{p1}) and (\ref{p2}), we notice that these two
transformations lead to the same variational state, where $\lambda =\sigma
\Delta _{R}/2$.

Following the polaron transformation, the Hamiltonian $\bar{H}_{\mathrm{%
polaron}}=U_{\mathrm{polaron}}^{\dagger }H_{\mathrm{SB}}U_{\mathrm{polaron}}$
becomes%
\begin{eqnarray}
\bar{H}_{\mathrm{polaron}} &=&\frac{\Delta }{2}(\sigma
_{+}e^{-2iR^{T}\lambda }+\mathrm{H.c.})+\frac{1}{4}R^{T}\mathbf{\varepsilon }%
R  \notag \\
&&-\frac{1}{2}\sigma _{z}R^{T}G+C_{0},
\end{eqnarray}%
where the matrix $\mathbf{\varepsilon }={\openone}_{2}\otimes
diag(\varepsilon _{k})$, the vector $G=(g_{k}+2\varepsilon _{k}\lambda
_{p,k},-2\varepsilon _{k}\lambda _{x,k})^{T}$, and $C_{0}=\lambda ^{T}%
\mathbf{\varepsilon }\lambda +g^{T}\lambda _{p}-\sum_{k}\varepsilon _{k}/2$
is defined by the vector $g^{T}=(g_{1},...,g_{k},...)$.

We remark that the Hamiltonian $\bar{H}_{\mathrm{polaron}}$ differs from the
previously considered setting of Bose/Fermi systems because it contains spin
operators. We can however proceed with our usual framework using a fermionic
representation of spin operators. We define $\sigma _{+}=c_{\uparrow
}^{\dagger }c_{\downarrow }$ and $\sigma _{z}=c_{\uparrow }^{\dagger
}c_{\uparrow }-c_{\downarrow }^{\dagger }c_{\downarrow }$. The covariance
matrix of the two-mode fermions is $\Gamma _{f}=\left\langle CC^{\dagger
}\right\rangle $, where $C=(c_{\uparrow },c_{\downarrow },c_{\uparrow
}^{\dagger },c_{\downarrow }^{\dagger })^{T}$. To describe the spin in the
two-dimensional Hilbert space, the four-dimensional fermionic space must be
restricted to the single occupation subspace. Then the pairing terms $%
\left\langle c_{\downarrow }c_{\uparrow }\right\rangle $ and $\left\langle
c_{\uparrow }c_{\downarrow }\right\rangle $ are not allowed and $\Gamma _{f}$
is block-diagonal.

In the next three subsections we present results for the ground state
properties and real-time dynamics in the even subspace using only one of the
transformations, since results for the other one should be identical.

\subsection{Parity transformation}

In this subsection, we derive the equations of motion for $\Delta _{R}$ and $%
\Gamma _{b}$ in the non-Gaussian state (\ref{p1}) given by the parity
transformation. Following the procedure in Appendix \ref{AppendixME}, we
obtain Eqs. (\ref{RGGSI}) and (\ref{RGGSR}) for the imaginary- and real-
time evolutions, which are equivalent to Eqs. (\ref{IMNGS}) and (\ref{RENGS}%
) with $O_{\Delta }=0$ and $O_{b}=0$.

As shown in Eq. (\ref{hY}), the vector $h_{\Delta }=h_{\Delta }^{t}=2\delta
E_{\mathrm{parity}}/\delta \Delta _{R}$ and the matrix $h_{b}=h_{b}^{t}=4%
\delta E_{\mathrm{parity}}/\delta \Gamma _{b}$ are determined by the
functional derivatives. The energy $E_{\mathrm{parity}}=\left\langle \bar{H}%
_{\mathrm{parity}}\right\rangle _{\mathrm{GS}}$ is%
\begin{eqnarray}
E_{\mathrm{parity}} &=&\frac{1}{4}\Delta _{R}^{T}\mathbf{\varepsilon }\Delta
_{R}+\frac{1}{4}tr(\mathbf{\varepsilon }\Gamma _{b})-\frac{1}{2}%
\sum_{k}\varepsilon _{k}  \notag \\
&&-\frac{\Delta }{2}e^{-\frac{1}{2}\Delta _{R}^{T}\Gamma _{b}^{-1}\Delta
_{R}}-\frac{1}{2}\sum_{k}g_{k}\left\langle x_{k}\right\rangle .  \label{E1SB}
\end{eqnarray}%
It follows from Eqs. (\ref{E1SB}) and (\ref{hY}) that%
\begin{eqnarray}
h_{\Delta } &=&\mathbf{\varepsilon }\Delta _{R}+\Delta e^{-\frac{1}{2}\Delta
_{R}^{T}\Gamma _{b}^{-1}\Delta _{R}}\Gamma _{b}^{-1}\Delta _{R}-\left(
\begin{array}{c}
g \\
0%
\end{array}%
\right) ,  \notag \\
h_{b} &=&\mathbf{\varepsilon }-\Delta e^{-\frac{1}{2}\Delta _{R}^{T}\Gamma
_{b}^{-1}\Delta _{R}}\Gamma _{b}^{-1}\Delta _{R}\Delta _{R}^{T}\Gamma
_{b}^{-1}.  \label{hsbp}
\end{eqnarray}%
Here, $(g,0)^{T}$ is a short hand notation for $%
(g_{1},g_{2},...g_{k},0,...,0)^{T}$. By solving Eqs. (\ref{RGGSI}) and (\ref%
{RGGSR}) with $h_{\Delta }$ and $h_{b}$ determined by Eq. (\ref{hsbp}), we
obtain $\Delta _{R}$ and $\Gamma _{b}$ in the ground state and the real-time
dynamics. In the ground state, the solution of $\Delta _{R}$ satisfies the
nonlinear equation%
\begin{equation}
\Delta _{R}=\frac{1}{\mathbf{\varepsilon }+\Delta e^{-\frac{1}{2}\Delta
_{R}^{T}\Gamma _{b}^{-1}\Delta _{R}}\Gamma _{b}^{-1}}\left(
\begin{array}{c}
g \\
0%
\end{array}%
\right)  \label{Dr}
\end{equation}%
obtained from the fixed point condition $h_{\Delta }=0$.

\subsection{Polaron transformation}

In this subsection, we derive the equations of motion for $\Delta _{R}$, $%
\Gamma _{b,f}$, and $\lambda $ in the non-Gaussian state defined by the
polaron transformation (\ref{pp2}) and (\ref{polaron}). We follow the
general procedure discussed in Appendix \ref{AppendixME} [see equations (\ref%
{IMNGS}) and (\ref{RENGS})] to determine time evolution of $\lambda $, $%
\Delta _{R}$, and $\Gamma _{b,f}$. Expressions for the vectors $O_{\Delta }$%
, $h_{\Delta }$ and matrices $O_{b,f}$, $h_{b,f}$ can be obtained using
functional derivatives as discussed in Eq. (\ref{OY}), where the mean-values
are%
\begin{equation}
\left\langle O\right\rangle _{\mathrm{GS}}=i\lambda ^{T}\sigma \partial
_{\tau }\lambda +i\Delta _{R}^{T}\partial _{\tau }\lambda \left\langle
\sigma _{z}\right\rangle _{\mathrm{GS}},  \label{Osb}
\end{equation}%
and $\left\langle \bar{H}_{\mathrm{polaron}}\right\rangle _{\mathrm{GS}%
}\equiv E_{\mathrm{polaron}}$:%
\begin{eqnarray}
E_{\mathrm{polaron}} &=&\Delta \text{Re}\left\langle \sigma
_{+}\right\rangle _{\mathrm{GS}}\left\langle e^{-2iR^{T}\lambda
}\right\rangle _{\mathrm{GS}}+\frac{1}{4}\Delta _{R}^{T}\mathbf{\varepsilon }%
\Delta _{R}  \notag \\
&&+\frac{1}{4}tr(\mathbf{\varepsilon }\Gamma _{b})-\frac{1}{2}\left\langle
\sigma _{z}\right\rangle _{\mathrm{GS}}\Delta _{R}^{T}G+C_{0}.  \label{Esb}
\end{eqnarray}%
The expectation value $\left\langle e^{-2iR^{T}\lambda }\right\rangle _{%
\mathrm{GS}}=e^{-2i\Delta _{R}^{T}\lambda }e^{-2\lambda ^{T}\Gamma
_{b}\lambda }$ follows from Eq. (\ref{NO}), and $\left\langle \sigma _{z,\pm
}\right\rangle _{\mathrm{GS}}$ can be easily expressed as linear
combinations of the matrix elements $(\Gamma _{f})_{ij}$.

The functional derivatives of $\left\langle O\right\rangle _{\mathrm{GS}}$
and $E_{\mathrm{polaron}}$ result in%
\begin{equation}
O_{\Delta }=2i\partial _{\tau }\lambda \left\langle \sigma _{z}\right\rangle
_{\mathrm{GS}},O_{b}=0,O_{f}=\tau _{z}\otimes i\Delta _{R}^{T}\partial
_{\tau }\lambda \tau _{z},  \label{Dsb}
\end{equation}%
and%
\begin{eqnarray}
h_{\Delta } &=&\mathbf{\varepsilon }\Delta _{R}+4\Delta \text{Im}%
\left\langle \sigma _{+}\right\rangle _{\mathrm{GS}}\left\langle
e^{-2iR^{T}\lambda }\right\rangle _{\mathrm{GS}}\lambda -\left\langle \sigma
_{z}\right\rangle _{\mathrm{GS}}G,  \notag \\
h_{b} &=&\mathbf{\varepsilon }-8\Delta \text{Re}\left\langle \sigma
_{+}\right\rangle _{\mathrm{GS}}\left\langle e^{-2iR^{T}\lambda
}\right\rangle _{\mathrm{GS}}\lambda \lambda ^{T},  \label{nsb} \\
h_{f} &=&\tau _{z}\otimes \frac{1}{2}[\Delta (\left\langle
e^{-2iR^{T}\lambda }\right\rangle _{\mathrm{GS}}\tau _{+}+\mathrm{H.c.}%
)-\Delta _{R}^{T}G\tau _{z}],  \notag
\end{eqnarray}%
where $\tau _{z,\pm }$ are Pauli matrices defined in the basis $%
C=(c_{\uparrow },c_{\downarrow },c_{\uparrow }^{\dagger },c_{\downarrow
}^{\dagger })^{T}$. The coefficients (\ref{Dsb}) and (\ref{nsb}) in the
normal ordering expansion lead to Eqs. (\ref{IMNGS}) and (\ref{RENGS}) for $%
\Delta _{R}$ and $\Gamma _{b}$.

The equations of motion for $\lambda $ can be obtained using the projection
on the tangent vector $U_{\mathrm{polaron}}U_{\mathrm{GS}}\left\vert
D_{k}\right\rangle $ that is equivalent to the projection of states (\ref%
{UL0}) and (\ref{UR0}) on the state $\left\vert D_{k}\right\rangle
=b_{k}^{\dagger }$:$U_{\mathrm{GS}}^{\dagger }\sigma _{z}U_{\mathrm{GS}}$:$%
\left\vert 0\right\rangle $, as shown in Appendix \ref{AppendixME}. The
projection leads to%
\begin{eqnarray}
\left\langle D_{k}\right\vert \delta O\left\vert 0\right\rangle
&=&-\left\langle D_{k}\right\vert \delta \bar{H}_{\mathrm{polaron}%
}\left\vert 0\right\rangle ,  \notag \\
\left\langle D_{k}\right\vert \delta O\left\vert 0\right\rangle
&=&-i\left\langle D_{k}\right\vert \delta \bar{H}_{\mathrm{polaron}%
}\left\vert 0\right\rangle ,  \label{Rsb}
\end{eqnarray}%
for the imaginary- and real- time evolutions, respectively, where the
operator $\delta O=iR^{T}S_{b}^{T}\partial _{\tau }\lambda \tilde{\sigma}%
_{z} $ and the cubic operator%
\begin{equation}
-\frac{1}{2}R^{T}S_{b}^{T}G\tilde{\sigma}_{z}-i\Delta \left\langle
e^{-2iR^{T}\lambda }\right\rangle _{\mathrm{GS}}R^{T}S_{b}^{T}\lambda \tilde{%
\sigma}_{+}+\mathrm{H.c.}
\end{equation}%
in $\delta \bar{H}_{\mathrm{polaron}}$ are determined by the normal ordered
operators $\tilde{\sigma}_{z,\pm }=$:$U_{\mathrm{GS}}^{\dagger }\sigma
_{z,\pm }U_{\mathrm{GS}}$:. Relation (\ref{Rsb}) then leads to the motion
equations%
\begin{widetext}
\begin{eqnarray}
(1-\left\langle \sigma _{z}\right\rangle _{\mathrm{GS}}^{2})\sigma \partial
_{\tau }\lambda  &=&2\Delta \text{Re}(\left\langle \sigma _{+}\right\rangle
_{\mathrm{GS}}\left\langle e^{-2iR^{T}\lambda }\right\rangle _{\mathrm{GS}%
})\sigma \lambda -2\Delta \left\langle \sigma _{z}\right\rangle _{\mathrm{GS}%
}\text{Im}(\left\langle \sigma _{+}\right\rangle _{\mathrm{GS}}\left\langle
e^{-2iR^{T}\lambda }\right\rangle _{\mathrm{GS}})\Gamma _{b}\lambda -\frac{1%
}{2}(1-\left\langle \sigma _{z}\right\rangle _{\mathrm{GS}}^{2})\Gamma _{b}G,
\notag \\
(1-\left\langle \sigma _{z}\right\rangle _{\mathrm{GS}}^{2})\partial
_{t}\lambda  &=&-2\Delta \text{Re}(\left\langle \sigma _{+}\right\rangle _{%
\mathrm{GS}}\left\langle e^{-2iR^{T}\lambda }\right\rangle _{\mathrm{GS}%
})\sigma \Gamma _{b}\lambda +2\Delta \left\langle \sigma _{z}\right\rangle _{%
\mathrm{GS}}\text{Im}(\left\langle \sigma _{+}\right\rangle _{\mathrm{GS}%
}\left\langle e^{-2iR^{T}\lambda }\right\rangle _{\mathrm{GS}})\lambda +%
\frac{1}{2}(1-\left\langle \sigma _{z}\right\rangle _{\mathrm{GS}}^{2})G.
\notag \\
&&  \label{LSB}
\end{eqnarray}%
\end{widetext}

The solution of Eqs. (\ref{IMNGS}), (\ref{RENGS}), and (\ref{LSB}) in the
even subspace has the following properties: (a) $\Delta _{R}=0$ and (b)%
\begin{equation}
\left\langle \sigma _{+}\right\rangle _{\mathrm{GS}}=\left\langle \sigma
_{-}\right\rangle _{\mathrm{GS}}=-\frac{1}{2},\left\langle \sigma
_{z}\right\rangle _{\mathrm{GS}}=0,
\end{equation}%
which imply that in Eq. (\ref{pp2}) the Gaussian state $\left\vert \Psi _{%
\mathrm{GS}}^{\mathrm{polaron}}\right\rangle _{s}$ of the impurity is $%
\left\vert -\right\rangle $ and the Gaussian state $\left\vert \Psi _{%
\mathrm{GS}}^{\mathrm{polaron}}\right\rangle _{b}$ of the bath is a squeezed
state, as we discussed in Sec. \ref{TwoNG}. The squeezing part of the
bosonic wavefucntion around the impurity is described by $\Gamma _{b}$ and $%
\lambda $, which obey equations of motion%
\begin{eqnarray}
\partial _{\tau }\Gamma _{b} &=&\sigma ^{T}h_{b}\sigma -\Gamma
_{b}h_{b}\Gamma _{b},  \notag \\
\partial _{\tau }\lambda &=&-\Delta e^{-2\lambda ^{T}\Gamma _{b}\lambda
}\lambda +\frac{1}{2}\sigma \Gamma _{b}G,  \label{hsbpi}
\end{eqnarray}%
and%
\begin{eqnarray}
\partial _{t}\Gamma _{b} &=&\sigma h_{b}\Gamma _{b}-\Gamma _{b}h_{b}\sigma ,
\notag \\
\partial _{t}\lambda &=&\Delta e^{-2\lambda ^{T}\Gamma _{b}\lambda }\sigma
\Gamma _{b}\lambda +\frac{1}{2}G,  \label{hsbpr}
\end{eqnarray}%
for the imaginary- and real- time evolution respectively.

In the ground state, the fixed point condition $\partial _{\tau }\lambda =0$
results in%
\begin{equation}
\sigma \lambda =-\frac{1}{2}\frac{1}{\Delta e^{-2\lambda ^{T}\Gamma
_{b}\lambda }\Gamma _{b}^{-1}+\varepsilon }\left(
\begin{array}{c}
g \\
0%
\end{array}%
\right) .  \label{L}
\end{equation}%
It immediately follows from Eqs. (\ref{Dr}) and (\ref{L}) that $\lambda
=\sigma \Delta _{R}/2$ is in agreement with the result of Sec. \ref{TwoNG}.

\begin{figure}[tbp]
\includegraphics[width=0.9\linewidth]{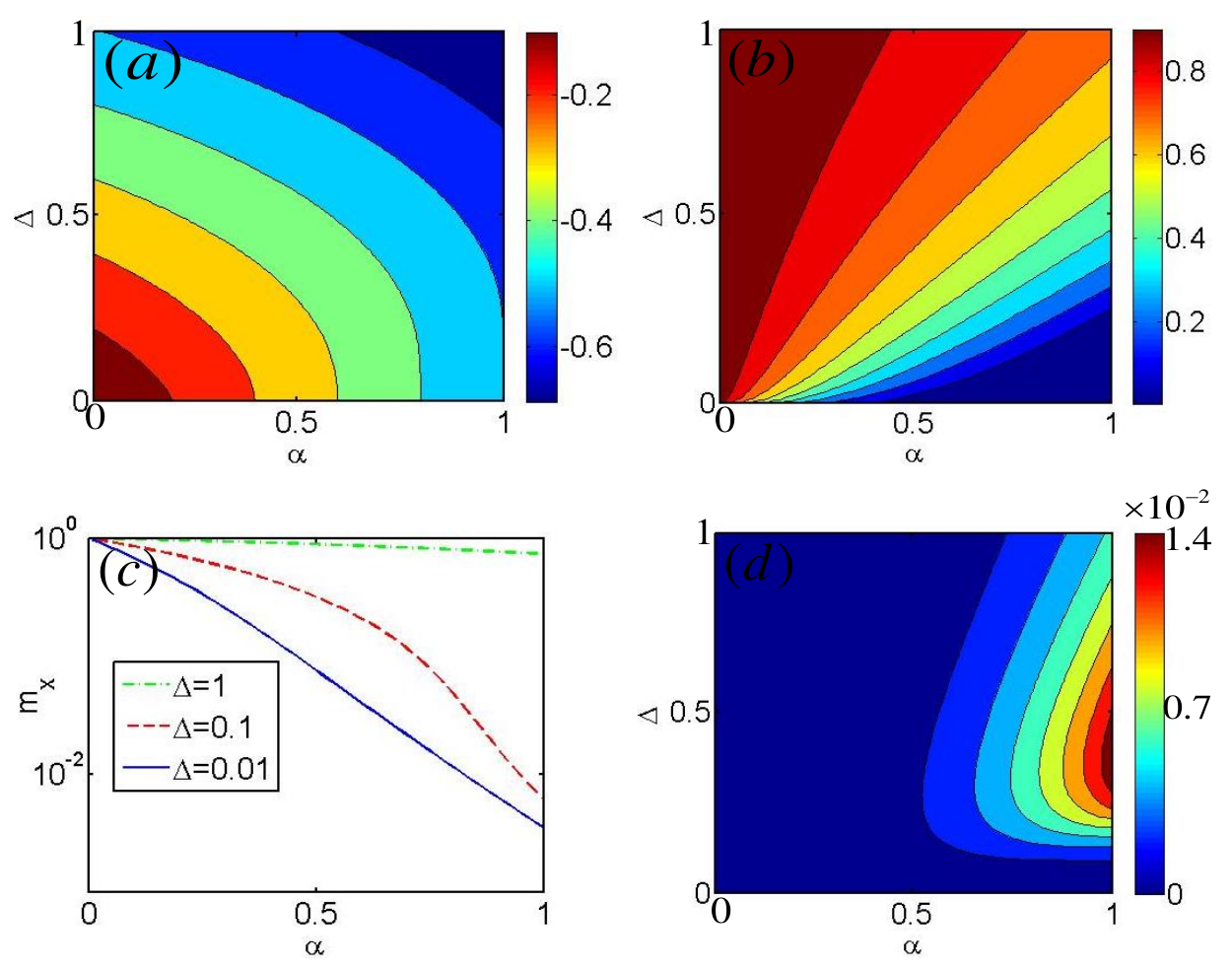}
\caption{The ground state energy, the magnetization, and the square norm $%
\mathcal{N}_{\mathrm{\perp }}$ in the $\protect\alpha $-$\Delta $ plane for
the spin-boson model with the Ohmic spectrum, where the frequency cut-off $%
\protect\omega _{c}$ is taken as the unit, and the system size is $N_b=200$.
(a) The ground state energy $E_{\mathrm{GS}}$; (b) The magnetization $m_{%
\mathrm{x}}$ along $x$-direction; (c) The magnetization along the cuts $%
\Delta =0.01$, $0.1$, and $1$; (d) The small norm $\mathcal{N}_{\mathrm{%
\perp }}$.}
\label{EGSB}
\end{figure}

\begin{figure}[tbp]
\includegraphics[width=0.9\linewidth]{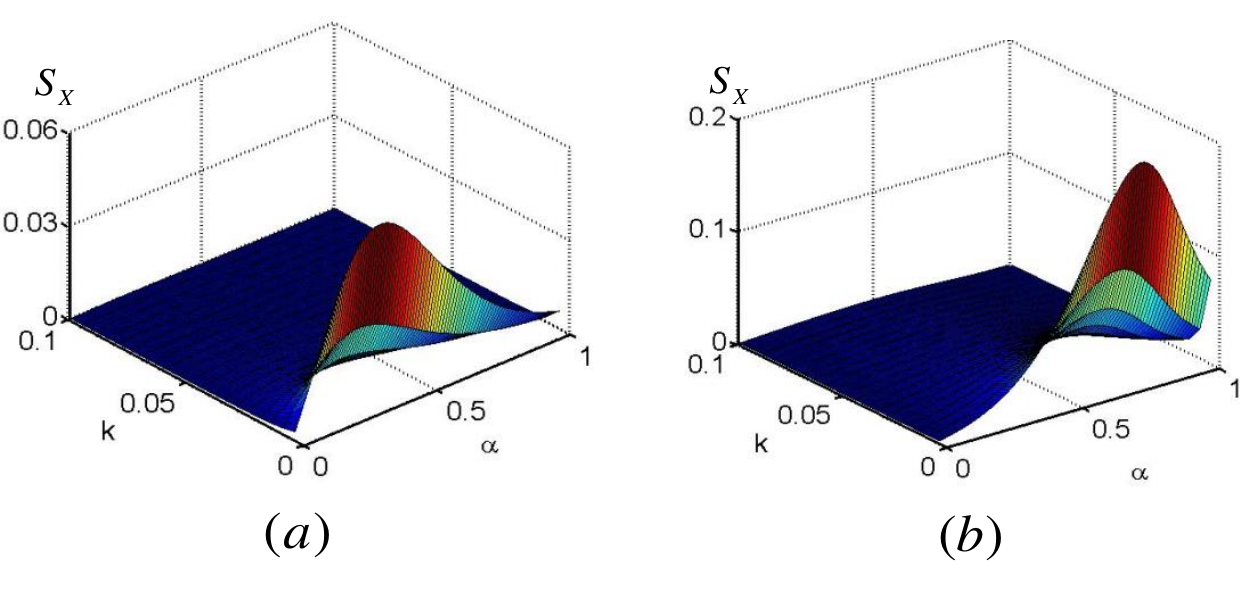}
\caption{The squeezing parameters $S_X$ for $\Delta=0.01$ (a) and $%
\Delta=0.1 $ (b). Here, system size $N_b=200$. }
\label{SBGx}
\end{figure}

\subsection{Numerical results}

In this subsection, we study the ground state and non-equilibrium dynamics
of the anisotropic Kondo model in different parameter regimes. We use the
transformation between the Kondo and spin-boson Hamiltonians $\left\vert
\Psi _{\mathrm{NGS}}^{\mathrm{K}}\right\rangle =U_{\gamma =1}\left\vert \Psi
_{\mathrm{NGS}}^{\mathrm{SB}}\right\rangle $ (see discussion above) to
translate the problems into the spin-boson model and analyze the latter. We
remind the readers that Kondo Hamiltonian is mapped to the spin-boson models
with Ohmic dissipation, i.e., $\varepsilon _{k}\equiv k$ and $g_{k}=\sqrt{%
2\alpha \omega _{c}k/N_{b}}\theta (\omega _{c}-k)$. The relation between the
interaction parameters in the two models is $\alpha =[1-J_{\parallel }/(4\pi
)]^{2}$ ($\alpha =1/2$ is the Toulouse point). To analyze the spin-boson
model we solve equations (\ref{IMNGS}), (\ref{RENGS}), and (\ref{LSB})
numerically, using the cut-off frequency $\omega _{c}$ as the unit of
energy. We used mode discretization as $k=\omega _{c}n_{k}/N_{b}$ where $%
n_{k}=1,2,..,N$ and $N_{b}$. Since the parity and polaron transformations
lead to equivalent variational states, numerical results obtained using Eqs.
(\ref{hsbp}), (\ref{hsbpi}), and (\ref{hsbpr}) give identical results. In
the numerical calculation we use the same energy level spacing $\omega
_{c}/N_{b}=2\pi /L$ for the bath field in the spin-boson and Kondo models,
and the short-distance cut-off $l_{c}$ in equation (\ref%
{Phi_definition_bosonization}) and the sharp frequency cut-off $\omega _{c}$
in equation (\ref{hsb}) are related via
\begin{equation}
\mathbf{\psi }(N_{b}+1)+\gamma _{0}=-\ln (1-e^{-\frac{\omega _{c}}{N_{b}}%
l_{c}}),
\end{equation}%
Here $\mathbf{\psi }(z)$ is the digamma function and $\gamma _{0}$ is Euler
constant.

\begin{figure}[tbp]
\includegraphics[width=0.9\linewidth]{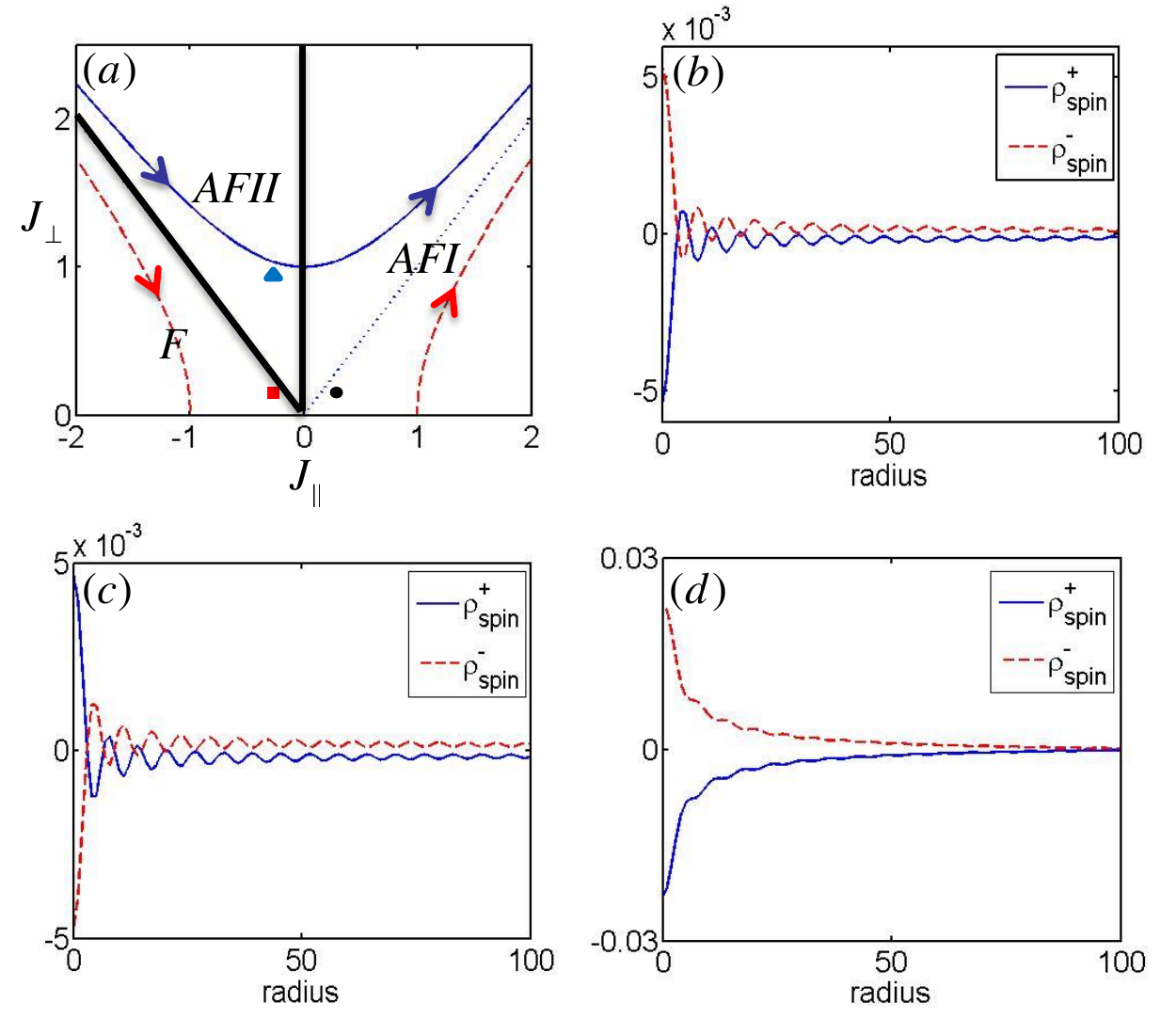}
\caption{(a) Phase diagram and RG flows of Kondo model. (b)-(d) The spin
density fluctuations display the transition between the antiferromagnetic
and ferromagnetic phases, where $N_b=200$: (b) $J_{\perp }=0.1$ and $%
J_{\parallel }=0.2$; (c) $J_{\perp }=0.1$ and $J_{\parallel }=-0.2$; (d) $%
J_{\perp }=1$ and $J_{\parallel }=-0.2$.}
\label{Kondo}
\end{figure}

\begin{figure*}[tbp]
\begin{center}
\includegraphics[width=0.7\linewidth]{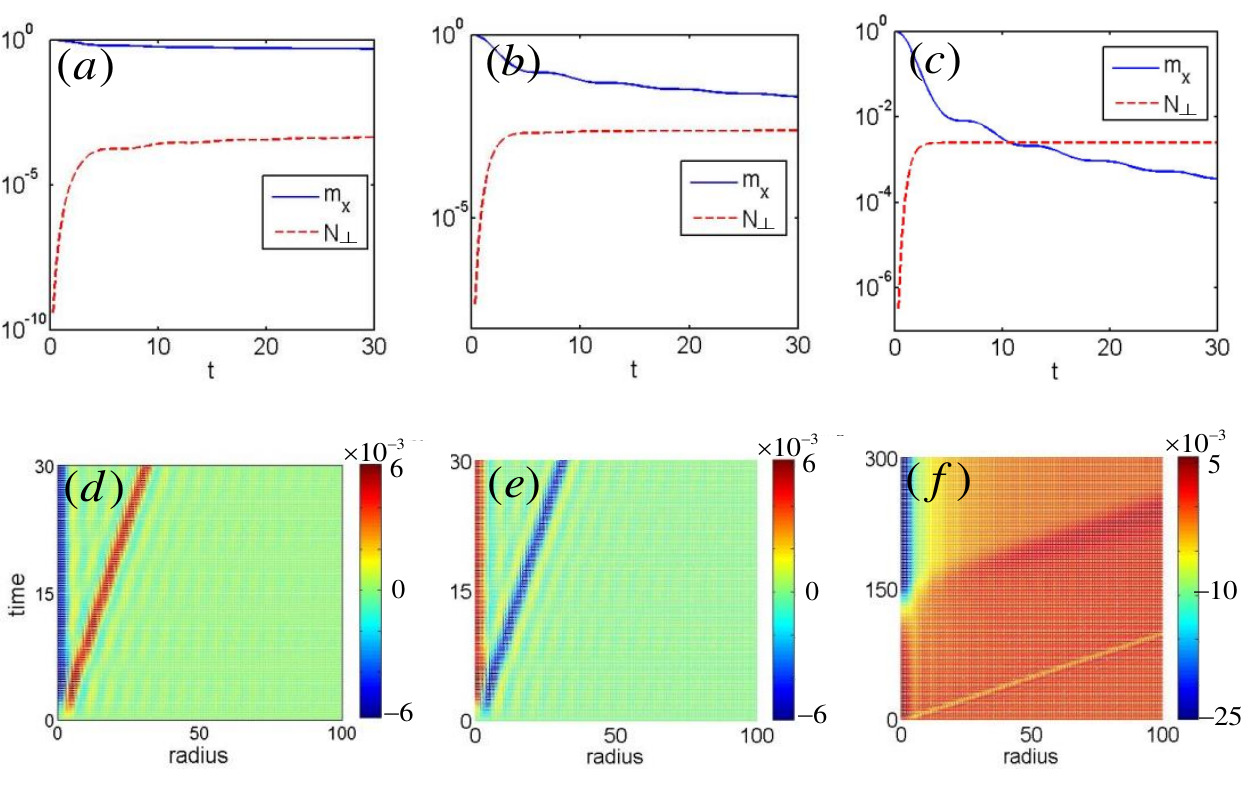}
\end{center}
\caption{(a)-(c) Time evolution of the magnetization $m_{x}$ and the square
norm $\mathcal{N}_{\mathrm{\perp }}$ for $\Delta =0.01$ and the initial
state $\left\vert -\right\rangle \left\vert 0\right\rangle _{b}$, where $%
\protect\alpha =0.1$, $0.5$, and $1.2$ in (a), (b), and (c), respectively;
system size $N_b=200$. (d)-(f) Time evolutions of spin density fluctuations
for the initial state $\left\vert -\right\rangle \left\vert \mathrm{FS}
\right\rangle$ in antiferromagnetic and ferromagnetic phases of Kondo model,
where $J_{\perp }=0.1$ and $J_{\parallel }=0.2$ (d), $J_{\perp }=0.1$ and $%
J_{\parallel }=-0.2$ (e), and $J_{\perp }=1$ and $J_{\parallel }=-0.2$ (f).}
\label{EGSBRT}
\end{figure*}

In Fig. \ref{EGSB}a-b, we show the ground state energy $E_{\mathrm{GS}}=E_{%
\mathrm{parity}}=E_{\mathrm{polaron}}$ and the magnetization
\begin{equation}
m_{x}=-\left\langle \sigma _{x}\right\rangle =e^{-2\lambda ^{T}\Gamma
_{b}\lambda }
\end{equation}%
in the $\alpha $-$\Delta $ parameter plane. In Fig. \ref{EGSB}c, the
magnetization $m_{x}$ is shown along the horizontal cuts $\Delta =0.01$, $%
0.1 $, and $1$. Compared with the Silbey transformation \cite{Silbey},
results for the magnetization $m_{x}$ are considerably improved and are in
excellent agreement with the NRG calculation \cite{spinco}. To check the
validity of the variational approach we can check the energy variance $%
\mathcal{N}_{\mathrm{\perp }}=\left\vert \left\vert \Psi _{\mathrm{\perp }%
}\right\rangle \right\vert ^{2}$, i.e., the square norm%
\begin{equation}
\mathcal{N}_{\mathrm{\perp }}=\frac{\Delta ^{2}}{8}(2-e^{-4y})-2\Delta
^{2}e^{-4y}(y+\frac{1}{4})^{2}  \label{Nperp}
\end{equation}%
of the state $\left\vert \Psi _{\mathrm{\perp }}\right\rangle $ orthogonal
to the tangent space in the limit $\tau \rightarrow \infty $, where $%
y=\lambda ^{T}\Gamma _{b}\lambda $. In Fig. \ref{EGSB}d, the small energy
variance $\mathcal{N}_{\mathrm{\perp }}<1.5\times 10^{-2}$ in the $\alpha $-$%
\Delta $ plane justifies the validity of the non-Gaussian variational state
in the even excitation subspace.

The magnetization $m_{x}$ in Figs. \ref{EGSB}b-c shows that for fixed $%
\Delta $, increasing the coupling constant $\alpha $ reduces the
magnetization. For a small coupling constant $\alpha \ll 1$, the state (\ref%
{p1}) in the even subspace can be approximated as $\left\vert -\right\rangle
\left\vert 0\right\rangle $. As $\alpha $ increases, the spin is entangled
with the bath degree of freedom. When the weights $\left\vert \left\vert
\mathrm{even}\right\rangle \right\vert ^{2}\sim \left\vert \left\vert
\mathrm{odd}\right\rangle \right\vert ^{2}$, the spin magnetization $%
m_{x}\sim 0$ due to the strong entanglement.

The main difference between the variational ansatz (\ref{p2}) and the
variational state $e^{S_{\mathrm{polaron}}}\left\vert -\right\rangle
\left\vert 0\right\rangle _{b}$ used by Silbey and collaborators (see e.g.
Ref. \cite{Silbey}) is the vacuum state of the bath degrees of freedom. The
imaginary time evolution allows us to minimize the energy with respect to
the covariance matrix with $\sim 3N_{b}^{2}$ variational parameters. Figure %
\ref{SBGx} shows the $k$-mode squeezing parameter $(S_{X})_{k}=\left\langle
x_{k}^{2}\right\rangle -1$, where $\Delta =0.01$ and $\Delta =0.1$ in Figs. %
\ref{SBGx}a and \ref{SBGx}b, respectively. The low frequency modes are
squeezed along the $p$-direction in phase space, and the high frequency
modes are in the vacuum state. As $\Delta $ increases, the peak position of $%
S_{X}$ shifts towards a larger $\alpha $, and the peak value increases.

The ground state%
\begin{equation}
\left\vert \Psi _{\mathrm{NGS}}^{\mathrm{K}}\right\rangle =\frac{1}{\sqrt{2}}%
(\left\vert \uparrow \right\rangle \left\vert \Psi _{\mathrm{sdw}%
}^{+}\right\rangle -\left\vert \downarrow \right\rangle \left\vert \Psi _{%
\mathrm{sdw}}^{-}\right\rangle )
\end{equation}%
of the Kondo model describes the spin density configuration in the fermionic
bath by $\left\vert \Psi _{\mathrm{sdw}}^{\pm }\right\rangle =e^{\pm i\sqrt{2%
}\phi _{s}(0)/2}\left\vert \Psi ^{\pm }\right\rangle $. The spin density
fluctuation around the impurity is characterized by%
\begin{eqnarray}
\rho _{\mathrm{spin}}^{\pm } &=&\frac{1}{\sqrt{2}\pi }\left\langle \Psi _{%
\mathrm{sdw}}^{\pm }\right\vert \partial _{x}\phi _{s}(x)\left\vert \Psi _{%
\mathrm{sdw}}^{\pm }\right\rangle  \notag \\
&=&\mp \frac{\omega _{c}}{\pi N_{b}}\sum_{q>0}[\sqrt{2n_{q}}\lambda
_{x,q}\sin qx+  \notag \\
&&(1+\sqrt{2n_{q}}\lambda _{p,q})\cos qx].
\end{eqnarray}

It is well-known that the anisotropic Kondo model exhibits a quantum phase
transition between the ferromagnetic and antiferromagnetic phases \cite{TLS}%
. In Fig. \ref{Kondo}a we show the phase diagram in the $J_{\perp }$-$%
J_{\parallel }$ parameter plane. We label the antiferromagnetic phase ($%
J_{\parallel }>0$) \textquotedblleft AFI\textquotedblright , the upper-left
triangular antiferromagnetic phase ($J_{\perp }>-J_{\parallel }>0$)
\textquotedblleft AFII\textquotedblright , and the lower-left triangular
ferromagnetic phase ($-J_{\parallel }>J_{\perp }>0$) \textquotedblleft
F\textquotedblright . The renormalization flows in different regions are
shown by the arrows.

In Figs. \ref{Kondo}b-c, the spin density distributions of the ground states
are shown for the \textquotedblleft AFI\textquotedblright\ phase (the black
dot in Fig. \ref{Kondo}a) and the \textquotedblleft F\textquotedblright\
phase (the red square in Fig. \ref{Kondo}a), which display the singlet and
triplet pairs of the impurity spin and the surrounding electrons. In Fig. %
\ref{Kondo}d, the spin density distributions for the \textquotedblleft
AFII\textquotedblright\ phase (the blue triangle) exhibit an
antiferromagnetic ground state.

In Figs. \ref{EGSBRT}a-c, for the system initially prepared in the state $%
\left\vert -\right\rangle \left\vert 0\right\rangle _{b}$, we show time
evolutions of the magnetization $m_{x}(t)$ for different coupling strength $%
\alpha =0.1$ (a), $0.5$ (b), and $1$ (c), with $\Delta =0.01$. The small
square norm $\mathcal{N}_{\mathrm{\perp }}<2\times 10^{-3}$ justifies the
validity of the variational state. The larger the coupling constant $\alpha $%
, the faster the magnetization $m_{x}(t)$ relaxes to zero, which agrees with
our intuition regarding the entanglement between the spin and the bosonic
bath described by the states (\ref{p1}) and (\ref{p2}).

Time evolution of the spin density configuration $\rho _{\mathrm{spin}%
}^{+}=-\rho _{\mathrm{spin}}^{-}$ for $J_{\perp }=0.1$ and $J_{\parallel
}=0.2$ (Fig. \ref{EGSBRT}d), $J_{\perp }=0.1$ and $J_{\parallel }=-0.2$
(Fig. \ref{EGSBRT}e), and $J_{\perp }=1$ and $J_{\parallel }=-0.2$ (Fig. \ref%
{EGSBRT}f) are shown in Figs. \ref{EGSBRT}d-f. The initial state $\left\vert
-\right\rangle \left\vert \mathrm{FS}\right\rangle $ describes the impurity
spin in the state $\left\vert -\right\rangle $ and the unperturbed Fermi
sea. Note that $b_{qs}\left\vert \mathrm{FS}\right\rangle =0$ indicates that
there are no spin density fluctuations in the initial state.

Figure \ref{EGSBRT}d shows that in the AFI regime, following the quench bath
electrons quickly screen the impurity spin. This is consistent with our
intuition of the spin singlet ground state in the AF Kondo model. Formation
of the screening cloud is accompanied by the spin wavepacket propagating
away from the impurity. Figure \ref{EGSBRT}e shows the spin dynamics in the
opposite regime of the ferromagnetic phase F. In this case, bath electrons
become co-aligned with the impurity spin, which is what we expect based on
the triplet ground state of the system. Notice again a single wavepacket
propagating away from the impurity. The AFII phase shows the most surprising
case of the dynamics. At short times, the electron bath develops a
polarization which is co-aligned with the impurity spin. However, at longer
times, the polarization cloud changes sign and we find the impurity spin
screened by the surrounding electrons. This polarization cloud dynamics is
accompanied by two wavepackets propagating away from the impurity. The first
one appears when the electrons develop a transient ferromagnetic cloud
around the impurity, and the second one when the final antiferromagnetic
cloud is formed. This two-stage spin dynamics is easily understood if we
consider the RG flow diagram in Fig. \ref{Kondo}a. Short time dynamics
corresponds to the high energy Hamiltonian characterized by the
ferromagnetic $J_{\parallel }$ interactions. At longer times we observe low
energy degrees of freedom, which correspond to the antiferromagentic $%
J_{\parallel }$ arising from the RG flow in Fig. \ref{Kondo}a.

\subsection{Summary of Section \protect\ref{SB}}

We now summarize the main results obtained in this section. We introduced
variational approach for describing the ground state and non-equilibrium
dynamics of the spin-boson model. This model is known to be equivalent to
the Kondo Hamiltonian, hence our results have direct implications for the
non-equilibrium dynamics in Kondo-related systems, such as transport through
a quantum dot \cite{Glazman}. We showed that variational approaches can be
introduced either utilizing the conserved parity operator or using a
Silbey-type polaron transformation. Surprisingly both approaches result in
the same variational family of wavefunctions. We used this class of
wavefunctions to analyze the ground state of the Kondo problem and found
excellent agreement with the results of earlier studies. Our variational
approach improves over earlier variational states introduced by Silbey and
collaborators by including squeezing between the bosonic bath modes, which
becomes significant for larger values of dissipation strength $\alpha $. We
applied our variational wavefunctions to study the non-equilibrium dynamics
of the Kondo model with a focus on the dynamical formation of electron spin
polarization following a rapid introduction of the impurity spin. In the
regime of antiferromagnetic interaction in $J_{\parallel }$ we found the
formation of the screening cloud with faster relaxation for larger $%
J_{\parallel }$. In the regime of ferromagnetic easy axis in $J_{\parallel }$
we observed the dynamics of surrounding electrons getting co-aligned with
the impurity spin. Our most surprising results were obtained in the regime
of ferromagnetic $J_{\parallel }$ with easy plane anisotropy. We observed a
transient ferromagnetic cloud formation, which was followed by the ultimate
formation of the screening cloud. These dynamics are in agreement with the
equilibrium RG flow diagram. We are not aware of earlier work on dynamics of
the Kondo in the latter regime. Its special challenge is the requirement of
analyzing both the low temperature and long time dynamics, which is crucial
for capturing the dynamical crossover. In all cases we examined the validity
of the non-Gaussian approximation by evaluating $\mathcal{N}_{\mathrm{\perp }%
}$ [see Eq (\ref{Nperp})].

\section{Superconducting and CDW phases in Holstein models \label{SCCDW}}

In this section, we investigate the quantum phase transition between the SC
and CDW phases in the Holstein model. In contrast to our discussion in
Section III, here we consider systems with a finite electron density. An
important feature of the Holstein model that will play a prominent role in
our analysis is that phonons interact with electrons locally, i.e., phonon
operators couple to the on-site energy of electrons. The system Hamiltonian
reads

\begin{eqnarray}
H &=&\sum_{nm}(\omega _{b})_{nm}b_{n}^{\dagger }b_{m}+\sum_{nm,\sigma
}t_{nm}c_{n\sigma }^{\dagger }c_{m\sigma }  \notag \\
&&-\mu \sum_{n,\sigma }c_{n\sigma }^{\dagger }c_{n\sigma }+g\sum_{n,\sigma
}c_{n\sigma }^{\dagger }c_{n\sigma }(b_{n}+b_{n}^{\dagger }).
\label{Holstein_H_general}
\end{eqnarray}

For the Einstein phonon $(\omega _{b})_{nm}=\omega _{0}\delta _{nm}$, the
Holstein model has been studied extensively using the Lang-Firsov type
polaron transformation $U=e^{S}$ characterized by a single variational
parameter
\begin{equation}
S=i\lambda \sum_{n}p_{n}c_{n\sigma }^{\dagger }c_{n\sigma }.  \label{Ss}
\end{equation}%
After this transformation, the effective electron-phonon interaction in the
Hamiltonian $\bar{H}=e^{-S}He^{S}$ is reduced, the effective hopping
strength is suppressed, reflecting the so-called polaronic dressing, and
there is explicit attractive interaction between electrons. The ground state
of $\bar{H}$ is then approximated using a vacuum state of phonons and a
Slater determinant state for electrons. Our goal is to introduce a broader
class of variational states, which can provide a better description of the
Holstein model (see also \cite{vonderLinden2004}). Firstly we point out that
the procedure outlined above is equivalent to analyzing a non-Gaussian state
\begin{equation}
\left\vert \Psi _{\mathrm{NGS}}\right\rangle =e^{S}\left\vert \Psi _{\mathrm{%
GS}}\right\rangle _{f}\left\vert 0\right\rangle _{b}\text{,}  \label{NGSMPs}
\end{equation}%
where $\left\vert \Psi _{\mathrm{GS}}\right\rangle _{f}$ is the Gaussian
state of electrons and $\left\vert 0\right\rangle _{b}$ is the vacuum state
of phonons. The parameter $\lambda $ can be obtained by the minimization of
the ground state energy $E=\left\langle \Psi _{\mathrm{NGS}}\right\vert
H\left\vert \Psi _{\mathrm{NGS}}\right\rangle $.

We notice that the non-Gaussian state (\ref{NGSMPs}) with the generating
function (\ref{Ss}) belongs to the family%
\begin{equation}
\left\vert \Psi _{\mathrm{NGS}}\right\rangle =e^{S}\left\vert \Psi _{\mathrm{%
GS}}\right\rangle _{f}\left\vert \Psi _{\mathrm{GS}}\right\rangle _{b}
\label{NGSMP}
\end{equation}%
with the general generating function%
\begin{equation}
S=i\sum_{n,m\sigma }(x_{n}\lambda _{n,m\sigma }^{x}+p_{n}\lambda _{n,m\sigma
}^{p})c_{m\sigma }^{\dagger }c_{m\sigma }.  \label{SN}
\end{equation}%
Since the state (\ref{NGSMP}) contains many variational parameters $\lambda
_{n,m\sigma }^{x,p}$, $\Delta _{R}$, and $\Gamma _{b,f}$, the brute force
minimization of $E$ may seem difficult and inefficient. However, a
variational approach utilizing wavefunction evolution in imaginary time
strongly reduces the difficulty of the problem. As we discuss below it is
possible to analyze the ground state of the Holstein model using the full
set of variational parameters in Eq. (\ref{SN}).

To find the ground state of the Holstein model with local electron-phonon
interaction, it suffices to limit the generating function to the form
\begin{equation}
S=i\sum_{n,m\sigma }p_{n}\lambda _{n,m\sigma }c_{m\sigma }^{\dagger
}c_{m\sigma }.  \label{SMP}
\end{equation}%
Notice that the last equation is a special case of Eq. (\ref{SN}) with $%
\lambda _{n,m\sigma }^{x}=0$ and $\lambda _{n,m\sigma }^{p}\equiv \lambda
_{n,m\sigma }$ (see discussion below for justification of setting $\lambda
^{x}=0$). In the next subsections, we derive the equations of motion for $%
\lambda _{n,m\sigma }$, $\Delta _{R}$, and $\Gamma _{b,f}$ in the imaginary
time evolution, which we use to find the non-Gaussian state (\ref{NGSMP})
with the minimal energy.

\subsection{Equations of motion for the variational parameters}

For the imaginary time evolution, the vectors $O_{\Delta }$, $h_{\Delta }$
and the matrices $O_{b,f}$, $h_{b,f}$ in the flow Eqs. (\ref{IMNGS}) and (%
\ref{IGf}) are determined by the functional derivatives of the average values%
\begin{eqnarray}
\left\langle O\right\rangle _{\mathrm{GS}} &=&\left\langle e^{-S}\partial
_{\tau }e^{S}\right\rangle _{\mathrm{GS}}  \notag \\
&=&i\sum_{l,n\sigma }\left\langle p_{l}\right\rangle _{\mathrm{GS}}\partial
_{\tau }\lambda _{l,n\sigma }\left\langle c_{n\sigma }^{\dagger }c_{n\sigma
}\right\rangle _{\mathrm{GS}}  \label{OSMP}
\end{eqnarray}%
and $E=\left\langle \bar{H}\right\rangle _{\mathrm{GS}}$.

The energy%
\begin{widetext}
\begin{eqnarray}
E &=&\sum_{nm,\sigma }t_{nm}\left\langle e^{-i\sum_{l}R_{l}w_{l,nm\sigma
}}\right\rangle _{\mathrm{GS}}\left\langle c_{n\sigma }^{\dagger }c_{m\sigma
}\right\rangle _{\mathrm{GS}}-\sum_{n,\sigma }\mu _{n,\sigma }\left\langle
c_{n\sigma }^{\dagger }c_{n\sigma }\right\rangle _{\mathrm{GS}}+\frac{1}{2}%
\sum_{n\sigma ,ms}V_{n\sigma ,ms}^{e}\left\langle c_{n\sigma }^{\dagger
}c_{ms}^{\dagger }\right\rangle _{\mathrm{GS}}\left\langle c_{ms}c_{n\sigma
}\right\rangle _{\mathrm{GS}}  \notag \\
&&+\frac{1}{2}\sum_{n\sigma ,ms}V_{n\sigma ,ms}^{e}\left\langle c_{n\sigma
}^{\dagger }c_{n\sigma }\right\rangle _{\mathrm{GS}}\left\langle
c_{ms}^{\dagger }c_{ms}\right\rangle _{\mathrm{GS}}-\frac{1}{2}\sum_{n\sigma
,ms}V_{n\sigma ,ms}^{e}\left\langle c_{n\sigma }^{\dagger
}c_{ms}\right\rangle _{\mathrm{GS}}\left\langle c_{ms}^{\dagger }c_{n\sigma
}\right\rangle _{\mathrm{GS}}  \notag \\
&&+\frac{1}{4}tr(\mathbf{\omega }\Gamma _{b})+\frac{1}{4}\Delta _{R}^{T}%
\mathbf{\omega }\Delta _{R}+\sum_{l,n\sigma }\delta g_{l,n\sigma
}\left\langle x_{l}\right\rangle _{\mathrm{GS}}\left\langle c_{n\sigma
}^{\dagger }c_{n\sigma }\right\rangle _{\mathrm{GS}}-\frac{1}{2}%
\sum_{n}\omega _{nn}  \label{EMP}
\end{eqnarray}
\end{widetext}is obtained by means of the Wick theorem, where the matrix $%
w_{l,nm\sigma }=(0_{N\times 2N^{2}},\lambda _{ln,\sigma }-\lambda
_{lm,\sigma })^{T}$ contains the $N\times 2N^{2}$-dimensional zero matrix $%
0_{N\times 2N^{2}}$, the site-dependent chemical potential $\mu _{n,\sigma
}=\mu -V_{n\sigma ,n\sigma }^{e}/2$, the effective electron-electron
interaction%
\begin{equation}
V_{n\sigma ,ms}^{e}=2(\lambda ^{T}\omega _{b}\lambda )_{n\sigma
,ms}-2g\lambda _{n,ms}-2g\lambda _{m,n\sigma },
\end{equation}%
$\mathbf{\omega }={\openone}_{2}\otimes \omega _{b}$, and the renormalized
electron-phonon interaction $\delta g_{l,n\sigma }=g\delta _{ln}-(\omega
_{b}\lambda )_{l,n\sigma }$. The average values like $\left\langle c_{\alpha
}^{\dagger }c_{\beta }\right\rangle $ and $\left\langle c_{\alpha }^{\dagger
}c_{\beta }^{\dagger }\right\rangle $ are elements of the covariance matrix $%
\Gamma _{f}$, and the phonon-dressed hopping strength%
\begin{eqnarray}
\tilde{t}_{nm\sigma } &\equiv &t_{nm}\left\langle
e^{-i\sum_{l}R_{l}w_{l,nm\sigma }}\right\rangle _{\mathrm{GS}}  \notag \\
&=&t_{nm}e^{-i(\Delta _{R}^{T}w)_{nm\sigma }}e^{-\frac{1}{2}(w^{T}\Gamma
_{b}w)_{nm\sigma ,nm\sigma }}
\end{eqnarray}%
is obtained through Eq. (\ref{NO}).

It follows from Eqs. (\ref{OY}) and (\ref{hY}) that the vectors%
\begin{equation}
O_{\Delta }=2i\sum_{n\sigma }\left\langle c_{n\sigma }^{\dagger }c_{n\sigma
}\right\rangle _{\mathrm{GS}}\left(
\begin{array}{c}
0 \\
\partial _{\tau }\lambda _{l,n\sigma }%
\end{array}%
\right) ,  \label{YSMP}
\end{equation}%
and%
\begin{eqnarray}
h_{\Delta } &=&\mathbf{\omega }\left\langle R\right\rangle +2\sum_{n\sigma
}\left\langle c_{n\sigma }^{\dagger }c_{n\sigma }\right\rangle _{\mathrm{GS}%
}\left(
\begin{array}{c}
\delta g_{l,n\sigma } \\
0_{N}%
\end{array}%
\right)  \notag \\
&&-2i\sum_{nm,\sigma }\tilde{t}_{nm\sigma }\left\langle c_{n\sigma
}^{\dagger }c_{m\sigma }\right\rangle _{\mathrm{GS}}w_{l,nm\sigma }
\label{YHMP}
\end{eqnarray}%
are obtained by the functional derivatives of $\left\langle O\right\rangle _{%
\mathrm{GS}}$ and $E$ with respect to $\Delta _{R}$. By the functional
derivatives of $E$ to $\Gamma _{b}$, we obtain the matrix $O_{b}=0$ and%
\begin{equation}
(h_{b})_{ll^{\prime }}=\mathbf{\omega }_{ll^{\prime }}-2\sum_{nm,\sigma }%
\tilde{t}_{nm\sigma }\left\langle c_{n\sigma }^{\dagger }c_{m\sigma
}\right\rangle _{\mathrm{GS}}w_{l,nm\sigma }w_{nm\sigma ,l^{\prime }}^{T}.
\label{OHMP}
\end{equation}

The functional derivative of $\left\langle O\right\rangle _{\mathrm{GS}}$
with respect to $\Gamma _{f}$ determines the matrix $O_{f}=\tau _{z}\otimes
o_{f}$, where $\tau _{z}$ is the Pauli matrix, and the diagonal matrix $%
o_{f} $ has nonzero elements $(o_{f})_{n\sigma ,n\sigma
}=i\sum_{l}\left\langle p_{l}\right\rangle _{\mathrm{GS}}\partial _{\tau
}\lambda _{l,n\sigma }$. The mean-field single particle Hamiltonian%
\begin{equation}
h_{f}=\left(
\begin{array}{cc}
\mathcal{E} & \Delta \\
\Delta ^{\dagger } & \mathcal{E}^{T}%
\end{array}%
\right)  \label{HHMP}
\end{equation}%
is determined by the functional derivative of $E$ with respect to $\Gamma
_{f}$, where the diagonal term
\begin{equation}
\mathcal{E}_{n\sigma ,ms}=\tilde{t}_{nm\sigma }\delta _{\sigma s}-\tilde{\mu}%
_{n,\sigma }\delta _{nm}\delta _{\sigma s}-V_{n\sigma ,ms}^{e}\left\langle
c_{ms}^{\dagger }c_{n\sigma }\right\rangle _{\mathrm{GS}}
\end{equation}%
contains the effective chemical potential%
\begin{equation}
\tilde{\mu}_{n,\sigma }=\mu _{n,\sigma }-\sum_{l}\delta g_{l,n\sigma
}\left\langle x_{l}\right\rangle _{\mathrm{GS}}-\sum_{ms}V_{n\sigma
,ms}^{e}\left\langle c_{ms}^{\dagger }c_{ms}\right\rangle _{\mathrm{GS}},
\end{equation}%
and the off-diagonal term is the order parameter $\Delta _{n\sigma
,ms}=V_{n\sigma ,ms}^{e}\left\langle c_{ms}c_{n\sigma }\right\rangle _{%
\mathrm{GS}}$. To obtain flow equations for the Gaussian part of the
wavefunction we use Eq. (\ref{IMNGS}) together with expressions (\ref{YSMP}%
)-(\ref{HHMP}).

Equations of motion for $\lambda $ can be obtained by taking the projection
of states (\ref{UL0}) and (\ref{UR0}) onto the states $\left\vert
D_{ln}\right\rangle =b_{l}^{\dagger }$:$U_{\mathrm{GS}}^{-1}c_{n\sigma
}^{\dagger }c_{n\sigma }U_{\mathrm{GS}}$:$\left\vert 0\right\rangle $, i.e.,%
\begin{equation}
\left\langle D_{ln}\right\vert \delta O\left\vert 0\right\rangle
=-\left\langle D_{ln}\right\vert \delta \bar{H}\left\vert 0\right\rangle .
\label{ph}
\end{equation}%
The state
\begin{equation}
\delta O\left\vert 0\right\rangle =i\sum_{l,m\sigma
}(R^{T}S_{b}^{T})_{l}\partial _{\tau }\lambda _{l,m\sigma }\text{:}U_{%
\mathrm{GS}}^{-1}c_{m\sigma }^{\dagger }c_{m\sigma }U_{\mathrm{GS}}\text{:}%
\left\vert 0\right\rangle
\end{equation}%
on the left hand side of Eq. (\ref{ph}) contains only the cubic operator
acting on the vaccum state, and the cubic terms in the state $\delta \bar{H}%
\left\vert 0\right\rangle $ on the right hand side of Eq. (\ref{ph}) is%
\begin{eqnarray}
&&[i\sum_{nm,\sigma }\tilde{t}_{nm\sigma }\text{:}U_{\mathrm{GS}%
}^{-1}c_{n\sigma }^{\dagger }c_{m\sigma }U_{\mathrm{GS}}\text{:}%
(R^{T}S_{b}^{T}w)_{nm\sigma }  \notag \\
&&-\sum_{l,n\sigma }(R^{T}S_{b}^{T})_{l}\delta \tilde{g}_{l,n\sigma }\text{:}%
U_{\mathrm{GS}}^{-1}c_{n\sigma }^{\dagger }c_{n\sigma }U_{\mathrm{GS}}\text{:%
}]\left\vert 0\right\rangle ,
\end{eqnarray}%
where the electron-phonon coupling matrix $\delta \tilde{g}=(\delta
g,0_{N\times 2N})^{T}$.

The projection (\ref{ph}) leads to the equation%
\begin{widetext}
\begin{eqnarray}
&&\sum_{ms}\partial _{\tau }\lambda _{l,ms}\left\langle c_{ms}^{\dagger
}c_{ms}c_{n\sigma }^{\dagger }c_{n\sigma }\right\rangle _{c}=-i\frac{1}{2}%
\sum_{l^{\prime }n^{\prime }ms}(\Gamma _{b})_{ll^{\prime }}w_{l^{\prime
},n^{\prime }ms}\tilde{t}_{n^{\prime }ms}\left\langle \{c_{n\sigma
}^{\dagger }c_{n\sigma },c_{n^{\prime }s}^{\dagger }c_{ms}\}\right\rangle
_{c}  \notag \\
&&+\sum_{l^{\prime }ms}(\Gamma _{b})_{ll^{\prime }}\delta \tilde{g}%
_{l^{\prime },ms}\left\langle c_{ms}^{\dagger }c_{ms}c_{n\sigma }^{\dagger
}c_{n\sigma }\right\rangle _{c}+\sum_{l^{\prime }m}\sigma _{ll^{\prime
}}w_{l^{\prime },nm\sigma }\text{Re}(\tilde{t}_{nm\sigma }\left\langle
c_{n\sigma }^{\dagger }c_{m\sigma }\right\rangle _{\mathrm{GS}}),
\label{dLMP}
\end{eqnarray}%
and the constraint
\begin{equation}
0=\sum_{l^{\prime }ms}(\Gamma _{b})_{l_{1}l^{\prime }}\delta \tilde{g}%
_{l^{\prime },ms}\left\langle c_{ms}^{\dagger }c_{ms}c_{n\sigma }^{\dagger
}c_{n\sigma }\right\rangle _{c}-i\frac{1}{2}\sum_{l^{\prime }n^{\prime
}ms}(\Gamma _{b})_{l_{1}l^{\prime }}w_{l^{\prime },n^{\prime }ms}\tilde{t}%
_{n^{\prime }ms}\left\langle \{c_{n\sigma }^{\dagger }c_{n\sigma
},c_{n^{\prime }s}^{\dagger }c_{ms}\}\right\rangle _{c}  \label{dLCMP}
\end{equation}%
\end{widetext}In the equations above the connected correlation function is
defined as $\left\langle A_{1}A_{2}\right\rangle _{c}=\left\langle
A_{1}A_{2}\right\rangle _{\mathrm{GS}}-\left\langle A_{1}\right\rangle _{%
\mathrm{GS}}\left\langle A_{2}\right\rangle _{\mathrm{GS}}$, and the indices
$1\leq l\leq N$, $1+N\leq l_{1}\leq 2N$. Note that Eq. (\ref{dLCMP}) has the
form of a constraint only because we set $\lambda _{x}=0$ in Eq. (\ref{SMP}%
). Otherwise, the left hand side of equation (\ref{dLCMP}) would contain the
$\partial _{\tau }\lambda ^{x}$ term.

The equation (\ref{dLMP}) determines the imaginary time flow of $\lambda
_{n,m}$. In the next subsection, we show that the constraint (\ref{dLCMP})
is automatically satisfied for the ground state in the SC and CDW phase of
the Holstein model. We emphasize that $\lambda _{n,m\sigma }^{x}=0$ is only
a special case of a more general class of transformations, which turns out
to be sufficient for analyzing the ground state of the Holstein model. A
special feature of the Holstein model which makes this simplification
possible is the local character of the electron-phonon interaction. In cases
of more general electron-phonon interacting systems, including the SSH
model, the ans\"{a}tz (\ref{NGSMP}) with the full generating function (\ref%
{SN}) should be applied. Then, the term containing the time derivative $%
\partial _{\tau }\lambda ^{x}$ appears on the left-hand side of Eq. (\ref%
{dLCMP}), which determines the variational state with the minimal ground
state energy.

\subsection{Transitions between superconducting and CDW phases}

\begin{figure}[tbp]
\includegraphics[width=0.9\linewidth]{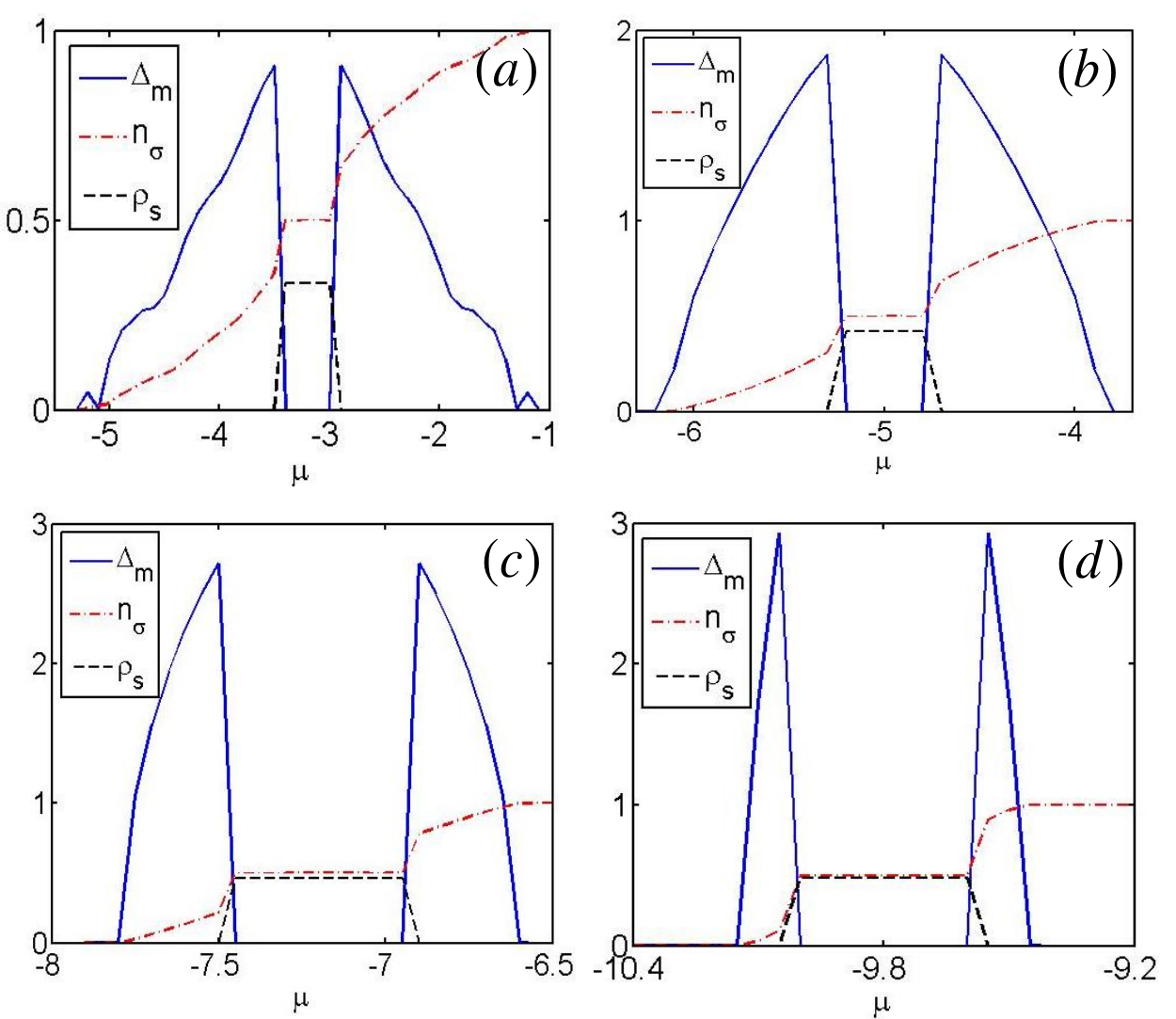}
\caption{The superconducting gap $\Delta$, the electron density $n _{\protect%
\sigma}$, and the staggered component $\protect\rho_s$ for the $10 \times 10$
lattice with the phonon frequency $\protect\omega _{0}=10t_{0}$, where the
hopping constant $t_{0}$ is taken as the unit. (a) $g=0.4\protect\omega _{0}$%
; (b) $g=0.5\protect\omega_{0}$; (c) $g=0.6\protect\omega_{0}$; (d) $g=0.7%
\protect\omega_{0}$.}
\label{PTMP}
\end{figure}

We now analyze the phase diagram of the 2D Holstein model by numerically
solving flow Eqs. (\ref{RENGS}) and (\ref{dLMP}). We consider Hamiltonian (%
\ref{Holstein_H_general}) with the nearest neighbor hopping $%
t_{nm}=-t_{0}\delta _{mn+e_{\alpha }}$, where $m=n+e_{\alpha }$ correspond
to nearest neighboring sites in the $\alpha =x,y$ directions. We assume an
Einstein model of dispersionless phonons with $\omega _{b,nm}=\omega
_{0}\delta _{nm}$.

The variational parameter $\lambda _{n,m}$ that do not break translational
symmetry only depends on the difference between sites $n$ and $m$, hence it
is convenient to introduce
\begin{equation}
\lambda _{nm,\sigma }=\frac{1}{N}\sum_{q}\lambda _{q}e^{iq(n-m)}.  \label{LT}
\end{equation}%
Note that the translational symmetry of $\lambda _{n,m}$ does not rule out
finite expectation values of $\lambda _{q}$ for $q\neq 0$. On the other
hand, the phonon displacements $R$ can only have finite expectation value at
$q=0$ in states which do not break translational invariance.

\begin{figure}[tbp]
\includegraphics[width=0.9\linewidth]{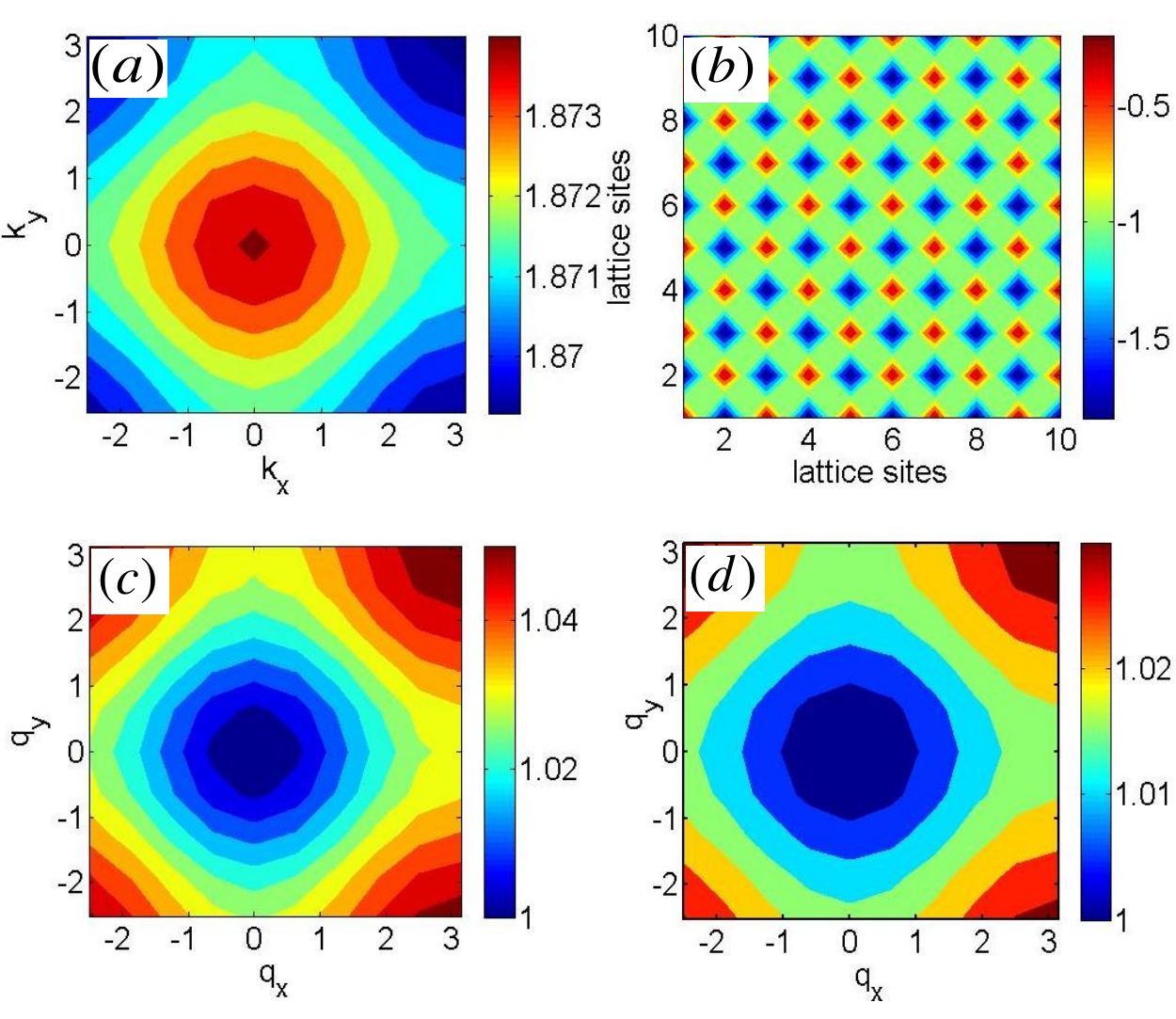}
\caption{The order parameter, the displacement and the covariant matrices of
phonons, where the system size is $10 \times 10$, the phonon frequency is $%
\protect\omega_{0}=10t_{0}$, the coupling $g=0.5\protect\omega _{0}$, and
the hopping constant $t_{0}$ is taken as the unit. (a) The superconducting
order parameter in the momentum space for $\protect\mu=-5.3$; (b) The
displacement of phonons in the CDW phase with $\protect\mu=-5$; (c)-(d) The
covariant matrix of phonons in the momentum space for the superconducting ($%
\protect\mu=-5.3$) and CDW ($\protect\mu=-5$) phases.}
\label{GX}
\end{figure}

Results of our analysis are presented in Figs. \ref{PTMP} and \ref{GX}. We
find that away from half-filling, i.e., when the electron density $n_{\sigma
}=\sum_{n}\left\langle c_{n\sigma }^{\dagger }c_{n\sigma }\right\rangle
/N\neq 0.5$, the system is in the superconducting phase, which preserves
translational symmetry. There is a uniform displacement of all local phonons
given by $d_{j}=\left\langle \Psi _{\mathrm{NGS}}\right\vert x_{j}\left\vert
\Psi _{\mathrm{NGS}}\right\rangle =-4gn_{\sigma }/\omega _{0}$. To describe
electronic correlations it is convenient to use electron operators in
momentum space $c_{p\sigma }=\sum_{n}c_{n\sigma }e^{-ipn}/\sqrt{N}$. The
self-consistency equation for the anomalous expectation value is
\begin{equation}
\Delta _{k}=\frac{2}{N}\sum_{p}\lambda _{k-p}(\omega _{0}\lambda
_{k-p}-2g)\left\langle c_{-p\downarrow }c_{p\uparrow }\right\rangle _{%
\mathrm{GS}}.
\end{equation}%
Note that $\Delta _{k}$ determines the quasiparticle gap for electrons.

When the system is half-filled, i.e., $n_{\sigma }=0.5$, CDW phase emerges.
While the CDW state breaks translational invariance we find that the
optimized values of $\lambda _{nm,\sigma }$ still only depend on the
difference between $n$ and $m$ and representation (\ref{LT}) applies. The
electron density
\begin{equation}
\rho _{j}=\left\langle c_{j\sigma }^{\dagger }c_{j\sigma }\right\rangle _{%
\mathrm{GS}}=\frac{1}{2}+e^{iQ_{\pi }j}\rho _{\mathrm{s}}
\end{equation}%
has a finite Fourier component at momentum $Q_{\pi }=(\pi ,\pi )$. The
staggered part of the density $\rho _{\mathrm{s}}=\sum_{k}\left\langle
c_{k-\pi \sigma }^{\dagger }c_{k\sigma }\right\rangle _{\mathrm{GS}}/N$ is
determined by the elements $\left\langle c_{k\sigma }c_{k-\pi \sigma
}^{\dagger }\right\rangle _{\mathrm{GS}}$ of the covariance matrix. The
phonon displacement
\begin{equation}
d_{j}=-2\frac{g}{\omega _{0}}+e^{iQ_{\pi }j}d_{\mathrm{s}}
\end{equation}%
shows that the phonon quadrature $\Delta _{R}$ has non-zero expectation
values not only at $q=0$, but also at $Q_{\pi }$. We define $\left\langle
x_{Q_{\pi }}\right\rangle _{\mathrm{GS}}=\sum_{j}e^{iQ_{\pi }n}\left\langle
x_{j}\right\rangle _{\mathrm{GS}}/\sqrt{N}$ then the staggered part $d_{%
\mathrm{s}}=\left\langle x_{Q_{\pi }}\right\rangle _{\mathrm{GS}}/\sqrt{N}%
-4\lambda _{Q_{\pi }}\rho _{\mathrm{s}}$ of $d_{j}$. Surprisingly we find
that in the CDW phase the phonon covariance matrix $\Gamma _{b,nm}$ still
depends on $n-m$ only, which we would generally expect only for
translationally invariant systems. Thus, in both the SC and CDW phases all
information about phonon covariance can be represented using
\begin{equation}
\Gamma _{b,nm}=\frac{1}{N}\sum_{q}e^{iq(n-m)}\Gamma _{b,q},
\end{equation}%
namely, it has the translational symmetry.

The figure \ref{PTMP} shows the transition between the SC and CDW phases for
$\omega _{0}=10t_{0}$ and $g/\omega _{0}=0.4$, $0.5$, $0.6$, and $0.7$,
where the hopping constant $t_{0}$ is taken as the unit of energy. The local
order parameter $\Delta =\sum_{k}\Delta _{k}/N$, the density $n_{\sigma }$,
and the staggered components $\rho _{\mathrm{s}}$ show that the transition
from the SC to the CDW phase takes place at half-filling.

For $\omega _{0}=10t_{0}$ and $g=0.5\omega _{0}$, the order parameter $%
\Delta _{k}$ in the SC phase ($\mu =-5.3$) and the displacement $d_{j}$ in
the CDW phase ($\mu =-5$) are shown in Figs. \ref{GX}a and \ref{GX}b
respectively. In Figs. \ref{GX}c and \ref{GX}d, we show the squeezing $%
(\Gamma _{b,q})_{xx}$ of phonons as a function of momentum $q$ in the SC and
CDW phases respectively.

In both phases, the constraint (\ref{dLCMP}) is satisfied because $%
\left\langle p_{l}\right\rangle _{\mathrm{GS}}=0$ and $(\Gamma
_{b,nm})_{xp}=\left\langle \{\delta x_{n},\delta p_{m}\}\right\rangle _{%
\mathrm{GS}}/2=0$. In Eq. (\ref{dLCMP}), the first term vanishes due to $%
(\Gamma _{b,nm})_{xp}=0$, and the second term vanishes due to $\left\langle
p_{l}\right\rangle _{\mathrm{GS}}=0$. As a result, Eq. (\ref{dLCMP}) is
automatically satisfied by the optimal variational parameters in the ground
state.

\subsection{Summary of Section \protect\ref{SCCDW}}

In this section we introduced a class of non-Gaussian states constructed
using a generalization of the Lang-Firsov polaron transformation. They
provide a useful variational family of states for analyzing many-body
electron-phonon systems. We studied the ground state of the Holstein model
using the imaginary time flow approach and found that the CDW phase exists
only at half-filling, and away from half-filling the system is always in the
superconducting state. We presented results for the SC and CDW order
parameters, phonon quadratures and covariance matrix, and the optimal values
of the polaron transformation parameters $\lambda _{n,m\sigma }$.


\section{Summary, Discussion, and Outlook \label{CO}}

In this section we review the main results of the paper and suggest several
promising directions for future studies.

We introduced a new family of variational wavefunctions to describe both the
ground state and non-equilibrium dynamics of interacting many-body systems.
The essence of our approach is combining generalized canonical
transformations with the Gaussian ans\"{a}tz for bosons and fermions. These
wavefunctions retain the simplicity of Gaussian wavefunctions yet they are
characterized by non-factorizable correlations due to the canonical
transformations which introduce entanglement between the fields. By allowing
time-dependence of both the canonical transformation parameters and of the
Gaussian wavefunctions our method goes beyond approaches based on standard
canonical transformations, such as Lang-Firsov for electron-phonon systems
or Silbey's partial polaron transformation for the spin bath problem \cite%
{Silbey}. We obtained explicit equations for the time evolution of the
variational parameters by analyzing the differential structure of the
variational wavefunctions manifold. Ground state can be found by solving the
imaginary time equations of motion until the system reaches a fixed point.
This fixed point can then be used to find collective modes in the system by
solving the linearized real time equations of motion. The full real time
dynamics can be used to calculate spectral functions of operators, or to
analyze out of equilibrium phenomena.

While the focus of our paper was on non-Gaussian states we devote one of the
subsections to reviewing interesting questions that can be studied using
time-dependent Gaussian states.

\subsection{Overview of results}

\textit{Single polaron in a lattice. } 
We considered lattice polaron problems in the cases of Holstein and
Su-Schrieffer-Heegger models. We used the LLP transformation to eliminate
the impurity degree of freedom and reduce the problem to interacting bosons.
We then considered a Gaussian state of phonons as an ansatz for the polaron
wavefunction. This provides an extension of earlier studies which were
limited to coherent variational states for phonons \cite{Lakhno}. In the
analysis of the ground state we found a phase transition for the SSH
polaron. We also calculated polaron spectral functions using the real time
dynamics. We showed the importance of the non-classical Gaussian part of the
phonon wavefunctions in both the ground state and non-equilibrium
calculations. Our results are in excellent agreement with the numerical
approaches using exact diagonalization, Diagrammatic Monte Carlo, and the
Bold Diagrammatic Monte Carlo.

\textit{Anisotropic Kondo problem and Ohmic bosonic bath model. }
We analyzed both the ground state and dynamics of the Ohmic bosonic bath
model, which is equivalent to the anisotropic Kondo model. We showed that
the problem can be studied using two seemingly different types of canonical
transformations, which however lead to the same class of wavefunctions. The
first one uses the parity conservation and the second one is a
generalization of the partial polaron transformation introduced by Silbey
and co-authors \cite{Silbey} (see also \cite{spinco,Hutchison} and
references therein for more recent work). Our work differs from the earlier
papers utilizing partial polaron transformation in that we allow the
parameters of the transformation to be time dependent and consider a general
Gaussian state of bosons. When applied to the analysis of the ground state
of the anisotropic Kondo problem we find that we correctly reproduce the
phase diagram known from RG calculations. What is more exciting is that we
can study real time dynamics in the regimes and at times which are not
accessible for any other technique. In particular, we considered the
relaxation of the impurity spin in the ferromagnetic easy plane case. The
real time dynamics in this case is particularly difficult to analyze since
the RG flow has several distinct regimes. At first, Kondo interactions flow
to smaller values, while staying ferromagnetic. Then, at lower energy
scales, the z-component of the interaction changes sign and starts flowing
to larger values. Tantalizingly, we observed that these crossovers appear in
the real time dynamics. At short times, the impurity spin gets dressed by
the co-alligned polarization of the surrounding fermions (ferromagnetic
screening) but then at longer times the polarization cloud switches into an
anti-aligned configuration (antiferromagnetic screening). To our knowledge,
this is the first analysis of the Kondo model dynamics in this regime.

\textit{Competition of superconductivity and charge density wave orders in
the Holstein model. } 
We analyzed the phase diagram of the Holstein model in the case when the
phonon frequency is relatively high: ten times the electron hopping. We
found a direct transition between the CDW phase at half filling and the SC
phase away from half-filling. This is consistent with the expectation that
for large phonon frequency the Holstein model should be similar to the
negative-U Hubbard model, in which there is a degeneracy between the CDW to
the SC phases at half-filling (the so-called C.N. Yang's SU(2) symmetry),
while the superconducting phase is favored at other electron concentrations.
In agreement with earlier studies we find that finite phonon frequency
breaks the degeneracy of the two phases at half-filling in favor of the CDW
phase.


\subsection{Interesting questions for time-dependent Gaussian states}

The equations of motion (\ref{RGGSI}) and (\ref{RGGSR}) in the imaginary-
and real- time provide a systematic way to study the ground state properties
and real-time dynamics in the subspace of variational Gaussian states. For
many-body systems that only contain fermions they agree with the generalized
Hartree-Fock-BCS mean field theory \cite{HFGS}. In the usual implementation
of the Hartree-Fock-BCS approximation one needs to solve a challenging
multi-parameter minimization problem. The formalism of imaginary time flow
presented in this paper makes the search for the optimal mean-field states
more efficient.

\textit{1. Competing phases}.--- The analysis of the superconducting state
is easy in fermionic systems with equal densities of the two spins
components and attraction in the $s$-wave channel only. BCS mean-field
theory assumes that the variational Gaussian state is determined by the
single order parameter $\Delta _{0}\sim \sum_{k}\left\langle c_{-k\downarrow
}c_{k\uparrow }\right\rangle $ \cite{BCS}. Due to the simple structure of
the order parameter, one can analyze properties of conventional weakly
coupled BCS regime as well as the BCS-BEC crossovers by solving
self-consistent equations for the gap and the chemical potnetial \cite%
{BCSBEC}. For systems with more complicated interactions, e.g., dipolar
Fermi gases \cite{dipole1,dipole2,dipole3,dipole4}, the mixture of order
parameters $\Delta _{l}\sim \sum_{k}w_{l}(k)\left\langle c_{-k\downarrow
}c_{k\uparrow }\right\rangle $ with different spatial and spin symmetries
may co-exist in the ground state \cite{dipole1}. Here the orthonomal
functions$\ w_{l}(k)$ describe the structure of electron pairing in momentum
space. In this case, the common approach is to guess which order parameters $%
\Delta _{l}$ will be present in the ground state, and solve the nonlinear
gap and Fermi occupation numbers equations self-consistently. This is often
a demanding task since nonlinear equations may have multiple non-trivial
self-consistent solutions \cite{dipole3}. Another competing instability in
dipolar fermions that has been previously discussed is the Pomeranchuk type
instabilities in the particle-hole channel \cite{fradkin,Zhang}. To find the
actual ground state one needs to compare different saddle points and
determine which of them provides a global minimum of the energy (or free
energy at finite temperature). The equations of motion in imaginary time (%
\ref{RGGSI}) provide a powerful alternative technique for identifying the
lowest energy mean-field state.

\textit{Inhomogeneous states}. Many interesting systems are characterized by
spatially inhomogeneous order parameters. One important example is the
Fulde-Ferrel-Larkin-Ovchinnikov superconducting phase \cite{FF,LO}, which
may appear when there is spin imbalance in the system. In the FFLO phase,
the condensed fermion pairs have non-zero center of mass momenta $Q$, i.e.,
the system develops non-vanishing pairing amplitudes $\Delta _{Q,l}\sim
\sum_{k}w_{l}(k)\left\langle c_{Q/2-k\downarrow }c_{Q/2+k\uparrow
}\right\rangle $ with $Q\neq 0$. Additional order parameters $\Delta _{Q,l}$
make the numerical solution of the nonlinear gap equations particularly
challenging. The difficulty of analyzing such states comes from the near
degeneracy of many configurations. At the quadratic level, states with the
same magnitude of the ordering wavevector are degenerate regardless of the
wavevector direction. One needs to consider effects of the coupling between
different components of the order parameter at different wavevectors \cite%
{rajagopal}, including higher harmonics, to determine the lowest energy
state. Other important cases of inhomogeneous phases include stripe phases
and frustrated phase separation in electron systems \cite{pryadko}; vortex
lattice states in superconductors, in which the pairing amplitude is
suppressed near vortex cores; systems with disorder, in which the order
parameter may be suppressed in the vicinity of impurities. Generalized
Gaussian states include all possible two-point correlation functions and
provide a powerful toolbox for finding optimal configurations.

\textit{2. Fluctuations}.--- When discussing mean-field Gaussian states one
usually separates two types of excitations: single-particle Bogoliubov
excitations described by Eq. (\ref{V1}) and collective excitations described
by Eqs. (\ref{V2b}) and (\ref{V2f}). For instance, in the SC phase
single-particle Bogoliubov excitations describe fermionic quasiparticles
which result from breaking up Cooper pairs. And the simplest example of a
collective excitation is a gapless mode describing\ the phase fluctuation of
the superconducting order parameter, which corresponds to the Goldstone mode
originating from the spontaneous breaking of the $U(1)$ symmetry. The
spectrum of Bogoliubov excitations can be directly obtained by diagonalizing
the mean field Hamiltonians $h_{f}$.

To describe collective excitations, one usually introduces a
Hubbard-Strantanovich (HS) field to represent the collective pairing field,
and integrates out the fermionic fields in order to obtain an effective
theory for the HS field \cite{Zaikin}. The low-energy effective theory of
the HS field then describes the linear Goldstone modes. In superconductors,
the Meissner effect arises from the external electromagnetic field coupling
to the low-energy HS field and acquiring a \textquotedblleft
mass\textquotedblright\ . When the system contains multiple order parameters
$\Delta _{l}$, many HS fields corresponding to order parameters with
different symmetries should be introduced. This makes the analysis of the
effective action of the coupled HS fields rather cumbersome. The Gaussian
state approach provides an efficient way to study the properties of
collective excitations by solving the linearized equations of motion (\ref%
{dteta}). The low energy spectrum of these collective modes is determined by
the eigenvalues of the matrix $\mathbf{L}$. In systems with spontaneous
breaking of a continuous symmetry these equations are guaranteed to give a
gapless Goldstone mode.

\textit{3. Real time dynamics}.--- In non-equilibrium superconductors and
superfluidities, one is often interested in analyzing the coherent evolution
of order parameters (such as the superconducting gaps) after sudden changes
in system parameters \cite{quench} or following an electromagnetic pulse
\cite{TES}. This dynamics is captured by Eq. (\ref{RGGSR}).

\textit{4. Open systems}.--- The Gaussian state ans\"{a}tz can be
generalized to study dynamics and steady state behavior in open systems \cite%
{qo}, such as optical parametric oscillators \cite{ROPO}. The real-time
evolution of the reduced density matrix $\rho _{s}$ for the system coupled
to the bath is governed by the master equation $\partial _{t}\rho _{s}=%
\mathcal{L}\rho _{s}$ in the Markovian limit \cite{QO}. The reduced density
matrix can be approximated by the Gaussian mixed state, and equations of
motion of $\Delta _{R}$ and $\Gamma _{b,m}$ are determined by%
\begin{eqnarray}
\partial _{t}\Delta _{R} &=&tr(\mathcal{L}\rho _{s}R),  \notag \\
\partial _{t}\Gamma _{b} &=&\frac{1}{2}tr(\mathcal{L}\rho _{s}\{\delta
R,\delta R^{T}\}),  \notag \\
\partial _{t}\Gamma _{m} &=&\frac{i}{2}tr(\mathcal{L}\rho _{s}[A,A^{T}]).
\end{eqnarray}

We expect that interesting new insight into phase transitions and
far-from-equilibrium dynamics of open systems can be obtained using
time-dependent variational Gaussian state approach.


\subsection{Possible extensions of the non-Gaussian state analysis}

Before concluding this paper we would like to outline several promising
directions in which our work can be extended.

\textit{Fractional Quantum Hall Effect and Topological Phases}. In Sec. II
C1 we discussed the canonical transformation equivalent to the flux
attachment procedure. We pointed out that one can consider a broader class
of transformations, e.g., when one first makes Wannier type orbitals as a
linear superposition of the original single particle states and then
performs flux attachment for such Wannier orbitals. One interesting question
to consider is the nature of the excitations described by our variational
wavefunctions. We expect that neutral excitations of composite fermions and
bosons correspond to fluctuations in the Gaussian state part, while
fluctuations in $\omega _{ij}^{b,f}$ have a more subtle topological nature
\cite{Laughlin,nonAbelian}. This class of wavefunctions should be useful for
studying FQHE states in lattices, including out of equilibrium situations
relevant to systems realized with cold atoms and photons.

\textit{Analysis of Fermionic Bogoliubov Quasiparticles}. In the current
paper we focused on bosonic degrees of freedom. For example, collective
excitations which we discussed in Sec. II A3 correspond to the Goldstone and
amplitude (Higgs) modes of the ordered phases, or the incoherent particle
hole excitations of Fermi systems. Fermionic quasiparticles should also be
readily available from our analysis using time dependent effective quadratic
Hamiltonian [see e.g., Eq. (E10)]. They can be used for analyzing
time-resolved photoemission spectroscopy in pump and probe experiments \cite%
{trps1,trps2}.

\textit{Non-equilibrium Dynamics of Electron Phonon Systems}. Recent
experiments demonstrated several intriguing phenomena in non-equilibrium
electron-phonon systems. This includes photo-induced superconducitivity \cite%
{TSCE1,TSCE2,DS,TSCT,DES}, the observation of the amplitude Higgs model
excited with light, pump and probe spectroscopy of CDW states. A special
feature of our formalism is that it allows to treat on equal footing
electron and phonon degrees of freedom. Hence it goes beyond the usual
approach of solving the time dependent BCS model \cite{TDBCS}. This will be
particularly important for analyzing systems in which the non-equilibrium
state of phonons plays an important role \cite{TSCE1,TSCE2}.

\textit{Systems with Competing and Intertwined Orders}. A ubiquitous feature
of many-body systems is the interplay of competing \cite{CompeOrder} or
intertwined \cite{InterOrder} orders. In this paper we discussed the
competition between superconductivity and CDW order, which is a common
feature in electron-phonon systems. Another general feature of strongly
correlated Fermi systems is the competition of superconducting phases with
different symmetries of the order parameter. A canonical example is the
competition between the $A$ and $B$ phases in superfluid $^{3}$He. The
analysis of Gaussian states is not sufficient to understand this transition
since it is important to consider the feedback from the quasiparticle
spectrum on the magnetic fluctuations mediating attraction between
quasiparticles \cite{LeggettHe3}. Similar questions about the interplay of
several types of fluctuations and the analysis beyond Gaussian states are
common in electron systems. One important problem is identifying the best
"hidden" order parameter for explaining the pseudogap phase in high Tc
cuprates \cite{HTcE,RVB,ZhangRice}. Candidates include simple spin and
charge density wave phases, as well as a more exotic d-density wave and
Amperian pairing states. In iron based superconductors it is important to
understand the competition between d-wave and extended s-wave \cite{dswave}
pairing symmetries, which strongly depends on the nature of the magnetic
fluctuations in these materials. The variational approach that we discussed
in this paper should be a useful tool for analyzing the interplay of several
order parameters. When variational wavefunctions evolve in the imaginary
time they find local energy minima. It is possible, however, that the system
has several local minima. In this case one needs to compare energies of
several locally stable states.

Our formalism can be a powerful tool for analyzing competing orders in
nonequilibrium systems, such as when system parameters are changing in time.
Examples include the competition between fermion pairing and ferromagnetism
near a Feshbach resonance in ultracold atoms \cite{pekker}, or pump and
probe experiments in solids \cite{gedik}.

\textit{Magnetic Polarons}. The problem of magnetic polarons in the
fermionic Hubbard t-J models plays an important role in the physics of
strongly correlated electron systems (see e.g., \cite{Varma,Manousakis}).
Here the goal is to understand the dynamics of a single charge carrier,
e.g., a hole, in the background of an antiferromagnetically ordered Mott
insulator. This system is reminiscent of the phonon-polaron problem, but
with a hole exciting the magnons rather than phonons. In the magnetic
polaron system, the hole hopping causes frustration in the antiferromagnetic
background and leads to more dramatic polaronic effects. The LLP
transformation presented earlier and generalized squeezed states can be used
to study the spectral functions of polarons \cite{Manousakis} which can be
measured in solid state systems using ARPES \cite{ARPES1,ARPES2}.

\textit{Electrons interacting with nearly critical fields. }An important
class of models in strongly correlated electron systems comes from
considering electrons coupled to fluctuating bosonic fields in the vicinity
of a Quantum Critical Point (QCP). Physically relevant cases include anti-
and ferromagnetic spin fluctuations, CDW and orbital nematic fluctuations
\cite{Berg}. For example, in the case of antiferromagnetic fluctuations an
effective model can be written as $\mathcal{H}=\mathcal{H}_{\mathrm{e}}+%
\mathcal{H}_{\mathrm{AF}}+\mathcal{H}_{\mathrm{int}}$: (a) The electron
hopping term $\mathcal{H}_{\mathrm{e}}=\sum_{nm\sigma }t_{nm}c_{n\sigma
}^{\dagger }c_{m\sigma }$; (b) The Hamiltonian
\begin{eqnarray}
\mathcal{H}_{\mathrm{AF}} &=&\sum_{q_{i}<\Lambda }\,[\frac{1}{2}\vec{\pi}_{q}%
\vec{\pi}_{-q}+\frac{(r+q^{2})}{2}\vec{\phi}_{q}\vec{\phi}_{-q}  \notag \\
&&+u(\vec{\phi}_{q1}\vec{\phi}_{q2})(\vec{\phi}_{q3}\vec{\phi}%
_{q_{2}})\delta _{\sum q_{i}}\,]
\end{eqnarray}%
describes the antiferromagnetic fluctuations by the vector field $\vec{\phi}%
_{q}$ and its conjugate momentum $\vec{\pi}_{q}$; (c) The interaction term $%
\mathcal{H}_{\mathrm{int}}=g\sum_{kq}c_{k+Q_{\pi }+q,\alpha }^{\dagger }\vec{%
\sigma}_{\alpha \beta }c_{k,\beta }\vec{\phi}_{q}$. Here, $\Lambda $ sets a
UV energy cut-off for magnetic fluctuations and $r$ controls the distance to
QCP. It is easy to see a considerable resemblance between this model and the
Holstein model (\ref{Holstein_H_general}) that we considered before. We
expect that the generalized polaron transformation of the type defined in
Eq. (\ref{SN}), together with the Gaussian state for electrons and bosons
can provide a good variational ans\"{a}tz for studying the ground state and
response functions of the system. The latter includes electron spectral
functions, optical conductivities, and spin response functions. An important
advantage of this method is that it allows to work directly with real time
and frequencies as we demonstrated in this paper.

\textit{Gauge fields}. We expect that variational non-Gaussian states can
also be applied in the study of QCD and lattice gauge theories \cite%
{LGT,KogutSusskind}. The simplest possible system to consider would be a one
dimensional Schwinger model in which Dirac fermions interacts with photons.
We can find not only the ground state but also analyze \textquotedblleft
emergent\textquotedblright\ elementary particles by solving the linearized
equations of motion around the steady state.

\textit{Open Systems}. Another interesting direction for extending our work
on non-Gaussian states is to consider open systems. Considering the density
matrix describing the system as a vector in super-space $\left\vert \rho
_{s}\right\rangle $ (see e.g., \cite{qo}), we can write the master equation
as
\begin{equation}
\partial _{t}\left\vert \rho _{s}\right\rangle =\mathcal{L}\left\vert \rho
_{s}\right\rangle ,
\end{equation}%
where $\mathcal{L}$ is the Lindblad super-operator that contains both the
Hamiltonian evolution and decoherence due to coupling to the bath. We expect
that the method of generalized Gaussian transformations can be extended to
the superspace thus allowing to explore a broader class of dynamical
phenomena \cite{DP1,DP2}.

\acknowledgments This project has been supported by the EU project SIQS. The authors
acknowledge Max-Planck-Harvard Research Center for Quantum Optics. The authors thank the useful discussions with Fabian Grusdt, Richard Schmidt,
Yulia E. Shchadilova, Valentin Kasper, Marton Kanasz-Nagy, Erez Zohar, Alejandro Gonzalez Tudela,
Yue Chang, Yinghai Wu, Chengyi Luo, XiaoLiang Qi, Pablo Sala, Jan von Delft, Peter Zoller, and Su Yi. ED
acknowledges support from Harvard-MIT CUA, NSF Grant No. DMR-1308435, AFOSR
Quantum Simulation MURI, AFOSR grant number FA9550-16-1-0323, the Humboldt
Foundation, and the Max Planck Institute for Quantum Optics.

\begin{widetext}

\appendix

\section{Imaginary and real time evolutions from the geometric point of view
\label{AppendixImaginary}}

In this Appendix, from the differential geometry perspective \cite{tdvp}, we
derive the general equations of motion (\ref{GME}) for variational
parameters and the constraint (\ref{C}) in the imaginary- and real- time
evolutions. We focus on the imaginary time evolution first, and equations of
motion in the real time evolution can be obtained following a similar
procedure.

The time derivative on the left hand side of Eq. (\ref{IMM}) is%
\begin{equation}
\partial _{\tau }\left\vert \Psi (\xi )\right\rangle =\sum_{j}d_{\tau }\xi
_{j}\left\vert \Psi _{j}\right\rangle ,  \label{LH}
\end{equation}%
where the states $\left\vert \Psi _{j}\right\rangle =\partial _{\xi
_{j}}\left\vert \Psi (\xi )\right\rangle $ span the tangent space of the
variational manifold $\left\vert \Psi (\xi )\right\rangle $ and $\partial
_{\tau }\left\vert \Psi (\xi )\right\rangle $ is a tangent vector, as
denoted by the dashed (red) arrow in Fig. \ref{schematic}. Note that $%
\left\vert \Psi _{j}\right\rangle $ might not be linearly independent. If
the rank $r_{\mathbf{G}}$ of the\ Gram matrix $\mathbf{G}_{ij}=\left\langle
\Psi _{i}\left\vert \Psi _{j}\right\rangle \right. $ is smaller than the
number of variational parameters, namely, some redundant parameters in $\xi $
have been introduced, one can always fix the value of some of the parameters
until the Gram matrix becomes invertible. Thus, in general, the tangent
vector $\partial _{\tau }\left\vert \Psi (\xi )\right\rangle $ is a
superposition of $r_{\mathbf{G}}$ linear independent vectors $\left\vert
\Psi _{j=1,...,r_{\mathbf{G}}}\right\rangle $.

The right hand side $\left\vert \mathbf{R}_{\Psi }\right\rangle
=-(H-E)\left\vert \Psi (\xi )\right\rangle $ in Eq. (\ref{IMM}) can be
decomposed into the vectors $\left\vert \Psi _{\parallel }\right\rangle =%
\mathbf{P}_{\xi }\left\vert \mathbf{R}_{\Psi }\right\rangle $ and $%
\left\vert \Psi _{\perp }\right\rangle =(\mathbf{1}-\mathbf{P}_{\xi
})\left\vert \mathbf{R}_{\Psi }\right\rangle $ in and orthogonal to the
tangent space, as denoted by the solid (black) arrows in Fig. \ref{schematic}%
, where $E=\left\langle \Psi (\xi )\right\vert H\left\vert \Psi (\xi
)\right\rangle $ is the average energy, and $\mathbf{P}_{\xi }$ is the
projector onto the tangent space.

By projecting the motion Eq. (\ref{IMM}) onto the tangent space, we obtain%
\begin{equation}
\sum_{j=1}^{r_{\mathbf{G}}}d_{\tau }\xi _{j}\left\vert \Psi
_{j}\right\rangle =\left\vert \Psi _{\parallel }\right\rangle ,  \label{MES}
\end{equation}%
which leads to the motion Eq. (\ref{GME}), i.e.,%
\begin{equation}
d_{\tau }\xi _{i}=\sum_{j=1}^{r_{\mathbf{G}}}\mathbf{G}_{ij}^{-1}\left\langle \Psi _{j}\left\vert \Psi _{\parallel }\right\rangle \right. .
\end{equation}%
We remark that the motion Eq. (\ref{GME}) minimizes the distance $\mathbf{d}%
_{0}=\left\vert \partial _{\tau }\left\vert \Psi (\xi )\right\rangle
-\left\vert \mathbf{R}_{\Psi }\right\rangle \right\vert =\left\vert
\left\vert \Psi _{\perp }\right\rangle \right\vert $ at each instant $\tau $.

It follows from Eq. (\ref{MES}) that the energy $E$ evolves as%
\begin{eqnarray}
d_{\tau }E &=&2\text{Re}\left\langle \Psi (\xi )\right\vert H\partial _{\tau
}\left\vert \Psi (\xi )\right\rangle =2\text{Re}\left\langle \Psi (\xi
)\right\vert H\left\vert \Psi _{\parallel }\right\rangle   \notag \\
&=&-2\left\langle \mathbf{R}_{\Psi }\right\vert \mathbf{P}_{\xi }\left\vert
\mathbf{R}_{\Psi }\right\rangle +2E\text{Re}\left\langle \Psi (\xi
)\right\vert \mathbf{P}_{\xi }\left\vert \mathbf{R}_{\Psi }\right\rangle .
\end{eqnarray}%
Since the state $\left\vert \Psi (\xi )\right\rangle $ is normalized, the
condition%
\begin{equation}
\partial _{\tau }\left\langle \Psi (\xi )\left\vert \Psi (\xi )\right\rangle
\right. =2\text{Re}\left\langle \Psi (\xi )\left\vert \Psi _{\parallel
}\right\rangle \right. =0
\end{equation}%
is always satisfied, which results in the monotonic decreasing behavior%
\begin{equation}
d_{\tau }E=-2\left\langle \mathbf{R}_{\Psi }\right\vert \mathbf{P}_{\xi
}\left\vert \mathbf{R}_{\Psi }\right\rangle \leq 0.
\end{equation}%
For the variational ground state in the limit $\tau \rightarrow \infty $, $%
\left\vert \Psi _{\parallel }\right\rangle =0$ and the energy $E$ stops
flowing, i.e., $d_{\tau }E=0$. The square norm $\left\vert \left\vert \Psi
_{\perp }\right\rangle \right\vert ^{2}=\langle \mathbf{R}_{\Psi }|\mathbf{R}%
_{\Psi }\rangle $ of the vector orthogonal to the tangent space is the
variance of the energy, which should be very small if the state we reach in
the limit $\tau \rightarrow \infty $ is close to the real ground state.

\begin{figure*}[tbp]
\begin{center}
\includegraphics[width=0.5\linewidth]{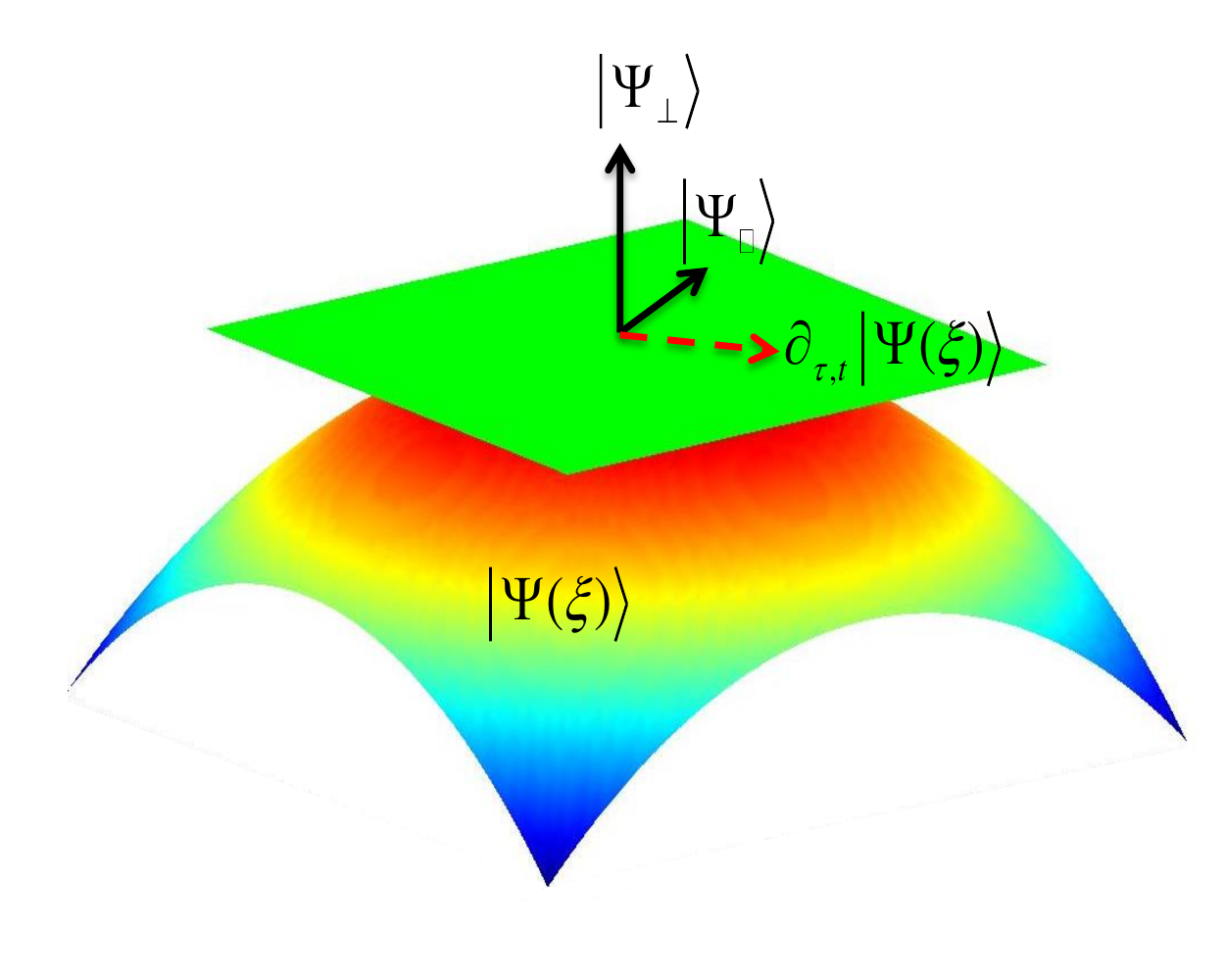}
\end{center}
\caption{The schematic of the non-Gaussian state manifold in the Hilbert
space, where the green plane denotes the tangent space. The Hamiltonian
generates the vector $\left\vert \Psi _{\parallel }\right\rangle +\left\vert
\Psi _{\perp }\right\rangle $.}
\label{schematic}
\end{figure*}

In the real time evolution, we project the Schr\"{o}dinger equation on the
tangent space as%
\begin{equation}
\partial _{t}\left\vert \Psi (\xi )\right\rangle =\mathbf{P}_{\xi
}\left\vert \mathbf{R}_{\Psi }\right\rangle ,  \label{PS}
\end{equation}%
where $\left\vert \mathbf{R}_{\Psi }\right\rangle =-iH\left\vert \Psi (\xi
)\right\rangle $. The projected Schr\"{o}dinger Eq. (\ref{PS}) results in
Eq. (\ref{GME}), where the derivative is taken with respect to the real time
$t$. It follows from the motion Eq. (\ref{GME}) that the total energy obeys%
\begin{equation}
d_{t}E=\left\langle d_{t}H\right\rangle +2\text{Re}\left\langle \Psi
_{\parallel }\right\vert H\left\vert \Psi (\{\xi \})\right\rangle .
\label{dtE}
\end{equation}%
The second term%
\begin{equation}
2\text{Re}\left\langle \Psi _{\parallel }\right\vert H\left\vert \Psi (\{\xi
\})\right\rangle =2\text{Re}i\left\vert \left\vert \Psi _{\parallel
}\right\rangle \right\vert ^{2}=0
\end{equation}%
in Eq. (\ref{dtE}) always vanishes, thus the condition (\ref{C}) is
satisfied.

The ground state $\left\vert \Psi _{\mathrm{G}}\right\rangle $ obtained from
the imaginary time evolution satisfies the relation%
\begin{equation}
\left\vert \Psi _{\parallel }\right\rangle =\mathbf{P}_{\xi }(H-E_{\mathrm{G}%
})\left\vert \Psi _{\mathrm{G}}\right\rangle =0.  \label{Imr}
\end{equation}%
It follows from Eq. (\ref{Imr}) that the real time evolution $\left\vert
\Psi _{\mathrm{G}}(t)\right\rangle =e^{-iE_{\mathrm{G}}t}\left\vert \Psi _{%
\mathrm{G}}\right\rangle $ of the variational ground state obeys the
projected Schr\"{o}dinger equation%
\begin{equation}
\partial _{t}\left\vert \Psi _{\mathrm{G}}(t)\right\rangle =-iE_{\mathrm{G}}%
\mathbf{P}_{\xi }\left\vert \Psi _{\mathrm{G}}(t)\right\rangle =-i\mathbf{P}%
_{\xi }H\left\vert \Psi _{\mathrm{G}}(t)\right\rangle .  \label{PSG}
\end{equation}

We consider the fluctuation $\xi =\xi _{\mathrm{G}}+\epsilon $ around the
ground state solution $\xi _{\mathrm{G}}$ of Eq. (\ref{GME}) in the limit $%
\tau \rightarrow \infty $, we expand the state%
\begin{equation}
\left\vert \Psi (\xi )\right\rangle =\left\vert \Psi _{\mathrm{G}%
}\right\rangle e^{-iE_{\mathrm{G}}t}+\sum_{j}\epsilon _{j}\left\vert \Psi
_{j}\right\rangle
\end{equation}%
to the linear order of $\epsilon $, where the vectors $\left\vert \Psi
_{j}\right\rangle =\left. \delta \left\vert \Psi (\xi )\right\rangle /\delta
\xi \right\vert _{\xi =\xi _{\mathrm{G}}}$ span the tangent space at $\xi _{%
\mathrm{G}}$.\ The projected Schrodinger Eq. (\ref{PS}) results in%
\begin{equation}
-iE_{\mathrm{G}}\mathbf{P}_{\xi }\left\vert \Psi _{\mathrm{G}%
}(t)\right\rangle +\sum_{j}d_{t}\epsilon _{j}\left\vert \Psi
_{j}\right\rangle =-i\mathbf{P}_{\xi }H\left\vert \Psi _{\mathrm{G}%
}(t)\right\rangle -i\sum_{j}\epsilon _{j}\mathbf{P}_{\xi }H\left\vert \Psi
_{j}\right\rangle ,  \label{dtf}
\end{equation}%
where the first terms on the right- and left- hand sides of Eq. (\ref{dtf})\
cancel each other due to the relation (\ref{PSG}). Finally, we obtain Eq. (%
\ref{GBB}) by projecting Eq. (\ref{dtf}) on the tangent vector $\left\vert
\Psi _{i}\right\rangle $.

\section{Tangent vector $\partial _{\protect\tau ,t}\left\vert \Psi _{%
\mathrm{NGS}}\right\rangle $ and $H\left\vert \Psi _{\mathrm{NGS}%
}\right\rangle $}

\label{Appendixtangent} In this Appendix, we prove that $H\left\vert \Psi _{%
\mathrm{NGS}}\right\rangle $ and the tangent vector $\partial _{\tau
}\left\vert \Psi _{\mathrm{NGS}}\right\rangle $ are composed of states with
the form (\ref{Texp}). The non-Gaussian states $\left\vert \Psi _{\mathrm{NGS%
}}\right\rangle =U_{S}\left\vert \Psi _{\mathrm{GS}}\right\rangle $ are
determined by the transformations $U_{S}=U_{3,4,5}$.

By moving the transformation $U_{S}$ to the most left side, we rewrite the
tangent vector as $\partial _{\tau }\left\vert \Psi _{\mathrm{NGS}%
}\right\rangle =U_{S}(u_{\mathrm{L}}+O)\left\vert \Psi _{\mathrm{GS}%
}\right\rangle $, where $u_{\mathrm{L}}=(\partial _{\tau }U_{\mathrm{GS}})U_{%
\mathrm{GS}}^{-1}$ and $O=U_{S}^{-1}\partial _{\tau }U_{S}$. The operator $%
u_{\mathrm{L}}$ can be obtained by the time derivatives $\partial _{\tau
}e^{i\frac{1}{2}R^{T}\sigma \Delta _{R}}$, $\partial _{\tau }e^{-i\frac{1}{4}%
R^{T}\xi _{b}R}$, and $\partial _{\tau }e^{i\frac{1}{4}A^{T}\xi _{m}A}$. The
derivative of the displacement operator is%
\begin{eqnarray}
\partial _{\tau }e^{i\frac{1}{2}R^{T}\sigma \Delta _{R}} &=&\partial _{\tau
}(e^{-\frac{1}{4}i\Delta _{x}^{T}\Delta _{p}}e^{\frac{1}{2}ix^{T}\Delta
_{p}}e^{-\frac{1}{2}ip^{T}\Delta _{x}})  \notag \\
&=&[\frac{1}{2}i\Delta _{p}^{T}\partial _{\tau }\Delta _{x}-\frac{1}{4}%
i\partial _{\tau }(\Delta _{x}^{T}\Delta _{p})+i\frac{1}{2}R^{T}\sigma
\partial _{\tau }\Delta _{R}]e^{i\frac{1}{2}R^{T}\sigma \Delta _{R}}  \notag
\\
&=&(\phi _{0}+i\frac{1}{2}R^{T}\sigma \partial _{\tau }\Delta _{R})e^{i\frac{%
1}{2}R^{T}\sigma \Delta _{R}},  \label{u1}
\end{eqnarray}%
where $\Delta _{R}=(\Delta _{x},\Delta _{p})^{T}$ and $\phi _{0}=-\frac{1}{4}%
i\Delta _{R}^{T}\sigma \partial _{\tau }\Delta _{R}$. In the first row of
Eq. (\ref{u1}), we used the canonical commutation relation $[x,p]=2i$ and
the Baker-Campbell-Hausdorff (BCH) formula $%
e^{O_{1}+O_{2}}=e^{-[O_{1},O_{2}]/2}e^{O_{1}}e^{O_{2}}$ for operators $%
O_{1}=ix^{T}\Delta _{p}/2$ and $O_{2}=-ip^{T}\Delta _{x}/2$. In the second
row, the displacement operator is moved to the most right side of other
operators by the relation $e^{ix\Delta _{p}/2}pe^{-ix\Delta _{p}/2}=p-\Delta
_{p}$. The derivative of the bosonic squeezing operator is%
\begin{eqnarray}
\partial _{\tau }e^{-i\frac{1}{4}R^{T}\xi _{b}R} &=&-i\frac{1}{4}%
\int_{0}^{1}due^{-iu\frac{1}{4}R^{T}\xi _{b}R}R^{T}(\partial _{\tau }\xi
_{b})Re^{iu\frac{1}{4}R^{T}\xi _{b}R}e^{-i\frac{1}{4}R^{T}\xi _{b}R}  \notag
\\
&=&-i\frac{1}{4}\int_{0}^{1}duR^{T}e^{u\xi _{b}\sigma }(\partial _{\tau }\xi
_{b})e^{-u\sigma \xi _{b}}Re^{-i\frac{1}{4}R^{T}\xi _{b}R}  \notag \\
&=&i\frac{1}{4}R^{T}\sigma (\partial _{\tau }S_{b})S_{b}^{-1}Re^{-i\frac{1}{4%
}R^{T}\xi _{b}R},  \label{u2}
\end{eqnarray}%
where in the first row of Eq. (\ref{u2}) we used the formula%
\begin{equation}
\partial _{\tau }e^{J(\tau )}=\int_{0}^{1}due^{uJ(\tau )}(\partial _{\tau
}J)e^{(1-u)J(\tau )}=\int_{0}^{1}due^{(1-u)J(\tau )}(\partial _{\tau
}J)e^{uJ(\tau )}  \label{dJ}
\end{equation}%
for any exponential operator $e^{J(\tau )}$, in the second row we used the
transformation%
\begin{equation}
e^{-iu\frac{1}{4}R^{T}\xi _{b}R}Re^{iu\frac{1}{4}R^{T}\xi _{b}R}=e^{-u\sigma
\xi _{b}}R,
\end{equation}%
and in the third row we calculated the integral%
\begin{equation}
\int_{0}^{1}due^{u\xi _{b}\sigma }(\partial _{\tau }\xi _{b})e^{-u\sigma \xi
_{b}}=-\sigma (\partial _{\tau }e^{\sigma \xi _{b}})e^{-\sigma \xi
_{b}}=-\sigma (\partial _{\tau }S_{b})S_{b}^{-1}
\end{equation}%
using Eq. (\ref{dJ}) and the property $e^{u\xi _{b}\sigma }\sigma
e^{-u\sigma \xi _{b}}=\sigma $ of the symplectic matrix $e^{-u\sigma \xi
_{b}}$.

The derivative of the fermionic squeezing operator is%
\begin{eqnarray}
\partial _{\tau }e^{i\frac{1}{4}A^{T}\xi _{m}A} &=&i\frac{1}{4}%
\int_{0}^{1}due^{iu\frac{1}{4}A^{T}\xi _{m}A}A^{T}(\partial _{\tau }\xi
_{m})Ae^{-iu\frac{1}{4}A^{T}\xi _{m}A}e^{i\frac{1}{4}A^{T}\xi _{m}A}  \notag
\\
&=&i\frac{1}{4}A^{T}\int_{0}^{1}due^{iu\xi _{m}}(\partial _{\tau }\xi
_{m})e^{-iu\xi _{m}}Ae^{i\frac{1}{4}A^{T}\xi _{m}A}  \notag \\
&=&\frac{1}{4}A^{T}(\partial _{\tau }U_{m})U_{m}^{T}Ae^{i\frac{1}{4}A^{T}\xi
_{m}A},  \label{u3}
\end{eqnarray}%
where Eq. (\ref{dJ}) is applied in the first and the third rows of Eq. (\ref%
{u3}), and in the second row we use the transformation%
\begin{equation}
e^{iu\frac{1}{4}A^{T}\xi _{m}A}Ae^{-iu\frac{1}{4}A^{T}\xi _{m}A}=e^{-iu\xi
_{m}}A
\end{equation}%
for fermions. The results (\ref{u1}), (\ref{u2}), and (\ref{u3}) show that $%
u_{\mathrm{L}}$ only contains the constant, linear, and quadratic terms of $R
$ and $A$.

In the next step, we analyze the structure of the operator $O$. For the
transformation $U_{3}$, the operator $O=U_{3}^{-1}\partial _{\tau }U_{3}$ is%
\begin{equation}
O=(\partial _{\tau }\lambda )P-\frac{1}{2\mathcal{N}}\partial _{\tau }%
\mathcal{N}.  \label{O3}
\end{equation}%
For the transformation $U_{4}$, the operator $O=U_{4}^{-1}\partial _{\tau
}U_{4}$ is%
\begin{eqnarray}
O &=&U_{4}^{-1}[(\partial _{\tau }e^{i\frac{1}{2}C^{\dagger }\bar{\xi}%
_{f}C})e^{i\frac{1}{2}\sum_{ij}\omega _{ij}^{f}\text{:}n_{i}^{f}n_{j}^{f}%
\text{:}}e^{i\sum_{ij}\bar{\omega}_{ij}R_{i}n_{j}^{f}}  \notag \\
&&+e^{i\frac{1}{2}C^{\dagger }\bar{\xi}_{f}C}(\partial _{\tau }e^{i\frac{1}{2%
}\sum_{ij}\omega _{ij}^{f}\text{:}n_{i}^{f}n_{j}^{f}\text{:}})e^{i\sum_{ij}%
\bar{\omega}_{ij}R_{i}n_{j}^{f}}  \notag \\
&&+e^{i\frac{1}{2}C^{\dagger }\bar{\xi}_{f}C}e^{i\frac{1}{2}\sum_{ij}\omega
_{ij}^{f}\text{:}n_{i}^{f}n_{j}^{f}\text{:}}(\partial _{\tau }e^{i\sum_{ij}%
\bar{\omega}_{ij}R_{i}n_{j}^{f}})].
\end{eqnarray}%
The derivative%
\begin{equation}
\partial _{\tau }e^{i\frac{1}{2}C^{\dagger }\bar{\xi}_{f}C}=e^{i\frac{1}{2}%
C^{\dagger }\bar{\xi}_{f}C}\frac{1}{2}C^{\dagger }\bar{U}_{f}^{\dagger
}(\partial _{\tau }\bar{U}_{f})C
\end{equation}%
in the first term can be obtained by the same procedure in Eq. (\ref{u3}),
where $\bar{U}_{f}=e^{i\bar{\xi}_{f}}$ is the unitary transformation and the
Gaussian transformation $e^{i\frac{1}{2}C^{\dagger }\bar{\xi}_{f}C}$ is
moved to the most left side of other operators. The derivative%
\begin{equation}
\partial _{\tau }e^{i\frac{1}{2}\sum_{ij}\omega _{ij}^{f}\text{:}%
n_{i}^{f}n_{j}^{f}\text{:}}=e^{i\frac{1}{2}\sum_{ij}\omega _{ij}^{f}\text{:}%
n_{i}^{f}n_{j}^{f}\text{:}}i\frac{1}{2}\sum_{ij}(\partial _{\tau }\omega
_{ij}^{f})\text{:}n_{i}^{f}n_{j}^{f}\text{:}  \label{d1}
\end{equation}%
in the second term is obtained by the commutation relation%
\begin{equation}
\lbrack e^{i\frac{1}{2}\sum_{ij}\omega _{ij}^{f}\text{:}n_{i}^{f}n_{j}^{f}%
\text{:}},i\frac{1}{2}\sum_{ij}(\partial _{\tau }\omega _{ij}^{f})\text{:}%
n_{i}^{f}n_{j}^{f}\text{:}]=0.  \label{d2}
\end{equation}%
The derivative%
\begin{equation}
\partial _{\tau }e^{i\sum_{ij}\bar{\omega}_{ij}R_{i}n_{j}^{f}}=e^{i\sum_{ij}%
\bar{\omega}_{ij}R_{i}n_{j}^{f}}[i\sum_{ij}n_{i}^{f}(\bar{\omega}^{T}\sigma
\partial _{\tau }\bar{\omega})_{ij}n_{j}^{f}+i\sum_{ij}(\partial _{\tau }%
\bar{\omega}_{ij})R_{i}n_{j}^{f}]  \label{d3}
\end{equation}%
in the third term is obtained by the same procedure in Eq. (\ref{u1}), where
the exponential operator is moved to the most left side. Eventially, the
operator $O$ becomes%
\begin{eqnarray}
O &=&\frac{1}{2}\bar{C}^{\dagger }\bar{U}_{f}^{\dagger }(\partial _{\tau }%
\bar{U}_{f})\bar{C}+i\sum_{i}(\bar{\omega}^{T}\sigma \partial _{\tau }\bar{%
\omega})_{ii}n_{i}^{f}  \notag \\
&&+i\sum_{ij}(\partial _{\tau }\bar{\omega}_{ij})R_{i}n_{j}^{f}+i\sum_{ij}(%
\frac{1}{2}\partial _{\tau }\omega ^{f}+\bar{\omega}^{T}\sigma \partial
_{\tau }\bar{\omega})_{ij}\text{:}n_{i}^{f}n_{j}^{f}\text{:},  \label{O4}
\end{eqnarray}%
where the operator $\bar{C}=(\bar{c},\bar{c}^{\dagger })^{T}$ is determined
by the transformation%
\begin{equation}
\bar{c}_{i}=e^{-i\sum_{ij}\bar{\omega}_{ij}R_{i}n_{j}^{f}}e^{-i\frac{1}{2}%
\sum_{ij}\omega _{ij}^{f}\text{:}n_{i}^{f}n_{j}^{f}\text{:}}c_{i}e^{i\frac{1%
}{2}\sum_{ij}\omega _{ij}^{f}\text{:}n_{i}^{f}n_{j}^{f}\text{:}}e^{i\sum_{ij}%
\bar{\omega}_{ij}R_{i}n_{j}^{f}}=e^{i\sum_{j}[(\omega ^{f}-\bar{\omega}%
^{T}\sigma \bar{\omega})_{ij}n_{j}^{f}+R_{j}\bar{\omega}_{ji}]}c_{i}.
\label{T4}
\end{equation}

For the transformation $U_{5}$, the operator $O=U_{5}^{-1}\partial _{\tau
}U_{5}$ is%
\begin{equation}
O=U_{5}^{-1}\partial _{\tau }(e^{i\frac{1}{2}C^{\dagger }\bar{\xi}_{f}C}e^{i%
\frac{1}{2}R^{T}\sigma \bar{\Delta}_{R}}e^{-i\frac{1}{4}R^{T}\bar{\xi}%
_{b}R})e^{J_{0}}+\partial _{\tau }J_{0},
\end{equation}%
where we used the commutation relation $[\partial _{\tau }J_{0},e^{J_{0}}]=0$
for the operator%
\begin{equation}
J_{0}=i\sum_{ij}(\frac{1}{2}\omega _{ij}^{f}\text{:}n_{i}^{f}n_{j}^{f}\text{:%
}+\frac{1}{2}\omega _{ij}^{b}\text{:}n_{i}^{b}n_{j}^{b}\text{:}+\omega
_{ij}^{bf}n_{i}^{b}n_{j}^{f}).
\end{equation}%
By the same procedure in Eqs. (\ref{u1}), (\ref{u2}), and (\ref{u3}), the
time derivative to the Gaussian transformation part is obtained, which leads
to%
\begin{eqnarray}
O &=&-\phi _{0}+i\frac{1}{2}\bar{R}^{T}\bar{S}_{b}^{T}\sigma \partial _{\tau
}\Delta _{R}+\frac{1}{2}\bar{C}^{\dagger }\bar{U}_{f}^{\dagger }(\partial
_{\tau }\bar{U}_{f})\bar{C}  \notag \\
&&+i\frac{1}{4}\bar{R}^{T}\bar{S}_{b}^{T}\sigma (\partial _{\tau }\bar{S}%
_{b})\bar{R}+\partial _{\tau }J_{0}.  \label{O5}
\end{eqnarray}%
Here, $\bar{S}_{b}=e^{\sigma \bar{\xi}_{b}}$, and the operators $\bar{C}=(%
\bar{c},\bar{c}^{\dagger })^{T}$ and $\bar{R}=(\bar{b}^{\dagger }+\bar{b},i(%
\bar{b}^{\dagger }-\bar{b}))^{T}$ are determined by the transformations%
\begin{eqnarray}
\bar{c}_{i} &=&e^{i\sum_{j}(w_{ij}^{f}n_{j}^{f}+w_{ji}^{bf}n_{j}^{b})}c_{i},
\notag \\
\bar{b}_{i} &=&e^{i\sum_{j}(w_{ij}^{b}n_{j}^{b}+w_{ij}^{bf}n_{j}^{f})}b_{i}.
\label{T5}
\end{eqnarray}%
The equations (\ref{O3}), (\ref{O4}), and (\ref{O5}) show that the operator $%
O$ has the form (\ref{ExpPL}). As a result, the tangent vector $\partial
_{\xi _{j}}\left\vert \Psi _{\mathrm{NGS}}\right\rangle $ is composed of
terms like those in Eq. (\ref{Texp}).

In the end of this section, we investigate the structure of the vector $%
\left\vert \mathbf{R}_{\Psi }\right\rangle $ determined by $H(R,C)\left\vert
\Psi _{\mathrm{NGS}}\right\rangle $. For the transformation $U_{3}$,%
\begin{equation}
H(R)\left\vert \Psi _{\mathrm{NGS}}\right\rangle =\frac{1}{\sqrt{\mathcal{N}}%
}[H(R)\cosh \lambda +PH(-R)\sinh \lambda ]\left\vert \Psi _{\mathrm{GS}%
}\right\rangle .  \label{H3}
\end{equation}%
For the transformations $U_{S}=U_{4,5}$, we move $U_{S}$ on the most left
side as%
\begin{equation}
H(R,C)\left\vert \Psi _{\mathrm{NGS}}\right\rangle =U_{S}\bar{H}%
(R,C)\left\vert \Psi _{\mathrm{GS}}\right\rangle
\end{equation}%
where $\bar{H}(R,C)=U_{S}^{\dagger }H(R,C)U_{S}$ is the Hamiltonian in the
rotating frame. The transformations $U_{S=4,5}$ act on the arguments of $%
H(R,C)$ as $\bar{H}(R,C)=H(R_{S},C_{S})$, where $R_{S}=U_{S}^{\dagger }RU_{S}
$ and $C_{S}=U_{S}^{\dagger }CU_{S}$. For $U_{S}=U_{4}$, the operators are%
\begin{eqnarray}
R_{S} &=&U_{4}^{\dagger }RU_{4}=R-2\sigma \bar{\omega}n^{f},  \notag \\
C_{S} &=&U_{4}^{\dagger }CU_{4}=\bar{U}_{f}\bar{C},  \label{RC4}
\end{eqnarray}%
where $\bar{C}$ is determined by Eq. (\ref{T4}). For $U_{S}=U_{5}$, the
operators are%
\begin{eqnarray}
R_{S} &=&U_{5}^{\dagger }RU_{5}=\bar{S}_{b}\bar{R}+\bar{\Delta}_{R},  \notag
\\
C_{S} &=&U_{5}^{\dagger }CU_{5}=\bar{U}_{f}\bar{C},  \label{RC5}
\end{eqnarray}%
where $\bar{R}$ and $\bar{C}$ are determined by Eq. (\ref{T5}).

For the Hamiltonian $H$ composed of the polynomials of $R$ and $C$, Eqs. (%
\ref{H3}), (\ref{RC4}), and (\ref{RC5}) show that the state $%
H(R,C)\left\vert \Psi _{\mathrm{NGS}}\right\rangle $ contains terms with the
form (\ref{Texp}).

\section{Mean values on bosonic Gaussian states}

\label{AppendixMB} In this Appendix, we show how to evaluate the mean values%
\begin{equation}
\left\langle e^{i\sum_{j}\gamma _{j}R_{j}}\text{poly}(R)\right\rangle _{%
\mathrm{GS}}  \label{m1}
\end{equation}%
and%
\begin{equation}
\left\langle e^{i\sum_{j}\beta _{j}n_{j}^{b}}\text{poly}(R)\right\rangle _{%
\mathrm{GS}}  \label{m2}
\end{equation}%
on the bosonic Gaussian state, which appear in Eq. (\ref{mExpPL}) for $%
U_{S}=U_{4,5}$.

We can use the normal ordering expansion%
\begin{eqnarray}
U_{\mathrm{GS}}^{\dagger }e^{iR^{T}\gamma }U_{\mathrm{GS}} &=&e^{i\Delta
_{R}^{T}\gamma }e^{iR^{T}S_{b}^{T}\gamma }  \notag \\
&=&e^{i\Delta _{R}^{T}\gamma }e^{-\frac{1}{2}\gamma ^{T}\Gamma _{b}\gamma }%
\text{:}e^{iR^{T}S_{b}^{T}\gamma }\text{:,}  \label{NO}
\end{eqnarray}%
to obtain $\left\langle e^{iR^{T}\gamma }\right\rangle _{\mathrm{GS}%
}=e^{i\Delta _{R}^{T}\gamma }e^{-\frac{1}{2}\gamma ^{T}\Gamma _{b}\gamma }$,
where $\gamma $ is the vector with the element $\gamma _{j}$. In the first
row of Eq. (\ref{NO}), the quadrature is displaced and squeezed by the
Gaussian transformation $U_{\mathrm{GS}}$, and in the second row the BCH
formula is used, where the normal ordering is defined with respect to the
vacuum state.

The mean value (\ref{m2}) contains the quadratic operators in the
exponential term. To evaluate this mean value, we introduce the Weyl
representation \cite{QO,Db}%
\begin{equation}
\rho _{\mathrm{GS}}=\int \frac{d^{2N_{b}}r}{(4\pi )^{N_{b}}}\chi (r)e^{-%
\frac{i}{2}\sum_{j}(\delta \hat{x}_{j}p_{j}-\delta \hat{p}_{j}x_{j})},
\label{dGS}
\end{equation}%
where the c-number vector $r=(x_{j},p_{j})$, and $\delta \hat{x}=\hat{x}%
-\Delta _{x}$ ($\delta \hat{p}=\hat{p}-\Delta _{p}$) describes the position
(momentum) fluctuation. For the bosonic Gaussian state, the characteristic
function is%
\begin{equation}
\chi (r)=tr[\rho _{\mathrm{GS}}e^{\frac{i}{2}\sum_{j}(\delta \hat{x}%
_{j}p_{j}-\delta \hat{p}_{j}x_{j})}]=e^{-\frac{1}{8}r^{T}\sigma ^{T}\Gamma
_{b}\sigma r}.
\end{equation}

As an example, we calculate the mean value%
\begin{equation}
A_{b}=\left\langle e^{i\sum_{i}\beta _{i}b_{i}^{\dagger
}b_{i}}b_{j_{1}}^{\dagger }...b_{j_{a}}^{\dagger
}b_{k_{1}}...b_{k_{b}}\right\rangle _{\mathrm{GS}}.
\end{equation}%
In terms of the density matrix (\ref{dGS}), the mean value becomes%
\begin{eqnarray}
A_{b} &=&\int \frac{d^{2N_{b}}r}{(4\pi )^{N_{b}}}\chi
(r)tr[e^{i\sum_{i}\beta _{i}b_{i}^{\dagger }b_{i}}b_{j_{1}}^{\dagger
}...b_{j_{a}}^{\dagger }b_{k_{1}}...b_{k_{b}}e^{-\frac{i}{2}\sum_{j}(\delta
\hat{x}_{j}p_{j}-\delta \hat{p}_{j}x_{j})}]  \notag \\
&=&\mathcal{F}_{J}\int \frac{d^{2N_{b}}r}{(4\pi )^{N_{b}}}\chi
(r)tr[e^{i\sum_{i}\beta _{i}b_{i}^{\dagger
}b_{i}}e^{\sum_{k}J_{k}b_{k}^{\dagger }}e^{\sum_{k}J_{k}^{\ast }b_{k}}e^{-%
\frac{i}{2}\sum_{j}(\delta \hat{x}_{j}p_{j}-\delta \hat{p}_{j}x_{j})}],
\end{eqnarray}%
where%
\begin{equation}
\mathcal{F}_{J}=\lim_{J\rightarrow 0}\frac{\delta }{\delta J_{j_{1}}}...%
\frac{\delta }{\delta J_{j_{a}}}\frac{\delta }{\delta J_{k_{1}}^{\ast }}...%
\frac{\delta }{\delta J_{k_{b}}^{\ast }}
\end{equation}%
denotes the functional derivative. By inserting the identity operator $%
I=\int d^{2}\mu \left\vert \mu \right\rangle \left\langle \mu \right\vert
/2\pi i$, we obtain%
\begin{eqnarray}
A_{b} &=&\mathcal{F}_{J}\int \frac{d^{2N_{b}}r}{(4\pi )^{N_{b}}}\chi (r)e^{-i%
\frac{1}{2}r^{T}\sigma \Delta _{R}}\int \frac{d^{2N_{b}}\mu }{(2\pi
i)^{N_{b}}}\times   \notag \\
&&\left\langle \{\mu _{j}\}\right\vert e^{\sum_{k}J_{k}b_{k}^{\dagger
}e^{i\beta _{k}}}e^{i\sum_{i}\beta _{i}b_{i}^{\dagger
}b_{i}}e^{\sum_{k}J_{k}^{\ast }b_{k}}e^{-\frac{i}{2}\sum_{j}(\hat{x}%
_{j}p_{j}-\hat{p}_{j}x_{j})}\left\vert \{\mu _{j}\}\right\rangle ,
\end{eqnarray}%
where the coherent state $\left\vert \mu \right\rangle =e^{-\left\vert \mu
\right\vert ^{2}/2}e^{\mu b^{\dagger }}\left\vert 0\right\rangle $. Using $%
b_{j}\left\vert \mu _{j}\right\rangle =\mu _{j}\left\vert \mu
_{j}\right\rangle $ and the relation%
\begin{equation}
e^{-\frac{i}{2}\sum_{j}(\hat{x}_{j}p_{j}-\hat{p}_{j}x_{j})}\left\vert \{\mu
_{j}\}\right\rangle =e^{\frac{1}{2}\sum_{j}(\nu _{j}^{\ast }\mu _{j}-\mu
_{j}^{\ast }\nu _{j})}e^{\sum_{j}[(\mu _{j}-\nu _{j})b_{j}^{\dagger }-(\mu
_{j}^{\ast }-\nu _{j}^{\ast })b_{j}]}\left\vert 0\right\rangle ,
\end{equation}%
we obtain%
\begin{eqnarray}
A_{b} &=&\mathcal{F}_{J}\int \frac{d^{2N_{b}}r}{(4\pi )^{N_{b}}}\chi (r)e^{-i%
\frac{1}{2}r^{T}\sigma \Delta _{R}}e^{-\sum_{j}J_{j}^{\ast }\nu _{j}-\frac{1%
}{2}\sum_{j}\nu _{j}^{\ast }\nu _{j}}  \notag \\
&&\int \frac{d^{2N_{b}}\mu }{(2\pi i)^{N_{b}}}\prod_{j}e^{-(1-e^{i\beta
_{j}})\mu _{j}^{\ast }\mu _{j}+(J_{j}^{\ast }+\nu _{j}^{\ast })\mu _{j}+\mu
_{j}^{\ast }(J_{j}-\nu _{j})e^{i\beta _{j}}},  \label{A2}
\end{eqnarray}%
where $\nu _{j}=(x_{j}+ip_{j})/2$.

In Eq. (\ref{A2}), the integrals over $r$ and $\mu $ are Gaussian integrals,
which can be evaluated analytically. The Gaussian integral over $\mu $ leads
to%
\begin{equation}
A_{b}=\prod_{j}(1-e^{i\beta _{j}})^{-1}\mathcal{F}_{J}e^{\sum_{j}J_{j}^{\ast
}\frac{e^{i\beta _{j}}}{1-e^{i\beta _{j}}}J_{j}}\int \frac{d^{2N_{b}}r}{%
(4\pi )^{N_{b}}}e^{-\frac{1}{8}r^{T}\sigma ^{T}(\Gamma _{b}+\frac{%
1+e^{i\beta }}{1-e^{i\beta }})\sigma r+\frac{1}{2}r^{T}(\Sigma \mathbf{J}%
-i\sigma \Delta _{R})},
\end{equation}%
where the vector $\mathbf{J}=(J_{j=1,...,N_{b}},J_{j=1,...,N_{b}}^{\ast
})^{T}$, the diagonal matrix $\beta ={\openone}_{2}\otimes diag(\beta _{j})$%
, and the matrix%
\begin{equation}
\Sigma =\left(
\begin{array}{cc}
\frac{e_{\beta }}{1-e_{\beta }} & -\frac{1}{1-e_{\beta }} \\
-i\frac{e_{\beta }}{1-e_{\beta }} & -i\frac{1}{1-e_{\beta }}%
\end{array}%
\right)
\end{equation}%
is define by $e_{\beta }=\exp [idiag(\beta _{j})]$ and the diagonal matrix $%
diag(\beta _{j})$ with elements $\beta _{j}$.

Redefining the variables $r=\sqrt{1-e^{i\beta }}\tilde{r}$, we rewrite%
\begin{eqnarray}
A_{b} &=&\mathcal{F}_{J}e^{\sum_{j}J_{j}^{\ast }\frac{e^{i\beta _{j}}}{%
1-e^{i\beta _{j}}}J_{j}}\int \frac{d^{2N_{b}}\tilde{r}}{(4\pi )^{N_{b}}}e^{-%
\frac{1}{8}\tilde{r}^{T}\sigma ^{T}\Gamma _{B}\sigma \tilde{r}+\frac{1}{2}%
\tilde{r}^{T}\sqrt{1-e^{i\beta }}(\Sigma \mathbf{J}-i\sigma \Delta _{R})}
\notag \\
&=&I_{G}e^{-\frac{1}{2}\Delta _{R}^{T}\sqrt{1-e^{i\beta }}\Gamma _{B}^{-1}%
\sqrt{1-e^{i\beta }}\Delta _{R}}\times   \notag \\
&&\mathcal{F}_{J}e^{\sum_{j}J_{j}^{\ast }\frac{e^{i\beta _{j}}}{1-e^{i\beta
_{j}}}J_{j}}e^{\frac{1}{2}\mathbf{J}^{\dagger }\Sigma ^{\dagger }\sigma ^{T}%
\sqrt{1-e^{i\beta }}\Gamma _{B}^{-1}\sqrt{1-e^{i\beta }}\sigma \Sigma
\mathbf{J}+i\Delta _{R}^{T}\sqrt{1-e^{i\beta }}\Gamma _{B}^{-1}\sqrt{%
1-e^{i\beta }}\sigma \Sigma \mathbf{J}},
\end{eqnarray}%
in terms of the Gaussian integral%
\begin{equation}
I_{G}=\int \frac{d^{2N_{b}}\tilde{r}}{(4\pi )^{N_{b}}}e^{-\frac{1}{8}\tilde{r%
}^{T}\sigma ^{T}\Gamma _{B}\sigma \tilde{r}},
\end{equation}%
where the symmetric matrix%
\begin{equation}
\Gamma _{B}=\sqrt{1-e^{i\beta }}\Gamma _{b}\sqrt{1-e^{i\beta }}+1+e^{i\beta
}=U_{\Gamma }^{T}d_{\Gamma }U_{\Gamma }
\end{equation}%
can be diagonalized by the positive-definite diagonal matrix $d_{\Gamma }$
and the unitary transformation $U_{\Gamma }$ \cite{Takagi}.

Preforming the Gaussian integral, we obtain%
\begin{eqnarray}
I_{G} &=&\int \frac{d^{2N_{b}}\tilde{r}}{(4\pi )^{N_{b}}}e^{-\frac{1}{8}%
\tilde{r}^{T}U_{\Gamma }^{T}d_{\Gamma }U_{\Gamma }\tilde{r}}=J(U_{\Gamma
}^{\dagger })\int \frac{d^{2N_{b}}r^{\prime }}{(4\pi )^{N_{b}}}e^{-\frac{1}{8%
}r^{\prime T}d_{\Gamma }r^{\prime }}  \notag \\
&=&J(U_{\Gamma }^{\dagger })\frac{1}{\sqrt{\det \frac{d_{\Gamma }}{2}}}=%
\frac{p_{0}}{\sqrt{\det \frac{\Gamma _{B}}{2}}}sign(\text{Re}\det U_{\Gamma
}),
\end{eqnarray}%
where $J(U_{\Gamma }^{\dagger })$ is the Jacobian for the change of integral
variables $r^{\prime }=U_{\Gamma }\tilde{r}$, and the sign $%
p_{0}=J(U_{\Gamma }^{\dagger })/\det U_{\Gamma }^{\dagger }$. Finally, the
mean value%
\begin{eqnarray}
A_{b} &=&\frac{s_{0}}{\sqrt{\det \frac{\Gamma _{B}}{2}}}e^{-\frac{1}{2}%
\Delta _{R}^{T}\sqrt{1-e^{i\beta }}\Gamma _{B}^{-1}\sqrt{1-e^{i\beta }}%
\Delta _{R}}\times   \notag \\
&&\mathcal{F}_{J}e^{\sum_{j}J_{j}^{\ast }\frac{e^{i\beta _{j}}}{1-e^{i\beta
_{j}}}J_{j}}e^{\frac{1}{2}\mathbf{J}^{\dagger }\Sigma ^{\dagger }\sigma ^{T}%
\sqrt{1-e^{i\beta }}\Gamma _{B}^{-1}\sqrt{1-e^{i\beta }}\sigma \Sigma
\mathbf{J}+i\Delta _{R}^{T}\sqrt{1-e^{i\beta }}\Gamma _{B}^{-1}\sqrt{%
1-e^{i\beta }}\sigma \Sigma \mathbf{J}}  \label{AJ}
\end{eqnarray}%
is obtained analytically, where%
\begin{equation}
s_{0}=p_{0}sign(\text{Re}\det U_{\Gamma }^{\dagger }).
\end{equation}

Here, we list the results several mean values%
\begin{equation}
\left\langle e^{i\sum_{i}\beta _{i}b_{i}^{\dagger }b_{i}}\right\rangle _{%
\mathrm{GS}}=\frac{s_{0}}{\sqrt{\det \frac{\Gamma _{B}}{2}}}e^{-\frac{1}{2}%
\Delta _{R}^{T}\sqrt{1-e^{i\beta }}\Gamma _{B}^{-1}\sqrt{1-e^{i\beta }}%
\Delta _{R}},  \label{B0}
\end{equation}%
\begin{eqnarray}
\left\langle e^{i\sum_{i}\beta _{i}b_{i}^{\dagger }b_{i}}b_{k}\right\rangle
_{\mathrm{GS}} &=&\left\langle e^{i\sum_{i}\beta _{i}b_{i}^{\dagger
}b_{i}}\right\rangle _{\mathrm{GS}}\frac{\delta }{\delta J_{k}^{\ast }}%
i\Delta _{R}^{T}\sqrt{1-e^{i\beta }}\Gamma _{B}^{-1}\sqrt{1-e^{i\beta }}%
\sigma \Sigma \mathbf{J}  \notag \\
&=&\left\langle e^{i\sum_{i}\beta _{i}b_{i}^{\dagger }b_{i}}\right\rangle _{%
\mathrm{GS}}\Delta _{R}^{T}\tilde{\Gamma}_{B}^{-1}\left(
\begin{array}{c}
\mathbf{1}_{k} \\
\mathbf{i}_{k}%
\end{array}%
\right) ,
\end{eqnarray}%
and%
\begin{eqnarray}
\left\langle e^{i\sum_{i}\beta _{i}b_{i}^{\dagger }b_{i}}b_{j}^{\dagger
}b_{k}\right\rangle _{\mathrm{GS}} &=&e^{i\beta _{j}}\left\langle
e^{i\sum_{i}\beta _{i}b_{i}^{\dagger }b_{i}}\right\rangle _{\mathrm{GS}}\{%
\frac{1}{1-e^{i\beta _{j}}}[\delta _{jk}-(\mathbf{1}_{j},-\mathbf{i}_{j})%
\tilde{\Gamma}_{B}^{-1}\left(
\begin{array}{c}
\mathbf{1}_{k} \\
\mathbf{i}_{k}%
\end{array}%
\right) ]  \notag \\
&&+(\mathbf{1}_{j},-\mathbf{i}_{j})(\tilde{\Gamma}_{B}^{-1})^{T}\Delta
_{R}\Delta _{R}^{T}\tilde{\Gamma}_{B}^{-1}\left(
\begin{array}{c}
\mathbf{1}_{k} \\
\mathbf{i}_{k}%
\end{array}%
\right) \},
\end{eqnarray}%
where $\tilde{\Gamma}_{B}=(1-e^{i\beta })\Gamma _{b}+(1+e^{i\beta })$, and
vectors $\mathbf{1}_{k}=(0,...,1_{k},...0)^{T}$, $\mathbf{i}%
_{k}=(0,...,i_{k},...0)^{T}$.

\section{Mean values on fermionic Gaussian states}

\label{AppendixMF} In this Appendix, we calculate the mean value%
\begin{equation}
\left\langle e^{i\sum_{j}\alpha _{j}n_{j}^{f}}\text{poly}(C)\right\rangle _{%
\mathrm{GS}}
\end{equation}%
on the fermionic Gaussian state. We introduce the fermionic Gaussian state
in the coherent representation as \cite{Df}%
\begin{equation}
\rho _{\mathrm{GS}}=\int d^{2N}\eta \chi _{N}(\eta )e^{\frac{1}{2}\eta
^{\ast }\eta }\int d^{2N}fe^{f\eta ^{\ast }-\eta f^{\ast }}\left\vert
f\right\rangle \left\langle -f\right\vert ,
\end{equation}%
by Grassmann numbers $\eta $ and $f$, where the characteristic function%
\begin{equation}
\chi _{N}(\eta )=tr[\rho _{\mathrm{GS}}e^{\eta c^{\dagger }-c\eta ^{\ast
}}]=\exp [i\frac{1}{8}\left( \eta _{1},\eta _{2}\right) \Gamma _{m}\left(
\begin{array}{c}
\eta _{1} \\
\eta _{2}%
\end{array}%
\right) ]  \label{Kf}
\end{equation}%
is determined by the covariance matrix $\Gamma _{m}$ and the real Grassmann
numbers $\eta _{1}=\eta ^{\ast }+\eta $ and $\eta _{2}=i(\eta ^{\ast }-\eta )
$.

We consider the mean value%
\begin{equation}
A_{f}=\left\langle e^{i\sum_{i}\alpha _{i}n_{i}^{f}}c_{j_{1}}^{\dagger
}...c_{j_{a}}^{\dagger }c_{k_{1}}...c_{k_{b}}\right\rangle _{\mathrm{GS}}.
\end{equation}%
In terms of the density matrix (\ref{Kf}), $A_{f}$ becomes%
\begin{equation}
A_{f}=\mathcal{F}_{J}\int d^{2N}\eta \chi _{N}(\eta )e^{\frac{1}{2}\eta
^{\ast }\eta }\int d^{2N}f\prod_{k}e^{-(1-e^{i\alpha _{k}})f_{k}^{\ast
}f_{k}+(J_{k}^{\ast }-\eta _{k}^{\ast })f_{k}+(J_{k}e^{i\alpha _{k}}-\eta
_{k})f_{k}^{\ast }}.  \label{Af}
\end{equation}%
In Eq. (\ref{Af}), the integrals over Grassmann numbers $\xi $ and $f$ are
Gaussian integrals, which can be calculated analytically.

The Gaussian integral over $f$ leads to%
\begin{eqnarray}
A_{f} &=&\prod_{j}(1-e^{i\alpha _{j}})\mathcal{F}_{J}e^{-\sum_{k}\frac{%
e^{i\alpha _{k}}J_{k}^{\ast }J_{k}}{1-e^{i\alpha _{k}}}}\times   \notag \\
&&\int d^{2N}\eta \exp [i\frac{1}{8}\left( \eta _{1},\eta _{2}\right)
(\Gamma _{m}-\frac{1+e^{i\alpha }}{1-e^{i\alpha }}\sigma )\left(
\begin{array}{c}
\eta _{1} \\
\eta _{2}%
\end{array}%
\right) +\frac{1}{2}(\eta _{1},\eta _{2})\Sigma _{F}\mathbf{J}],
\end{eqnarray}%
where $\alpha ={\openone}_{2}\otimes diag(\alpha _{j})$, and%
\begin{equation}
\Sigma _{F}=\left(
\begin{array}{cc}
\frac{e_{\alpha }}{1-e_{\alpha }} & -\frac{1}{1-e_{\alpha }} \\
-i\frac{e_{\alpha }}{1-e_{\alpha }} & -i\frac{1}{1-e_{\alpha }}%
\end{array}%
\right)
\end{equation}%
is defined by $e_{\alpha }=\exp [idiag(\alpha _{j})]$ and the diagonal
matrix $diag(\alpha _{j})$ with elements $\alpha _{j}$. Defining the new
Grassmann variable $\eta =\sqrt{1-e_{\alpha }}\tilde{\eta}$, we obtain%
\begin{equation}
A_{f}=(-\frac{1}{2})^{N}s_{f}\text{Pf}(\Gamma _{F})\mathcal{F}_{J}e^{\sum_{k}%
\frac{J_{k}^{\ast }J_{k}}{1-e^{-i\alpha _{k}}}-\frac{1}{2}i\mathbf{J}%
^{\dagger }\Sigma _{F}^{\dagger }\sqrt{1-e^{i\alpha }}\Gamma _{F}^{-1}\sqrt{%
1-e^{i\alpha }}\Sigma _{F}\mathbf{J}},
\end{equation}%
where Pf$(\Gamma _{F})$\ denotes the Pfaffian of the anti-symmetric matrix%
\begin{equation}
\Gamma _{F}=\sqrt{1-e^{i\alpha }}\Gamma _{m}\sqrt{1-e^{i\alpha }}%
-(1+e^{i\alpha })\sigma .
\end{equation}%
The sign $s_{f}$ is $(-1)^{N/2}$ for the even $N$ and $(-1)^{(N-1)/2}$ for
the odd $N$.

By taking the derivatives to $J$ and $J^{\ast }$, we obtain the average
values, e.g.,%
\begin{equation}
\left\langle e^{i\sum_{i}\alpha _{i}n_{i}^{f}}\right\rangle _{\mathrm{GS}}=(-%
\frac{1}{2})^{N}s_{f}\text{Pf}(\Gamma _{F}),
\end{equation}%
and%
\begin{equation}
\left\langle e^{i\sum_{i}\alpha _{i}n_{i}^{f}}c_{j}^{\dagger
}c_{k}\right\rangle _{\mathrm{GS}}=\frac{1}{4}ie^{i\alpha _{j}}\left\langle
e^{i\sum_{i}\alpha _{i}n_{i}^{f}}\right\rangle _{\mathrm{GS}}(\mathbf{1}_{k},%
\mathbf{i}_{k})(\Gamma _{m}+\sigma )\frac{1}{1+\frac{1}{2}(1-e^{i\alpha
})(\sigma \Gamma _{m}-1)}\left(
\begin{array}{c}
\mathbf{1}_{j} \\
-\mathbf{i}_{j}%
\end{array}%
\right) .
\end{equation}

\section{Equations of motion of $\Delta _{R}$ and $\Gamma _{b,m}$}

\label{AppendixME} In this Appendix, we explicitly derive the equations of
motion for $\Delta _{R}$ and $\Gamma _{b,m}$ in the non-Gaussian state
determined by the unitary transformation $U_{S}=U_{4,5}$. We focus on the
case $\bar{\xi}_{f,b}=\bar{\Delta}_{R}=0$. In the Gaussian limit $U_{S}=I$,
these equations reproduce the results (\ref{RGGSI}) and (\ref{RGGSR}) in
Sec. \ref{Gaussian state}.

For $U_{S}=U_{4,5}$ with $\bar{\xi}_{f,b}=\bar{\Delta}_{R}=0$, the tangent
vector $\partial _{\tau }\left\vert \Psi _{\mathrm{NGS}}\right\rangle $ has
the form $U_{S}$\textit{poly}$(R,C)\left\vert \Psi _{\mathrm{GS}%
}\right\rangle $. We shall construct the orthogonal tangent vectors
explicitly, such that the Gram matrix becomes diagonal. To orthogonalize the
tangent vectors, we move the unitary operators $U_{S}$ and $U_{\mathrm{GS}}$
to the most left side in the tangent vector $\partial _{\tau }\left\vert
\Psi _{\mathrm{NGS}}\right\rangle =U_{S}U_{\mathrm{GS}}U_{\mathrm{L}%
}\left\vert 0\right\rangle $. The relations (\ref{u1}), (\ref{u2}), and (\ref%
{u3}) obtained in Appendix \ref{Appendixtangent} give rise to%
\begin{eqnarray}
U_{\mathrm{L}} &=&\tilde{\phi}+i\frac{1}{2}R^{T}S_{b}^{T}\sigma \partial
_{\tau }\Delta _{R}+i\frac{1}{4}\text{:}R^{T}S_{b}^{T}\sigma (\partial
_{\tau }S_{b})R\text{:}  \notag \\
&&+\frac{1}{4}\text{:}A^{T}U_{m}^{T}(\partial _{\tau }U_{m})A\text{:}+U_{%
\mathrm{GS}}^{\dagger }OU_{\mathrm{GS}}-\left\langle O\right\rangle _{%
\mathrm{GS}},  \label{UL}
\end{eqnarray}%
\ where the imaginary number%
\begin{equation}
\tilde{\phi}=-\phi _{0}+i\frac{1}{4}tr[S_{b}^{T}\sigma (\partial _{\tau
}S_{b})\Gamma _{b}]+i\frac{1}{4}tr[U_{m}^{T}(\partial _{\tau }U_{m})\Gamma
_{m}]+\left\langle O\right\rangle _{\mathrm{GS}},
\end{equation}%
the term $O=U_{S}^{-1}\partial _{\tau }U_{S}$ higher than the quadratic
order, and all operators in $U_{\mathrm{L}}$ is normal ordered with respect
to the vacuum state.

To express the higher order term $U_{\mathrm{GS}}^{\dagger }OU_{\mathrm{GS}}$
in the normal ordering form, we employ the following theorem for the
arbitrary operator $\Xi $: the normal ordering expansion%
\begin{eqnarray}
U_{\mathrm{GS}}^{\dagger }\Xi U_{\mathrm{GS}} &=&\left\langle \Xi
\right\rangle _{\mathrm{GS}}+\frac{1}{2}R^{T}S_{b}^{T}\Xi _{\Delta }+\frac{1%
}{4}\text{:}R^{T}S_{b}^{T}\Xi _{b}S_{b}R\text{:}  \notag \\
&&+i\frac{1}{4}\text{:}A^{T}U_{m}^{T}\Xi _{m}U_{m}A\text{:}+\delta \Xi
\label{BEX}
\end{eqnarray}%
of the operator $U_{\mathrm{GS}}^{\dagger }\Xi U_{\mathrm{GS}}$ is
determined by the Wick theorem, where $\left\langle \Xi \right\rangle _{%
\mathrm{GS}}$ is the average value of $\Xi $ on the Gaussian state $U_{%
\mathrm{GS}}\left\vert 0\right\rangle $, and the operator $\delta \Xi $
contains the finite higher order normal ordered terms, e.g., the cubic and
quartic terms. The expansion coefficients%
\begin{equation}
\Xi _{\Delta }=2\frac{\delta \left\langle \Xi \right\rangle _{\mathrm{GS}}}{%
\delta \Delta _{R}},\Xi _{b}=4\frac{\delta \left\langle \Xi \right\rangle _{%
\mathrm{GS}}}{\delta \Gamma _{b}},\Xi _{m}=4\frac{\delta \left\langle \Xi
\right\rangle _{\mathrm{GS}}}{\delta \Gamma _{m}}  \label{OY}
\end{equation}%
are the functional derivatives of the aveage value $\left\langle \Xi
\right\rangle _{\mathrm{GS}}$ with respect to $\Delta _{R}$ and $\Gamma
_{b,m}$. Since the analytic result $\left\langle \Xi \right\rangle _{\mathrm{%
GS}}$ is obtained by the approach in Appendices \ref{AppendixMB} and \ref%
{AppendixMF}, the normal ordering expansion (\ref{BEX}) can be determined
analytically by Eq. (\ref{OY}).

By applying the result (\ref{BEX}) on the operator $U_{\mathrm{GS}}^{\dagger
}OU_{\mathrm{GS}}$, we obtain the normal ordering expansion%
\begin{eqnarray}
U_{\mathrm{GS}}^{\dagger }OU_{\mathrm{GS}} &=&\left\langle O\right\rangle _{%
\mathrm{GS}}+\frac{1}{2}R^{T}S_{b}^{T}O_{\Delta }+\frac{1}{4}\text{:}%
R^{T}S_{b}^{T}O_{b}S_{b}R\text{:}  \notag \\
&&+i\frac{1}{4}\text{:}A^{T}U_{m}^{T}O_{m}U_{m}A\text{:}+\delta O,
\label{BO}
\end{eqnarray}%
where $O_{\Delta }$, $O_{b}$, and $O_{m}$ are obtained by replacing $\Xi $
to $O$ in Eq. (\ref{OY}).

Applying $U_{\mathrm{L}}$ on the vacuum state, we obtain%
\begin{equation}
U_{\mathrm{L}}\left\vert 0\right\rangle =\sum_{n=0}^{2}\left\vert
L_{n}\right\rangle +\delta O\left\vert 0\right\rangle   \label{UL0}
\end{equation}%
by Eqs. (\ref{UL}) and (\ref{BEX}), where $\left\vert L_{0}\right\rangle =%
\tilde{\phi}\left\vert 0\right\rangle $, the linear term%
\begin{equation}
\left\vert L_{1}\right\rangle =\frac{1}{2}R^{T}S_{b}^{T}(i\sigma \partial
_{\tau }\Delta _{R}+O_{\Delta })\left\vert 0\right\rangle ,
\end{equation}%
and the quadratic terms%
\begin{eqnarray}
\left\vert L_{2}\right\rangle  &=&[\frac{1}{4}\text{:}R^{T}S_{b}^{T}(i\sigma
\partial _{\tau }S_{b}+O_{b}S_{b})R\text{:}  \notag \\
&&+\frac{1}{4}\text{:}A^{T}U_{m}^{T}(\partial _{\tau }U_{m}+iO_{m}U_{m})A%
\text{:}]\left\vert 0\right\rangle .
\end{eqnarray}%
Since the operators in $U_{\mathrm{L}}$ are normal ordered, the states $%
\left\vert L_{n}\right\rangle $ and $\delta O\left\vert 0\right\rangle $ in
Eq. (\ref{UL0}) only contain the creation operators acting on the vacuum
state. As a result, the states $\left\vert L_{n=0,1,2}\right\rangle $ and $%
\delta O\left\vert 0\right\rangle $ form the orthogonal basis in the tangent
space, which describe the different numbers $n=0,1,2,..$ of excitations on
the vacuum state.

Moving the unitary operators $U_{S}$ and $U_{\mathrm{GS}}$ to the most left
side, we obtain the state $\left\vert \mathbf{R}_{\Psi }\right\rangle
=U_{S}U_{\mathrm{GS}}\left\vert \Psi _{\mathrm{R}}\right\rangle $, where%
\begin{equation}
\left\vert \Psi _{\mathrm{R}}\right\rangle =\left\{
\begin{array}{c}
-(U_{\mathrm{R}}-\left\langle \bar{H}\right\rangle _{\mathrm{GS}})\left\vert
0\right\rangle \text{, \ \ imaginary time evolution} \\
-iU_{\mathrm{R}}\left\vert 0\right\rangle \text{, \ \ real time evolution}%
\end{array}%
\right. ,
\end{equation}%
and $U_{\mathrm{R}}=U_{\mathrm{GS}}^{\dagger }\bar{H}U_{\mathrm{GS}}$ is
deterimned by the Hamiltonian $\bar{H}=U_{S}^{\dagger }HU_{S}$ in the
rotating frame. The normal ordering expansion%
\begin{eqnarray}
U_{\mathrm{R}} &=&\left\langle \bar{H}\right\rangle _{\mathrm{GS}}+\frac{1}{2%
}R^{T}S_{b}^{T}h_{\Delta }+\frac{1}{4}\text{:}R^{T}S_{b}^{T}h_{b}S_{b}R\text{%
:}  \notag \\
&&+i\frac{1}{4}\text{:}A^{T}U_{m}^{T}h_{m}U_{m}A\text{:}+\delta \bar{H}
\label{UR}
\end{eqnarray}%
of $U_{\mathrm{R}}$ are obtained by Eq. (\ref{BEX}), where the expansion
coefficients%
\begin{equation}
h_{\Delta }=2\frac{\delta \left\langle \bar{H}\right\rangle _{\mathrm{GS}}}{%
\delta \Delta _{R}},h_{b}=4\frac{\delta \left\langle \bar{H}\right\rangle _{%
\mathrm{GS}}}{\delta \Gamma _{b}},h_{m}=4\frac{\delta \left\langle \bar{H}%
\right\rangle _{\mathrm{GS}}}{\delta \Gamma _{m}}  \label{hY}
\end{equation}%
follow from Eq. (\ref{OY}), and $\delta \bar{H}$ contains the higher order
normal ordered operators.

Applying $U_{\mathrm{R}}$ on the vacuum state, we obtain%
\begin{equation}
U_{\mathrm{R}}\left\vert 0\right\rangle =\sum_{n=0}^{2}\left\vert \tilde{L}%
_{n}\right\rangle +\delta \bar{H}\left\vert 0\right\rangle ,  \label{UR0}
\end{equation}%
where $\left\vert \tilde{L}_{0}\right\rangle =\left\langle \bar{H}%
\right\rangle _{\mathrm{GS}}\left\vert 0\right\rangle $, the linear term $%
\left\vert \tilde{L}_{1}\right\rangle =R^{T}S_{b}^{T}h_{\Delta }\left\vert
0\right\rangle /2$, the quadratic term%
\begin{equation}
\left\vert \tilde{L}_{2}\right\rangle =(\frac{1}{4}\text{:}%
R^{T}S_{b}^{T}h_{b}S_{b}R\text{:}+i\frac{1}{4}\text{:}%
A^{T}U_{m}^{T}h_{m}U_{m}A\text{:})\left\vert 0\right\rangle ,
\end{equation}%
and $\delta \bar{H}\left\vert 0\right\rangle $ are the orthogonal to each
other.

The equations of motion for $\Delta _{R}$ and $\Gamma _{b,m}$ are obtained
by the comparison of tangent vectors $\left\vert L_{1,2}\right\rangle $ and $%
\left\vert \tilde{L}_{1,2}\right\rangle $. For the imaginary- and real- time
evolutions, we have $\left\vert L_{1,2}\right\rangle =-\left\vert \tilde{L}%
_{1,2}\right\rangle $ and $\left\vert L_{1,2}\right\rangle =-i\left\vert
\tilde{L}_{1,2}\right\rangle $, respectively. For the imaginary time
evolution, the relations $\left\vert L_{1,2}\right\rangle =-\left\vert
\tilde{L}_{1,2}\right\rangle $ give rise to%
\begin{equation}
R^{T}S_{b}^{T}(i\sigma \partial _{\tau }\Delta _{R}+O_{\Delta })\left\vert
0\right\rangle =-R^{T}S_{b}^{T}h_{\Delta }\left\vert 0\right\rangle ,
\label{L1}
\end{equation}%
\begin{equation}
\text{:}R^{T}S_{b}^{T}(i\sigma \partial _{\tau }S_{b}+O_{b}S_{b})R\text{:}%
\left\vert 0\right\rangle =-\text{:}R^{T}S_{b}^{T}h_{b}S_{b}R\text{:}%
\left\vert 0\right\rangle ,  \label{Lb2}
\end{equation}%
and%
\begin{equation}
\text{:}A^{T}U_{m}^{T}(\partial _{\tau }U_{m}+iO_{m}U_{m})A\text{:}%
\left\vert 0\right\rangle =-i\text{:}A^{T}U_{m}^{T}h_{m}U_{m}A\text{:}%
\left\vert 0\right\rangle .  \label{Lf2}
\end{equation}%
Here, since all terms are normal ordered, only the creation operators
survive in Eqs. (\ref{L1})-(\ref{Lf2}). The left and right hand sides of Eq.
(\ref{L1}) become%
\begin{eqnarray}
&&R^{T}S_{b}^{T}(i\sigma \partial _{\tau }\Delta _{R}+O_{\Delta })\left\vert
0\right\rangle   \notag \\
&=&(b^{\dagger },ib^{\dagger })S_{b}^{T}(i\sigma \partial _{\tau }\Delta
_{R}+O_{\Delta })\left\vert 0\right\rangle   \notag \\
&=&-(b^{\dagger },ib^{\dagger })\sigma S_{b}^{-1}\sigma (i\sigma \partial
_{\tau }\Delta _{R}+O_{\Delta })\left\vert 0\right\rangle   \notag \\
&=&i(b^{\dagger },ib^{\dagger })S_{b}^{-1}\sigma (i\sigma \partial _{\tau
}\Delta _{R}+O_{\Delta })\left\vert 0\right\rangle ,  \label{L1l}
\end{eqnarray}%
and%
\begin{equation}
-R^{T}S_{b}^{T}h_{\Delta }\left\vert 0\right\rangle =-(b^{\dagger
},ib^{\dagger })S_{b}^{T}h_{\Delta }\left\vert 0\right\rangle ,  \label{L1r}
\end{equation}%
where $b^{\dagger }=(b_{1}^{\dagger },...,b_{N_{b}}^{\dagger })$ and we used
the relation $S_{b}^{T}\sigma S_{b}=\sigma $. By comparing the real and
imaginary parts in Eqs. (\ref{L1l}) and (\ref{L1r}), we obtain the motion
equation%
\begin{equation}
\partial _{\tau }\Delta _{R}=-\Gamma _{b}h_{\Delta }-i\sigma O_{\Delta }.
\end{equation}

The left and right hand sides of Eq. (\ref{Lb2}) are%
\begin{equation}
\text{:}R^{T}S_{b}^{T}(i\sigma \partial _{\tau }S_{b}+O_{b}S_{b})R\text{:}%
\left\vert 0\right\rangle =(b^{\dagger },ib^{\dagger })S_{b}^{T}(i\sigma
\partial _{\tau }S_{b}+O_{b}S_{b})\left(
\begin{array}{c}
b^{\dagger } \\
ib^{\dagger }%
\end{array}%
\right) \left\vert 0\right\rangle ,  \label{Lb2l}
\end{equation}%
and%
\begin{equation}
-\text{:}R^{T}S_{b}^{T}h_{b}S_{b}R\text{:}\left\vert 0\right\rangle =\frac{1%
}{2}i(b^{\dagger },ib^{\dagger })(S_{b}^{T}h_{b}S_{b}\sigma -\sigma
S_{b}^{T}h_{b}S_{b})\left(
\begin{array}{c}
b^{\dagger } \\
ib^{\dagger }%
\end{array}%
\right) \left\vert 0\right\rangle .  \label{Lb2r}
\end{equation}%
Here, we notice that the motion equation of $S_{b}$ can not be uniquely
determined, since the variational parameters in Eq. (\ref{GS}) has some
gauge degrees of freedoms, as discussed in Sec. \ref{Gaussian state}.
However, the motion equation of $\Gamma _{b}$ for each equivalent class is
uniquely determined, namely, $\Gamma _{b}$ is gauge invariant. Thus, we can
choose the motion equation for one $S_{b}$ in the equivalent class, and
derive the motion equation of $\Gamma _{b}$.

By comparing the right hand sides in Eqs. (\ref{Lb2l}) and (\ref{Lb2r}), we
obtain%
\begin{equation}
\partial _{\tau }S_{b}=-\frac{1}{2}\sigma h_{b}S_{b}\sigma -\frac{1}{2}%
\Gamma _{b}h_{b}S_{b}-i\sigma O_{b}S_{b},  \label{dS}
\end{equation}%
where $\partial _{\tau }(S_{b}\sigma S_{b}^{T})=0$ maintains the
symplecticity of $S_{b}$. The motion Eq. (\ref{dS}) of $S_{b}$ leads to%
\begin{equation}
\partial _{\tau }\Gamma _{b}=-\sigma h_{b}\sigma -\Gamma _{b}h_{b}\Gamma
_{b}+i\Gamma _{b}O_{b}\sigma -i\sigma O_{b}\Gamma _{b}.
\end{equation}

The left and right hand sides of Eq. (\ref{Lf2}) are%
\begin{equation}
\text{:}A^{T}U_{m}^{T}(\partial _{\tau }U_{m}+iO_{m}U_{m})A\text{:}%
\left\vert 0\right\rangle =\text{:}(c^{\dagger },ic^{\dagger
})U_{m}^{T}(\partial _{\tau }U_{m}+iO_{m}U_{m})\left(
\begin{array}{c}
c^{\dagger } \\
ic^{\dagger }%
\end{array}%
\right) \text{:}\left\vert 0\right\rangle ,  \label{Lf2l}
\end{equation}%
and%
\begin{equation}
-i\text{:}A^{T}U_{m}^{T}h_{m}U_{m}A\text{:}\left\vert 0\right\rangle =\frac{1%
}{2}\text{:}(c^{\dagger },ic^{\dagger })(\sigma
U_{m}^{T}h_{m}U_{m}-U_{m}^{T}h_{m}U_{m}\sigma )\left(
\begin{array}{c}
c^{\dagger } \\
ic^{\dagger }%
\end{array}%
\right) \text{:}\left\vert 0\right\rangle ,  \label{Lf2r}
\end{equation}%
where $c^{\dagger }=(c_{1}^{\dagger },...,c_{N_{b}}^{\dagger })$. Similar to
the bosonic case, the motion equation of $U_{m}$ is not unique, however, the
motion equation of $\Gamma _{b}$ can be determined uniquely in each
equivalent case. By comparing the right hand sides in Eqs. (\ref{Lf2l}) and (%
\ref{Lf2r}), we obtain%
\begin{equation}
\partial _{\tau }U_{m}=-\frac{1}{2}h_{m}U_{m}\sigma -\frac{1}{2}\Gamma
_{m}h_{m}U_{m}-iO_{m}U_{m},
\end{equation}%
which leads to%
\begin{equation}
\partial _{\tau }\Gamma _{m}=-h_{m}-\Gamma _{m}h_{m}\Gamma _{m}+i[\Gamma
_{m},O_{m}].
\end{equation}%
Here, $\partial _{\tau }(U_{m}U_{m}^{T})=0$ justifies the orthogonality of $%
U_{m}$.

The imaginary-time evolution equations of motion are summerized as follows%
\begin{eqnarray}
\partial _{\tau }\Delta _{R} &=&-\Gamma _{b}h_{\Delta }-i\sigma O_{\Delta },
\notag \\
\partial _{\tau }\Gamma _{b} &=&\sigma ^{T}h_{b}\sigma -\Gamma
_{b}h_{b}\Gamma _{b}+i\Gamma _{b}O_{b}\sigma -i\sigma O_{b}\Gamma _{b},
\notag \\
\partial _{\tau }\Gamma _{m} &=&-h_{m}-\Gamma _{m}h_{m}\Gamma _{m}+i[\Gamma
_{m},O_{m}].  \label{IMNGS}
\end{eqnarray}%
By the samilar procedure, the real-time equations of motion are%
\begin{eqnarray}
\partial _{t}\Delta _{R} &=&\sigma h_{\Delta }^{t},  \notag \\
\partial _{t}\Gamma _{b} &=&\sigma h_{b}^{t}\Gamma _{b}-\Gamma
_{b}h_{b}^{t}\sigma ,  \notag \\
\partial _{t}\Gamma _{m} &=&[h_{m}^{t},\Gamma _{m}],  \label{RENGS}
\end{eqnarray}%
where $h_{\Delta }^{t}=h_{\Delta }-iO_{\Delta }$, $h_{b}^{t}=h_{b}-iO_{b}$,
and $h_{m}^{t}=h_{m}-iO_{m}$. In Appendix \ref{AppendixME2}, we provide
another approach to derive Eqs. (\ref{IMNGS}) and (\ref{RENGS}).

We notice that in the motion Eqs. (\ref{IMNGS}) and (\ref{RENGS}), $%
O_{\Delta }$, $O_{b}$, and $O_{m}$ still contain the time derivatives of the
variational parameters in $U_{S}$, e.g., $\partial _{\tau }\bar{\omega}$, $%
\partial _{\tau }\omega ^{f}$, $\partial _{\tau }\omega ^{b}$, and $\partial
_{\tau }\omega ^{bf}$. The equations of motion of these rest variational
parameters can be obtained by the procedure shown in Sec. \ref%
{VariationalPrinciple} (see Secs. \ref{SB} and \ref{SCCDW} as examples).

From the equations of motion, one can use the relation (\ref{Gf}) to derive
the equations of motion%
\begin{equation}
\partial _{\tau }\Gamma _{f}=\{\Gamma _{f},h_{f}\}-2\Gamma _{f}h_{f}\Gamma
_{f}+[\Gamma _{f},O_{f}]  \label{IGf}
\end{equation}%
and%
\begin{equation}
\partial _{t}\Gamma _{f}=i[\Gamma _{f},h_{f}^{t}]  \label{RGf}
\end{equation}%
for the correlation matrix $\Gamma _{f}$ in the imaginary- and real- time
evolutions, respectively, where in the Dirac fermion basis the matrices $%
h_{f}^{t}=h_{f}-iO_{f}$ and%
\begin{equation}
h_{f}=i\frac{1}{2}W_{m}^{\dagger }h_{m}W_{m},O_{f}=i\frac{1}{2}%
W_{m}^{\dagger }O_{m}W_{m}.  \label{hfOf}
\end{equation}

For Gaussian states, i.e., $U_{S}=I$, the quantities $O_{\Delta
}=O_{b}=O_{m}=0$ and the motion Eqs. (\ref{IMNGS})-(\ref{RENGS}) result in
Eqs. (\ref{RGGSI}) and (\ref{RGGSR}) in Sec. \ref{Gaussian state}.

In Appendix \ref{AppendixImaginary}, we show that the relation (\ref{C}) is
guaranteed for the general variational ans\"{a}tz. Here, we show the
condition (\ref{C}) for Gaussian states by using the motion Eqs. (\ref{RGGSI}%
) and (\ref{RGGSR}) directly. Since the Gaussian state energy $E(t,\Delta
_{R},\Gamma _{b,m})$ is the function of $t$, $\Delta _{R}$ and $\Gamma _{b,m}
$, the time derivative of the energy has the form%
\begin{eqnarray}
\frac{dE}{dt} &=&\left\langle \frac{\partial H}{\partial t}\right\rangle
+\sum_{j}\frac{\delta E}{\delta \Delta _{R,j}}\partial _{t}\Delta _{R,j}
\notag \\
&&+\sum_{ij}(\frac{\delta E}{\delta \Gamma _{b,ij}}\partial _{t}\Gamma
_{b,ij}+\frac{\delta E}{\delta \Gamma _{m,ij}}\partial _{t}\Gamma _{m,ij}).
\end{eqnarray}%
The motion Eqs. (\ref{RGGSI}) and (\ref{RGGSR}) lead to%
\begin{eqnarray}
\sum_{j}\frac{\delta E}{\delta \left\langle R_{j}\right\rangle }\partial
_{t}\left\langle R_{j}\right\rangle  &=&\frac{1}{2}h_{\Delta }^{T}\sigma
h_{\Delta }=0,  \notag \\
\sum_{ij}\frac{\delta E}{\delta \Gamma _{b,ij}}\partial _{t}\Gamma _{b,ij}
&=&\frac{1}{4}tr[h_{b}\sigma h_{b},\Gamma _{b}]=0,  \notag \\
\sum_{ij}\frac{\delta E}{\delta \Gamma _{m,ij}}\partial _{t}\Gamma _{m,ij}
&=&-\frac{1}{4}tr(h_{m}[h_{m},\Gamma _{m}])=0,
\end{eqnarray}%
which eventually give rise to the constraint (\ref{C}), i.e.,%
\begin{equation}
\frac{dE}{dt}=\left\langle \frac{\partial H}{\partial t}\right\rangle .
\end{equation}

\section{Alternative derivation of motion Eqs. (\protect\ref{IMNGS}) and (%
\protect\ref{RENGS})}

\label{AppendixME2} In this Appendix, we derive the motion Eqs. (\ref{IMNGS}%
) and (\ref{RENGS}) from another point of view. For the imaginary time
evolution, $\left\vert L_{1,2}\right\rangle =-\left\vert \tilde{L}%
_{1,2}\right\rangle $ leads to $\sum_{j=1,2}\left\vert L_{j}\right\rangle
=-\sum_{j=1,2}\left\vert \tilde{L}_{j}\right\rangle $, i.e.,%
\begin{eqnarray}
&&[\frac{1}{2}R^{T}S_{b}^{T}(i\sigma \partial _{\tau }\Delta _{R}+O_{\Delta
})+\frac{1}{4}\text{:}R^{T}S_{b}^{T}(i\sigma \partial _{\tau
}S_{b}+O_{b}S_{b})R\text{:}+\frac{1}{4}\text{:}A^{T}U_{m}^{T}(\partial
_{\tau }U_{m}+iO_{m}U_{m})A\text{:}]\left\vert 0\right\rangle  \notag \\
&&=-[\frac{1}{2}R^{T}S_{b}^{T}h_{\Delta }+\frac{1}{4}\text{:}%
R^{T}S_{b}^{T}h_{b}S_{b}R\text{:}+i\frac{1}{4}\text{:}%
A^{T}U_{m}^{T}h_{m}U_{m}A\text{:}]\left\vert 0\right\rangle .  \label{MEI}
\end{eqnarray}%
The left hand side of Eq. (\ref{MEI}) can be written as%
\begin{equation}
U_{\mathrm{GS}}^{-1}\partial _{\tau }\left\vert \Psi _{\mathrm{GS}%
}\right\rangle +[\frac{1}{2}R^{T}S_{b}^{T}O_{\Delta }+\frac{1}{4}\text{:}%
R^{T}S_{b}^{T}O_{b}S_{b}R\text{:}+i\frac{1}{4}\text{:}%
A^{T}U_{m}^{T}O_{m}U_{m}A\text{:}]\left\vert 0\right\rangle .
\end{equation}%
By multiplying $U_{\mathrm{GS}}^{-1}$ on both sides of Eq. (\ref{MEI}), we
obtain the motion equation%
\begin{equation}
\partial _{\tau }\left\vert \Psi _{\mathrm{GS}}\right\rangle =-[\frac{1}{2}%
\delta R^{T}(h_{\Delta }+O_{\Delta })+\frac{1}{4}\text{:}\delta
R^{T}(h_{b}+O_{b})\delta R\text{:}+i\frac{1}{4}\text{:}A^{T}(h_{m}+O_{m})A%
\text{:}]\left\vert \Psi _{\mathrm{GS}}\right\rangle .  \label{MEW}
\end{equation}

The equation (\ref{MEW}) gives the motion equation%
\begin{eqnarray}
\partial _{\tau }\rho _{\mathrm{GS}} &=&-\{\frac{1}{2}\delta R^{T}h_{\Delta
}+\frac{1}{4}\text{:}\delta R^{T}h_{b}\delta R\text{:}+i\frac{1}{4}\text{:}%
A^{T}h_{m}A\text{:},\rho _{\mathrm{GS}}\}  \notag \\
&&-[\frac{1}{2}\delta R^{T}O_{\Delta }+\frac{1}{4}\text{:}\delta
R^{T}O_{b}\delta R\text{:}+i\frac{1}{4}\text{:}A^{T}O_{m}A\text{:},\rho _{%
\mathrm{GS}}]  \label{drho}
\end{eqnarray}%
of the density matrix $\rho _{\mathrm{GS}}=\left\vert \Psi _{\mathrm{GS}%
}\right\rangle \left\langle \Psi _{\mathrm{GS}}\right\vert $.

By taking the trace of operators $R$, $\{\delta R,\delta R^{T}\}/2$, and $%
i[A,A^{T}]/2$ on both sides of Eq. (\ref{drho}) as%
\begin{eqnarray}
\partial _{\tau }\Delta _{R} &=&tr[R\partial _{\tau }\rho _{\mathrm{GS}}],
\notag \\
\partial _{\tau }\Gamma _{b} &=&tr[\frac{1}{2}\{\delta R,\delta
R^{T}\}\partial _{\tau }\rho _{\mathrm{GS}}],  \notag \\
\partial _{\tau }\Gamma _{m} &=&tr[\frac{i}{2}[A,A^{T}]\partial _{\tau }\rho
_{\mathrm{GS}}],
\end{eqnarray}%
we can reproduce Eq. (\ref{IMNGS}) by the Wick theorem. Following the
samilar procedure, we can obtain the real-time motion Eqs. (\ref{RENGS}) in
the real time evolution.

\section{Equations of motion of $\protect\theta _{0}$ and $\Lambda _{1}$
\label{AppendixWN}}

In this Appendix, we derive the motion Eq. (\ref{TL}) from the projected Schr%
\"{o}dinger Eq. (\ref{SESP}). The evolution of $U_{\mathrm{GS}}(t)$ obeys%
\begin{equation}
i\partial _{t}U_{\mathrm{GS}}(t)=\bar{H}_{\mathrm{MF}}U_{\mathrm{GS}}(t),
\label{WN1}
\end{equation}%
where $U_{\mathrm{GS}}(t)=e^{iR^{T}\sigma \Delta _{R}/2}V_{\mathrm{GS}}(t)$
is determined by%
\begin{equation}
V_{\mathrm{GS}}(t)=e^{i\theta _{0}(t)}e^{b^{\dagger }\Lambda _{1}b^{\dagger
}}e^{b^{\dagger }\Lambda _{2}b}e^{b\Lambda _{3}b},
\end{equation}%
and $\Lambda _{1,3}$ are symmetric matrices.

It follows from Eq. (\ref{WN1}) that the motion equation of $V_{\mathrm{GS}%
}(t)$ is%
\begin{equation}
i\partial _{t}V_{\mathrm{GS}}(t)=(\frac{1}{4}R^{T}h_{b}R+\delta E_{k})V_{%
\mathrm{GS}}(t),  \label{WN2}
\end{equation}%
where $\delta E_{k}=E_{k}-\Delta _{R}^{T}h_{\Delta }/4-tr(h_{b}\Gamma _{b})/4
$, and the last term in $\delta E_{k}$ originates from removing the normal
ordering of the first quadratic term in Eq. (\ref{WN2}).

The left- hand side of Eq. (\ref{WN2}) is%
\begin{eqnarray}
i\partial _{t}V_{\mathrm{GS}}(t) &=&-\partial _{t}\theta _{0}(t)e^{i\theta
_{0}(t)}e^{b^{\dagger }\Lambda _{1}b^{\dagger }}e^{b^{\dagger }\Lambda
_{2}b}e^{b\Lambda _{3}b}+b^{\dagger }i\partial _{t}\Lambda _{1}b^{\dagger
}e^{i\theta _{0}(t)}e^{b^{\dagger }\Lambda _{1}b^{\dagger }}e^{b^{\dagger
}\Lambda _{2}b}e^{b\Lambda _{3}b}  \notag \\
&&+ie^{i\theta _{0}(t)}e^{b^{\dagger }\Lambda _{1}b^{\dagger }}(\partial
_{t}e^{b^{\dagger }\Lambda _{2}b})e^{b\Lambda _{3}b}+ie^{i\theta
_{0}(t)}e^{b^{\dagger }\Lambda _{1}b^{\dagger }}e^{b^{\dagger }\Lambda
_{2}b}(b\partial _{t}\Lambda _{3}b)e^{b\Lambda _{3}b}.  \label{dV}
\end{eqnarray}%
Using Eq. (\ref{dJ}) in Appendix \ref{Appendixtangent}, we obtain the time
derivative in the third term of Eq. (\ref{dV}) as%
\begin{equation}
\partial _{t}e^{b^{\dagger }\Lambda _{2}b}=b^{\dagger }(\partial
_{t}e^{\Lambda _{2}})e^{-\Lambda _{2}}be^{b^{\dagger }\Lambda _{2}b}.
\end{equation}%
In Eq. (\ref{dV}), all exponential operators can be moved to the most right
side by the relations%
\begin{eqnarray}
e^{b^{\dagger }\Lambda _{1}b^{\dagger }}b &=&(b-2\Lambda _{1}b^{\dagger
})e^{b^{\dagger }\Lambda _{1}b^{\dagger }},  \notag \\
e^{b^{\dagger }\Lambda _{2}b}b &=&e^{-\Lambda _{2}}be^{b^{\dagger }\Lambda
_{2}b},
\end{eqnarray}%
which leads to%
\begin{eqnarray}
i\partial _{t}V_{\mathrm{GS}}(t) &=&\{-\partial _{t}\theta
_{0}-2itr[e^{-\Lambda _{2}^{T}}(\partial _{t}\Lambda _{3})e^{-\Lambda
_{2}}\Lambda _{1}]  \notag \\
&&+ib^{\dagger }[(\partial _{t}e^{\Lambda _{2}})e^{-\Lambda _{2}}-4\Lambda
_{1}e^{-\Lambda _{2}^{T}}(\partial _{t}\Lambda _{3})e^{-\Lambda _{2}}]b
\notag \\
&&+ib^{\dagger }[\partial _{t}\Lambda _{1}-2(\partial _{t}e^{\Lambda
_{2}})e^{-\Lambda _{2}}\Lambda _{1}+4\Lambda _{1}e^{-\Lambda
_{2}^{T}}(\partial _{t}\Lambda _{3})e^{-\Lambda _{2}}\Lambda _{1}]b^{\dagger
}  \label{VL} \\
&&+ibe^{-\Lambda _{2}^{T}}(\partial _{t}\Lambda _{3})e^{-\Lambda _{2}}b\}V_{%
\mathrm{GS}}(t).  \notag
\end{eqnarray}

The right-hand side of Eq. (\ref{WN2}) is%
\begin{equation}
(\frac{1}{4}R^{T}h_{b}R+\delta E_{k})V_{\mathrm{GS}}(t)=(b^{\dagger }\mathbf{%
\omega }_{b}b+\frac{1}{2}b^{\dagger }\mathbf{\varpi }b^{\dagger }+\frac{1}{2}%
b\mathbf{\varpi }^{\dagger }b+\frac{1}{2}tr\mathbf{\omega }_{b}+\delta
E_{k})V_{\mathrm{GS}}(t),  \label{VR}
\end{equation}%
where the matrices $\mathbf{\omega }_{b}$ and $\mathbf{\varpi }$ are defined
by%
\begin{equation}
\left(
\begin{array}{cc}
\mathbf{\omega }_{b} & \mathbf{\varpi } \\
\mathbf{\varpi }^{\dagger } & \mathbf{\omega }_{b}^{T}%
\end{array}%
\right) =\frac{1}{2}W_{b}^{\dagger }h_{b}W_{b}.
\end{equation}%
Comparing Eqs. (\ref{VL}) and (\ref{VR}), we obtain the equations of motion%
\begin{eqnarray}
\partial _{t}\theta _{0} &=&-\delta E_{k}-\frac{1}{2}tr\mathbf{\omega }%
_{b}-tr(\mathbf{\varpi }^{\dagger }\Lambda _{1}),  \notag \\
i\partial _{t}\Lambda _{1} &=&\frac{1}{2}\mathbf{\varpi }+\mathbf{\omega }%
_{b}\Lambda _{1}+\Lambda _{1}\mathbf{\omega }_{b}^{T}+2\Lambda _{1}\mathbf{%
\varpi }^{\dagger }\Lambda _{1},  \notag \\
i\partial _{t}e^{\Lambda _{2}} &=&\mathbf{\omega }_{b}e^{\Lambda
_{2}}+2\Lambda _{1}\mathbf{\varpi }^{\dagger }e^{\Lambda _{2}},  \notag \\
i\partial _{t}\Lambda _{3} &=&\frac{1}{2}e^{\Lambda _{2}^{T}}\mathbf{\varpi }%
^{\dagger }e^{\Lambda _{2}}.
\end{eqnarray}

\end{widetext}


\begin{thebibliography}{999}
\bibitem{footnote} Note that the minimum does not necessarily has to be
global.
\end{thebibliography}
\end{document}